%% file: main.tex
\documentclass[10pt,journal,compsoc]{IEEEtran}

\usepackage{cite}
\usepackage{multirow}
\usepackage{subfigure}
\usepackage{amssymb}
\usepackage{xspace}
\usepackage{epsf}
\usepackage{setspace}
\usepackage{caption}
\usepackage{comment}
\usepackage{color}
\usepackage{balance}
\usepackage{url,epsfig,array}
\usepackage{amsmath}
\usepackage{epstopdf}
\usepackage{cases}
\usepackage{bm}
\usepackage{tikz}
\usepackage{threeparttable}
\usepackage{booktabs}

\usepackage{xcolor}

\usepackage{graphicx}
\usepackage{textcomp}

\def\BibTeX{{\rm B\kern-.05em{\sc i\kern-.025em b}\kern-.08em
    T\kern-.1667em\lower.7ex\hbox{E}\kern-.125emX}}

\usepackage{colortbl}
\usepackage{array}

\usepackage{graphicx}
\usepackage{epstopdf}
\usepackage{cite}
\usepackage{times}
\usepackage{dsfont}
\usepackage{cite}
\usepackage{times}
\usepackage{pifont}
\usepackage{multirow}   
\usepackage{tikz}
\usetikzlibrary{shapes.geometric,calc}
   
\usepackage{amsthm}
\usepackage{graphicx}
\usepackage{subfigure}
\usepackage{ifpdf}
\usepackage{epsfig}
\usepackage{latexsym}
\usepackage{amsfonts}
\usepackage{amssymb}
\usepackage{paralist}
\usepackage{comment}
\usepackage{xspace}
\usepackage{mathrsfs}
\usepackage{amssymb}
\usepackage{setspace}

\usepackage{amsmath,amssymb,amsfonts}

\usepackage{soul}
\sethlcolor{yellow}
\soulregister{\cite}7 
\soulregister{\citep}7 
\soulregister{\citet}7 
\soulregister{\ref}7 
\soulregister{\pageref}7 

\usepackage{subeqnarray}
\usepackage{amssymb}
\usepackage{url,epsfig,array}
\usepackage{leftidx}
\usepackage{amsmath}
\usepackage[T1]{fontenc}
\usepackage{aecompl}

\usepackage{calligra}
\usepackage{amsmath} 
\usepackage[linesnumbered, ruled]{algorithm2e}

\usepackage{makecell}

\newcommand*{\circled}[1]{\lower.7ex\hbox{\tikz\draw (0pt, 0pt)%
    circle (.5em) node {\makebox[1em][c]{\small #1}};}}

\def\ie{\textit{i.e.}\xspace}
\def\etal{\textit{et al.}\xspace}

\def\eg{\textit{e.g.}\xspace}

\def\st{\xspace\textbf{s.t.}\xspace}

\begin{document}
\title{Towards Communication-Efficient Decentralized Federated Graph Learning over Non-IID Data}
 \author{
	Shilong Wang,~
	Jianchun Liu,~\IEEEmembership{Member,~IEEE,~ACM,}
	Hongli Xu,~\IEEEmembership{Member,~IEEE,}~
    Chenxia Tang,\\
    Qianpiao Ma,~
    Liusheng Huang,~\IEEEmembership{Member,~IEEE}
 	\IEEEcompsocitemizethanks{
 		\IEEEcompsocthanksitem S. Wang, J. Liu, H. Xu, C. Tang, and L. Huang are with the School of Computer Science and Technology, University of Science and Technology of China, Hefei, Anhui, China, 230027, and also with Suzhou Institute for Advanced Research, University of Science and Technology of China, Suzhou, Jiangsu, China, 215123. E-mails: shilongwang@mail.ustc.edu.cn; jcliu17@ustc.edu.cn; xuhongli@ustc.edu.cn; tomorrowdawn@mail.ustc.edu.cn; lshuang@ustc.edu.cn.
            \IEEEcompsocthanksitem Q. Ma is with the School of Computer Science and Engineering, Nanjing University of Science and Technology, Nanjing, Jiangsu, China, 211111. E-mails: maqianpiao@njust.edu.cn.
 	}
 }

\markboth{IEEE Transactions on xxx, Vol., No., Nov. 2024}%
{Shell \MakeLowercase{\textit{et al.}}: Bare Advanced Demo of IEEEtran.cls for Journals}

\IEEEtitleabstractindextext{
\input{content/abstract.tex}
}

\maketitle
\IEEEdisplaynontitleabstractindextext
\IEEEpeerreviewmaketitle

\vspace{-2mm}
\section{Introduction}\label{sec:intro}
\input{content/intro2.tex}

\vspace{-1mm}
\section{Background and Motivation}\label{sec:background}
\input{content/background1}
\vspace{-2mm}
\section{Framework Design}\label{sec:framework}
\input{content/framework1}



\vspace{-1mm}
\section{Performance Evaluation}\label{sec:evaluation}
\input{content/evaluation.tex}

\section{Related Works}\label{sec:relatedworks}
\input{content/related}

\vspace{-1mm}
\section{Potential Limitations and Future Work}\label{sec:limitations}
\input{content/limitations}

\vspace{-1mm}
\section{Conclusions}\label{sec:conclusion}
\input{content/conclusion.tex}


\bibliographystyle{IEEEtran}
\bibliography{content/refs.bib}

\end{document}

%% file: content/abstract.tex
\begin{abstract}
Decentralized Federated Graph Learning (DFGL) overcomes potential bottlenecks of the parameter server in FGL by establishing a peer-to-peer (P2P) communication network among workers. 
However, while extensive cross-worker communication of graph node embeddings is crucial for DFGL training, it introduces substantial communication costs.
Most existing works typically construct sparse network topologies or utilize graph neighbor sampling methods to alleviate the communication overhead in DFGL.
Intuitively, integrating these methods may offer promise for doubly improving communication efficiency in DFGL. However, our preliminary experiments indicate that directly combining these methods leads to significant training performance degradation if they are jointly optimized.
To address this issue, we propose \textsc{Duplex}, a unified framework that jointly optimizes network topology and graph sampling by accounting for their coupled relationship, thereby significantly reducing communication cost while enhancing training performance in DFGL. To overcome practical DFGL challenges, \eg, statistical heterogeneity and dynamic network environments, \textsc{Duplex} introduces a learning-driven algorithm to adaptively determine optimal network topologies and graph sampling ratios for workers. Experimental results demonstrate that \textsc{Duplex} reduces completion time by 20.1\%--48.8\% and communication costs by 16.7\%--37.6\% to achieve target accuracy, while improving accuracy by 3.3\%--7.9\% under identical resource budgets compared to baselines.
\end{abstract}

%% file: content/intro2.tex
In many real-world applications such as e-commerce recommendation systems \cite{zhang2024review} and social network analysis \cite{yokotani2025predicting}, graph-structured data, comprising interconnected nodes and edges, has emerged as a fundamental representation for modeling complex relationships. For instance, industry leaders like Amazon and Alibaba represent user preferences for items as user-item interaction graphs to enable personalized recommendations \cite{sun2021pain, wu2024optimizing}.
To effectively process such data, Graph Convolutional Networks (GCNs) \cite{kipf2016semi, hamilton2017inductive, ren2024phase} have been proposed, demonstrating remarkable success in graph learning tasks. Fully realizing the potential of GCNs requires training on large volumes of graph data, which are continuously generated by users across a variety of edge devices (\eg, smartphones and laptops). Such data are usually privacy-sensitive, thereby raising significant concerns when collected for training purposes \cite{huang2024keystrokesniffer, liu2025adaptive}. 

To address these privacy issues, Federated Graph Learning (FGL) \cite{xie2021federated, zhang2021subgraph} has emerged as the de facto paradigm for collaboratively training GCNs across geo-distributed devices (\ie, workers) without exposing raw graph data. Traditional FGL frameworks typically adopt a Prameter Server (PS) architecture, in which massive workers communicate directly with the PS through wide area networks (WANs) \cite{liu2023finch, liu2024enhancing}.
As the number of workers skyrockets, the ingress bandwidth of the PS becomes a major bottleneck, as all workers must deliver their local
models to the server. As an alternative, Decentralized Federated Graph Learning (DFGL) \cite{pei2021decentralized, liu2021glint, liu2025accelerating, liu2022s} establishes a peer-to-peer (P2P) communication network among workers, enabling them to exchange local models with their neighbors. Since there is no need to transmit local models from workers to the PS, the risk of a single point failure at the PS is eliminated, and system scalability is notably improved.

However, the practical deployment of DFGL systems presents formidable challenges due to the unique characteristics of distributed graph learning.
For example, the Amazon product co-purchasing graph \cite{hu2020open} consists of over 3 million nodes and 60 million edges, representing products and their co-purchasing relationships, respectively. In a DFGL framework, each worker (representing a merchant on Amazon) maintains a localized subgraph that encapsulates multiple products. 
These subgraphs include both \emph{internal edges}, which connect products within a single worker, and \emph{external edges} that span across different workers, indicating that users purchase multiple products simultaneously.
The presence of external edges necessitates the exchange of hidden states of product nodes (\ie, node embeddings) among workers during local GCN training (as detailed in Section \ref{subsec:DFGL}). However, the stark mismatch between the enormous size of exchanged node embeddings and the limited bandwidth available on workers poses significant challenges in achieving efficient DFGL. For instance, federated training of the popular GraphSage model \cite{hamilton2017inductive} on the Amazon co-purchasing graph incurs hundreds of gigabytes of network traffic among workers. In contrast, typical WANs provide limited bandwidth ranging from $5$ to $20$ Mbps for workers \cite{zhou2024accelerating}, which is markedly inferior to that available in data centers (\eg, over 10 Gbps \cite{goyal2017accurate}). Consequently, the transmission of massive embedding data among workers often leads to prohibitively long communication times, thereby hindering the overall scalability and efficiency of DFGL systems.

Some recent arts have proposed various solutions to alleviate the high communication cost in DFGL. One common approach is network topology construction \cite{liu2021glint, liu2022s}. For example, Liu \etal \cite{liu2022s} introduced a novel framework named S-Glint, which constructs a sparse network topology for communication by selecting several neighbors with the highest performance contributions for each worker. By reducing the total number of communication links in DFGL, the overall communication overhead is mitigated. 
In addition to network topology construction, graph neighbor sampling \cite{chen2021fedgraph, du2022federated, li2024historical} is another approach for alleviating communication overhead. Specifically, graph neighbor sampling randomly selects a subset of nodes, rather than all nodes, in the graph data for model training, thereby reducing the number of node embeddings exchanged among workers. For instance, Du \etal \cite{du2022federated} analyzed the optimal sampling interval (\ie, the number of rounds between two adjacent sampling actions) to achieve the optimal trade-off between convergence rate and communication time. 


Intuitively, integrating network topology construction and graph neighbor sampling offers potential for doubly enhancing communication efficiency in DFGL.
This potential motivates us to investigate the benefits of combining these two distinct techniques. However, our preliminary experiments in Section \ref{subsec:joint_optimize} indicate that directly integrating these techniques may lead to significant degradation in training performance and yield suboptimal communication efficiency if they are not jointly optimized. In DFGL, these techniques are inherently coupled, that is, adjustments in one directly influence the other. For instance, under-sampling graph data in sparsely connected workers may lead to under-utilized bandwidth, while over-sampling in densely connected workers can result in communication redundancy. Furthermore, graph sampling affects the quality and diversity of information shared between workers. Without considering the communication network (\eg, its topology), isolated sampling strategies may fail to preserve critical information required for model convergence. To address these issues, we propose \textsc{Duplex}, a unified framework that jointly optimizes the network topology and graph sampling, thereby improving both communication efficiency and training performance in DFGL.
Specifically, \textsc{Duplex} accounts for the coupling between network topology construction and graph neighbor sampling by formulating the decision-making process as a coordinated configuration $\langle \mathbf{A}, \mathbf{R} \rangle$, where $\mathbf{A}$ is the adjacency matrix of the network topology and $\mathbf{R}$ denotes the set of graph sampling ratios (\ie, the proportions of graph nodes sampled for training) for workers.

However, achieving optimal communication efficiency and training performance (\eg, model accuracy and convergence rate) with \textsc{Duplex} proves challenging due to several practical issues (Section \ref{sec:challenges}). \textbf{First}, statistical heterogeneity in DFGL adversely affects the training performance of \textsc{Duplex}. In real-world scenarios, user data on workers are typically generated under different contexts, such as personal usage patterns and geographic variations. Consequently, the local graph data across workers are non-Independent and Identically Distributed (non-IID) \cite{wang2022accelerating, liu2022enhancing}. 
In non-IID scenarios, the local model gradients computed on individual workers constitute biased estimates of the theoretically optimal global gradient computed over the collective data, resulting in a notable deterioration of training performance.
\textbf{Second}, workers in DFGL usually connect to the network via wireless links, whose quality tends to fluctuate over time due to worker mobility and link instability \cite{waqas2018mobility}. Besides, concurrent applications on a worker may compete for limited bandwidth, further leading to dynamic fluctuations in the available network resources \cite{wang2022accelerating}. Due to varying network conditions, optimal configurations for network topology and sampling ratios may evolve as training progresses, necessitating adaptive adjustments rather than static solutions.
To address these challenges, \textsc{Duplex} introduces consensus distance, a metric that quantifies the discrepancy among local model parameters, to guide the decision-making process under non-IID conditions. Moreover, \textsc{Duplex} leverages a learning-driven algorithm based on deep reinforcement learning (DRL) to adaptively determine the optimal coordinated configuration $\langle \mathbf{A}, \mathbf{R} \rangle$ in response to dynamic network conditions.
Our contributions are summarized as follows:
\begin{itemize}
    \item We propose \textsc{Duplex}, an efficient DFGL framework that focuses on the joint optimization of network topology and graph sampling, aiming to reduce communication costs and improve training performance on non-IID graph data.
    \item 
    We propose an effective algorithm to adaptively determine the optimal network topology and sampling ratios for workers, simultaneously considering dynamic network conditions and heterogeneous data distributions among workers.
    \item The extensive experimental results demonstrate that \textsc{Duplex} can reduce the completion time by 20.1\%-48.8\%
    and communication cost by 16.7\%-37.6\% to reach the target accuracy, while improving model accuracy by 3.3\%-7.9\% under the same resource budgets, compared to the baselines.
\end{itemize}


%% file: content/background1.tex
In this section, we begin by outlining the fundamentals of GCNs and DFGL. Next, we identify key bottlenecks in DFGL and focus on the impact of network topology and graph sampling ratios on training efficiency and model accuracy. Furthermore, we demonstrate that optimizing these components in isolation leads to suboptimal performance, thereby motivating the need for a joint optimization approach. Finally, we discuss the challenges in achieving optimal performance for \textsc{Duplex}, particularly in dynamic network environments with non-IID graph data.

\subsection{Graph Convolutional Networks}
Graph-structured data (\eg, a citation network) represents entities (\eg, documents) and their relationships (\eg, citations) in a non-Euclidean space. Formally, such data can be represented as an undirected graph $G = (V,E)$, where $V$ is a set of nodes representing entities and $E$ is a set of edges representing relationships between nodes. Each node $v_i\in V$ is associated with a raw feature vector $x_i$. 
Graphs are irregular and non-Euclidean structures, posing challenges for traditional neural networks designed for grid-like data (\eg, images and sequences). This necessitates specialized architectures such as GCNs, which generalize deep learning to graph domains by exploiting inductive biases tied to graph topology.

GCNs operate through iterative message-passing mechanisms \cite{kipf2016semi, hamilton2017inductive}, where nodes aggregate information from their neighborhoods to refine their representations. Specifically, a GCN typically consists of $L$ graph convolutional layers. Let $h^{l}_i\in \mathbb{R}^{d_l}$ denote the feature vector (\ie, embedding) of node $v_i$ at layer $l$. The initial node feature vector $h^{0}_i$ is typically set to $x_i$. Each graph convolutional layer $l \in (1, 2, \dots, L)$ is responsible for transforming $\{h^{l-1}_i \mid v_i\in V\}$ into $\{h^{l}_i \mid v_i\in V\}$ through the graph convolution (GC) operation, which involves two stages: (1) Aggregation: For each node $v_i$, gather features from its neighbors $\mathcal{N}(v_i)$. (2) Update: Update the embedding of node $v_i$ by combining its previous state $h^{l-1}_i$ with the aggregated neighborhood features. Mathematically, the GC operation at layer $l$ is formalized as follows:
\begin{equation}
    \begin{gathered}
    \mathscr{E}_i^l = \text{AGG}(\{h^{l-1}_j  \mid \forall v_j\in\mathcal{N}(v_i)\}),\\
    h^{l}_i = \mathbf{U}^l(h^{l-1}_i \Vert \mathscr{E}_i^l),
    \end{gathered}\label{eq:gc}
\end{equation}
where $\text{AGG}(\cdot)$ is the aggregation function (\eg, element-wise mean or sum of vectors), $\mathbf{U}^l(\cdot)$ is the state update function at layer $l$ (\eg, a single-layer perceptron followed by a nonlinear transformation), and $\Vert$ denotes concatenation. Stacking $L$ such layers enables nodes to integrate information from $L$-hop neighborhoods.

After propagating through $L$ graph convolutional layers, the GCN makes
predictions for downstream tasks as follows:
\begin{equation}\label{eq:predict}
    \hat{y}_{\mathcal{S}} = \mathbf{W}(\{h_i^L \mid \forall v_i\in \mathcal{S}\}),
\end{equation}
where $\mathbf{W}(\cdot)$ can be a concatenation function followed by a multi-layer perceptron, and $\mathcal{S}\subseteq V$ is a node set used for prediction.
Given a mini-batch of nodes $\mathcal{B}$ uniformly sampled from $G$, the training loss of a GCN is defined as:
\begin{equation}\label{loss_of_GCN}
    \mathcal{L}_\mathcal{B} = F(\omega; y_{\mathcal{B}}, \hat{y}_{\mathcal{B}}),
\end{equation}
where $\omega$ denotes the learnable parameters in the GCN, $y_{\mathcal{B}}$ is the set of labels for nodes in $\mathcal{B}$, and $F(\cdot)$ can be any feasible loss function (\eg, Cross Entropy and L2-Norm). For brevity, we denote 
$F(\omega; y_{\mathcal{B}}, \hat{y}_{\mathcal{B}})$ simply as $F(\omega; \mathcal{B})$ hereafter.

\begin{figure}[t]\centering
    \includegraphics[width=0.435\textwidth]{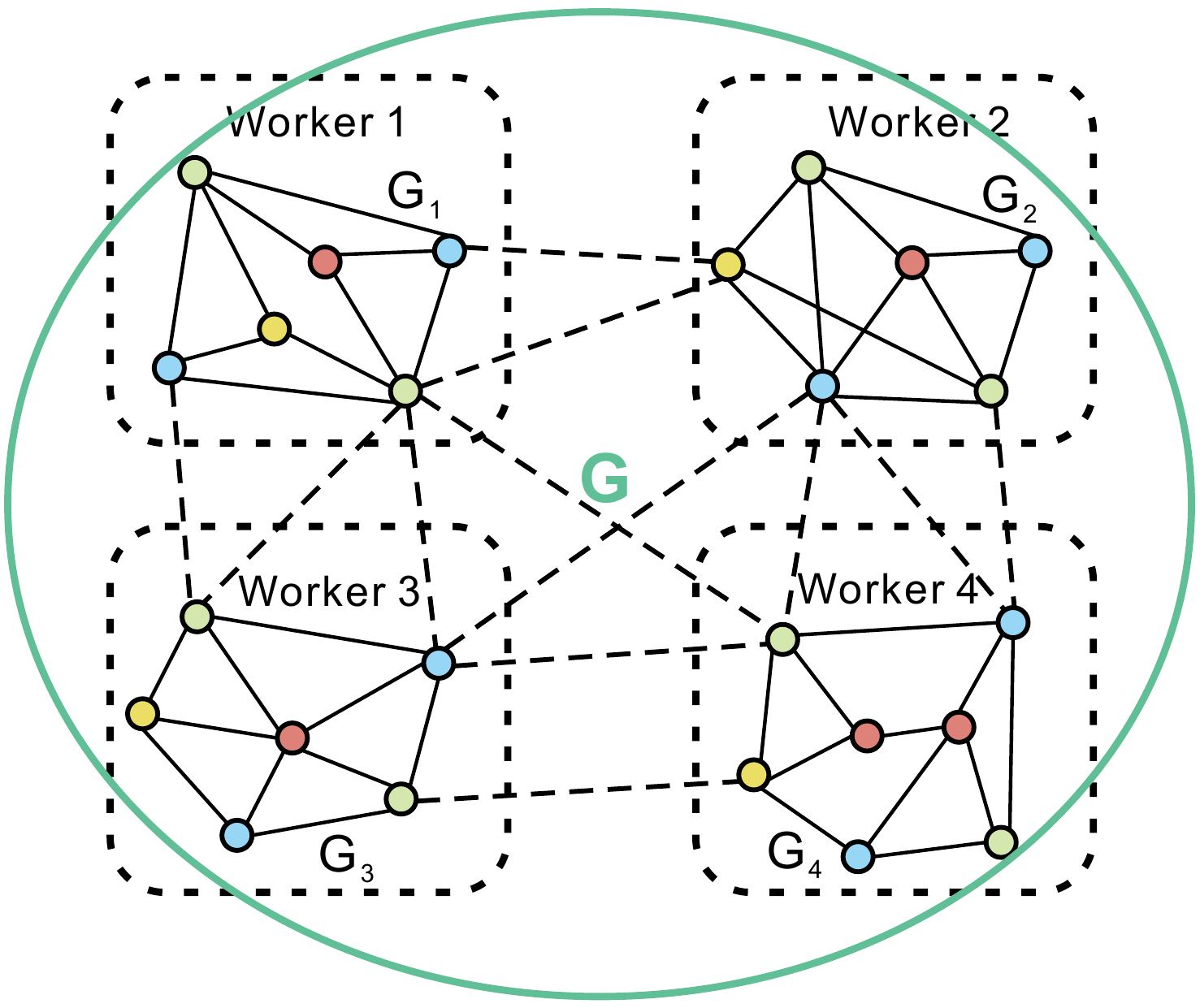}
    \caption{Illustration of local subgraphs on workers, where the dashed lines between nodes represent the external edges across different subgraphs. The colors of nodes represent different node classes.}\label{fig:dfgl}
\end{figure}

\subsection{Decentralized Federated Graph Learning}\label{subsec:DFGL}
In a network with a set of workers $\mathcal{M}=\{1,2,...,m \}$, each worker $i \in \mathcal{M}$ trains a local GCN model on its local subgraph $G_{i}$, which 
is assumed to be partitioned from an underlying global graph $G$, as shown in Fig. \ref{fig:dfgl}. 
Unlike other data types, such as images, that are typically independent across  different workers, graph data inherently contain strong inter-dependencies manifested through external edges linking nodes across distinct subgraphs. Consequently, when performing GC operations, each worker must exchange node embeddings with others due to the existence of external edges. For instance, updating the embedding of a yellow node on worker 2 requires the aggregating features from adjacent blue and green nodes that reside on worker 1.
Neglecting such cross-worker interactions can lead to significant reductions in both the accuracy and convergence speed of the learned GCN models \cite{liu2021glint, chen2021fedgraph, du2022federated}.
Moreover, in practical implementations of DFGL, the graph data across different workers are characterized by heterogeneous distributions due to factors such as varying personal usage patterns and geographic locations.
For example, different companies often maintain extensive social network data that exhibit substantial heterogeneity (\eg, diverse node classes
and varying connection rules),  stemming from different data collection methods
and objectives.

\begin{figure*}[t]\centering
    \begin{minipage}[t]{0.325\linewidth}\centering
        \subfigure[Model accuracy ($\alpha=10.0$).]{
            \label{fig:motiv_acc_epoch}
            \includegraphics[width=0.7\textwidth]{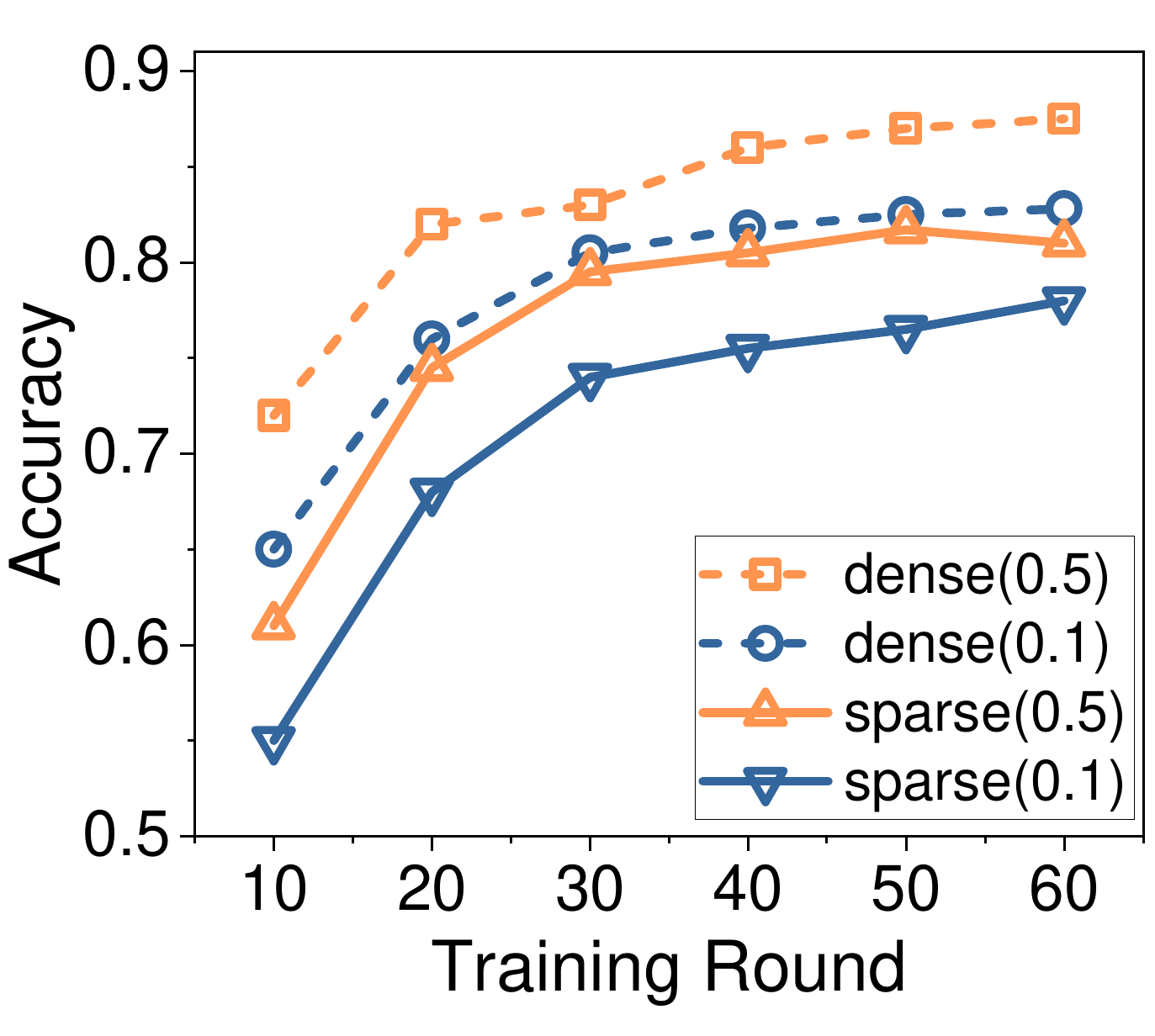}
        }
    \end{minipage}
    \begin{minipage}[t]{0.325\linewidth}\centering
        \subfigure[Training Time ($\alpha=10.0$).]{
            \label{fig:motiv_time_epoch}
            \includegraphics[width=0.7\textwidth]{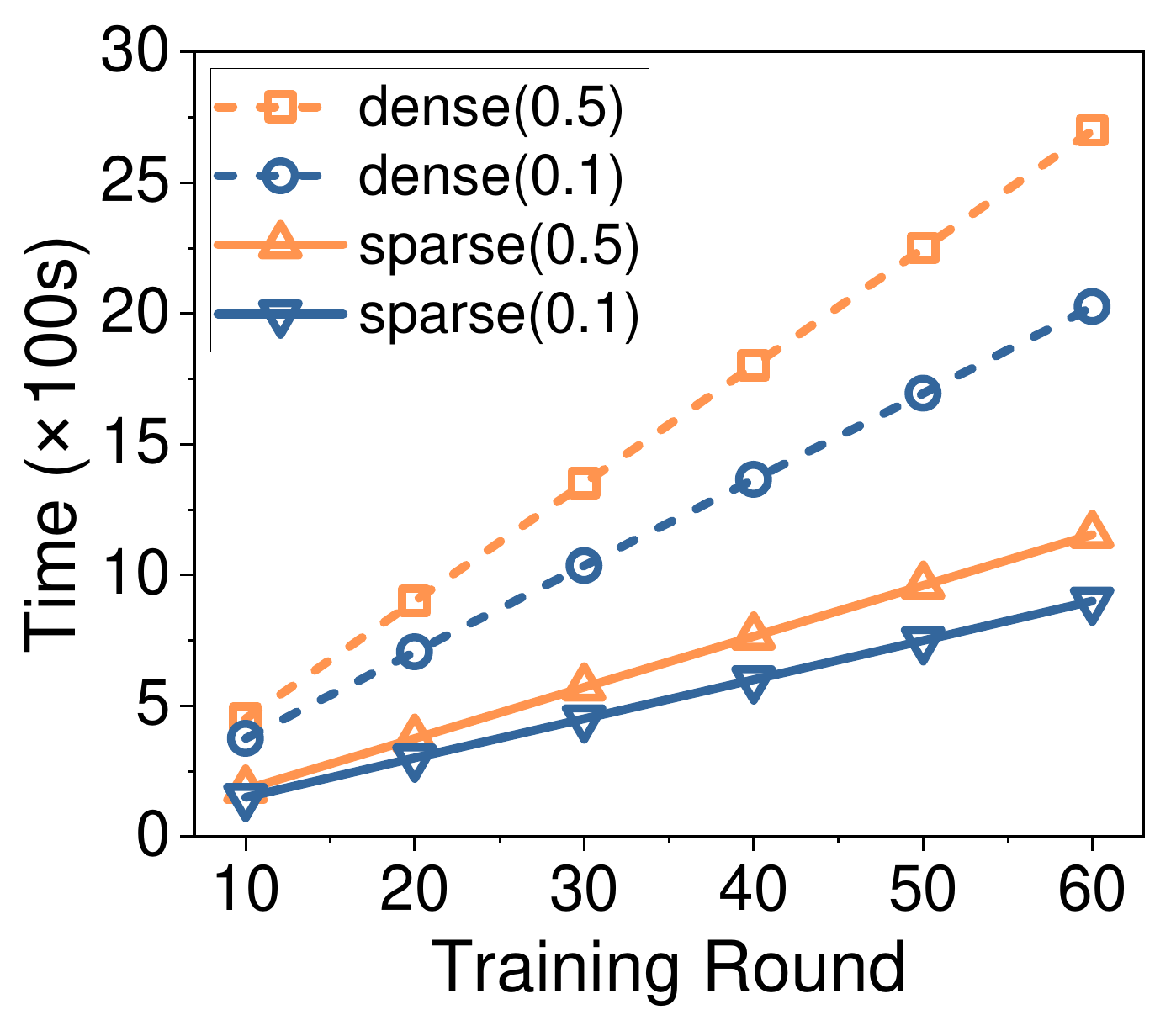}
        }
    \end{minipage}
    \begin{minipage}[t]{0.325\linewidth}\centering
        \subfigure[Communication Cost ($\alpha=10.0$).]{
            \label{fig:motiv_comm_acc}
            \includegraphics[width=0.7\textwidth]{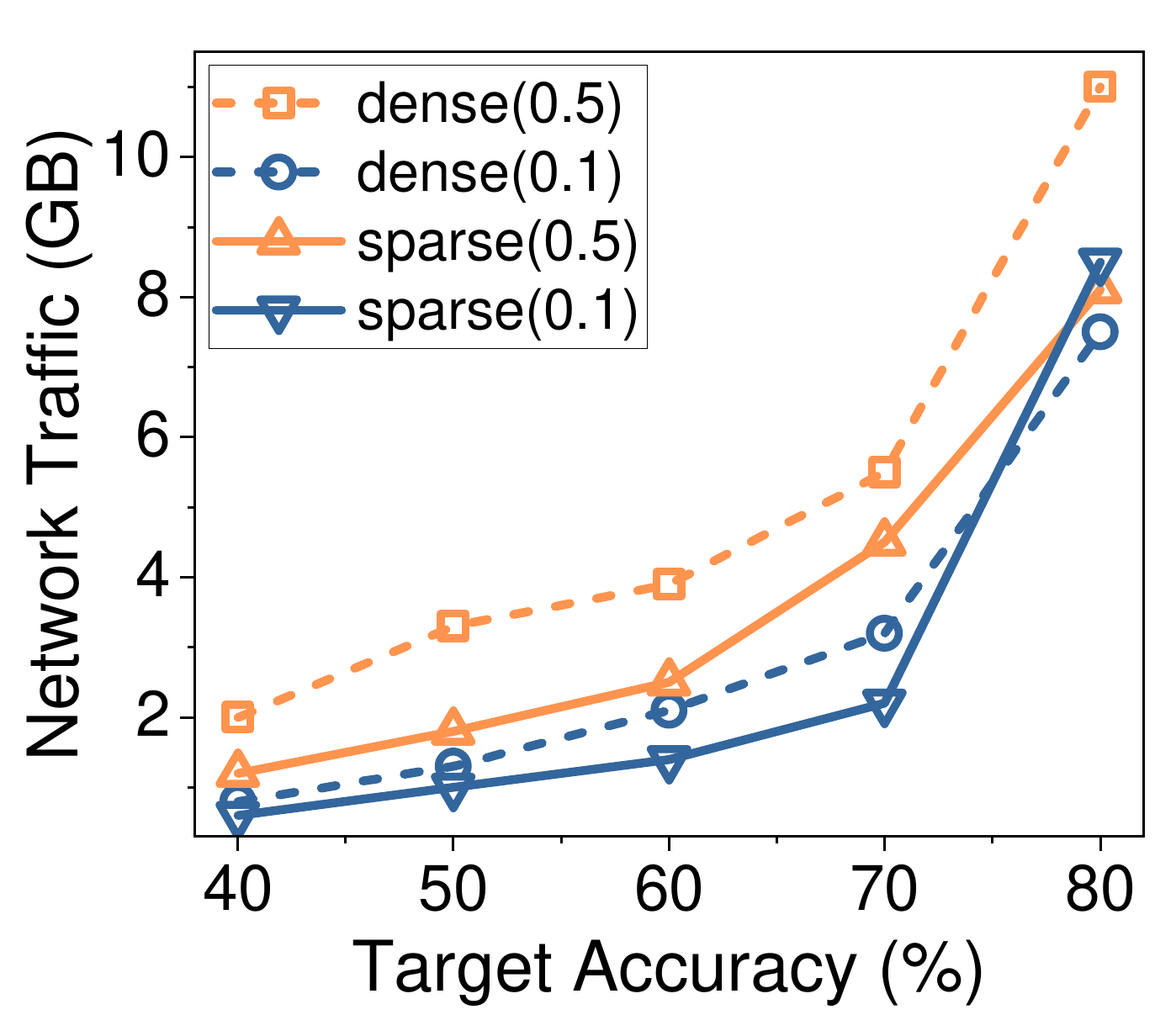}
        }
    \end{minipage}
    \caption{
    Impact of network topology/graph sampling ratio on accuracy and resource consumption of model training.}\label{fig:topology}
\end{figure*}
Different from FGL, which relies on a PS to aggregate local models in each training round, DFGL employs a decentralized approach in which workers directly exchange their local GCN model parameters with neighboring workers after local training. This design effectively mitigates the risk of communication bottlenecks associated with a central PS.
Let $f_{i}(\omega_{i}):=\mathbb{E}_{\mathcal{B}_{i}\sim G_{i}}F_{i}(\omega_{i};\mathcal{B}_{i})$ denote the local objective function of worker $i$, where $\mathcal{B}_{i}$ represents a mini-batch of nodes randomly sampled from the local subgraph $G_{i}$, and $\omega_{i}$ is the local GCN model parameters of worker $i$. In general, the training process of DFGL is formulated as the optimization of the following objective function \cite{wang2019matcha}:
\begin{equation}\label{objective}
    f^{\ast}:=\underset{\{\omega_{i} \mid i \in \mathcal{M}\}}{\min}[\frac{1}{m}\sum\limits_{i\in \mathcal{M}}f_{i}(\omega_{i})].
\end{equation}
This setting covers the important cases of empirical risk minimization in DFGL.
\subsection{Motivations for Framework Design}\label{subsec:motivations}

Network topology and the number of graph node embeddings exchanged across workers during training affect both training performance and communication efficiency in DFGL. To gain a deeper understanding, we conduct several sets of experiments on 10 workers using the Reddit dataset
\cite{hamilton2017inductive}, where each worker trains a two-layer GCN model \cite{kipf2016semi}. Concretely, we set the number of neighbors per worker to 2 and 9 to simulate sparse and dense network topologies, respectively. We establish several configurations by combining the two network topologies (\ie, sparse and dense) with different graph sampling ratios (\eg, 0.1, 0.5 and 1.0), which represent the proportions of aggregated neighboring node features for each node during the GC operation. The bandwidth for each worker fluctuates randomly between $1$ and $20$ Mbps \cite{liao2024mergesfl, liao2024parallelsfl}.
To simulate real-world non-IID graph data, we follow the prior work \cite{he2021fedgraphnn} and partition Reddit using the Dirichlet distribution $Dir(\alpha)$. In particular, we independently sample a subgraph $G_{i}\sim Dir_{i}(\alpha)$ from the global graph $G$ and allocate $G_{i}$ to each worker $i$, where $\alpha$ determines the degree of non-IID data. A lower value of $\alpha$ generates a higher label distribution shift.

\begin{table}[t]
    \centering
    \caption{Resource (time and network traffic) consumption for local computing and exchanging local models/node embeddings among workers.} \label{table:communication_bottleneck}
    \resizebox{88mm}{!}{
        \centering
        \begin{tabular}{c|c|c|c}
            \hline
            Resource & \makecell[c]{Local\\Computing} & \makecell[c]{Exchanging\\Local Models} & \makecell[c]{Exchanging\\Node Embeddings} \\
            
            \hline 

            Time & 313 s & 115 s & 3160 s \\
            \hline

            Network Traffic & N/A & 0.91 GB & 15.2 GB \\
            
            
            \hline
        \end{tabular}
    }
\end{table}
\subsubsection{Communication Bottleneck in DFGL}\label{sec:comm_bottleneck}
The first set of experiments involves training the GCN model on a dense network topology with a graph sampling ratio of $1.0$. We record the breakdown of total network traffic on all workers and the time required to reach a target training accuracy of 80\%.
The results in Table \ref{table:communication_bottleneck} show that the communication time for exchanging node embeddings among workers is over $10\times$ longer than the local computing time. Besides, since the size of a single GCN model is only approximately 0.5-2MB, the network traffic related to exchanging models among workers is negligible compared to that required for exchanging graph node embeddings. Therefore, it is critical to reduce both the communication time and the network traffic associated with exchanging graph node embeddings during training, as these represent major efficiency bottlenecks in DFGL.


\subsubsection{Impact of Network Topology and Sampling Ratios}
We further investigate the impact of network topology and graph sampling ratios on training efficiency and model accuracy. The experimental results in Fig. \ref{fig:motiv_acc_epoch} indicate that a denser network topology combined with with a higher graph sampling ratio improves model accuracy. For example, on the dense topology, training with a sampling ratio of 0.5 enhances the accuracy by 5.1\% compared to using a ratio of 0.1.
On the other hand, under the same sampling ratio (\eg, 0.5), the accuracy on the dense topology increases by 6.5\% compared to that on the sparse topology.
However, these accuracu improvements come at a cost: resource consumption, such as training time and network traffic, increases significantly with denser network topologies and higher sampling ratios, as shown in Figs. \ref{fig:motiv_time_epoch} and \subref{fig:motiv_comm_acc}.
Therefore, it is imperative for DFGL frameworks to optimize network topologies and graph sampling strategies, as they critically mediate the trade-off between training efficiency and model performance.

\begin{figure}[t]\centering
    \begin{minipage}[t]{0.485\linewidth}\centering
        \subfigure[Model accuracy ($\alpha=10.0$).]{
            \includegraphics[width=1.0\textwidth]{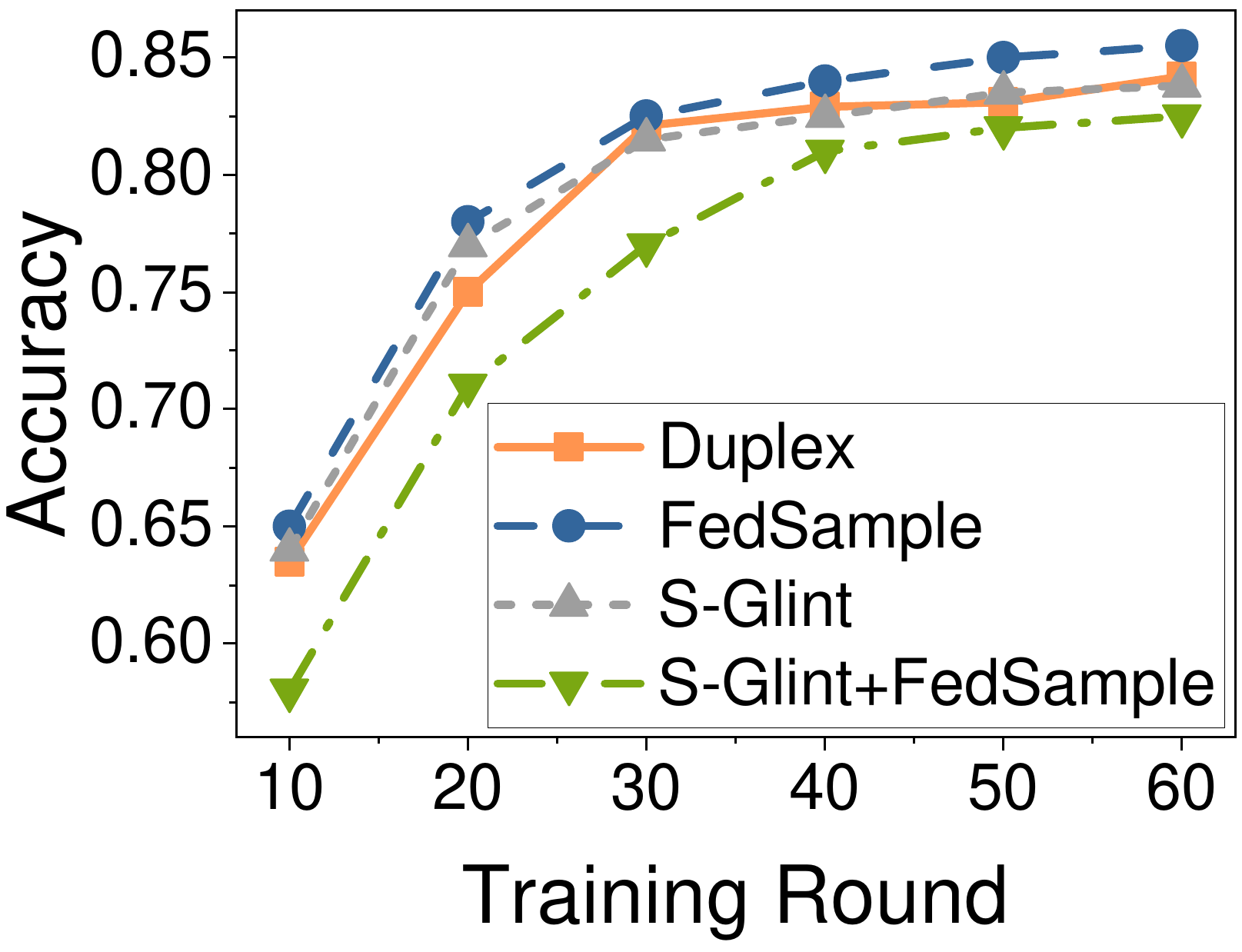}\label{fig:joint-acc}
        }
    \end{minipage}
    \begin{minipage}[t]{0.485\linewidth}\centering
        \subfigure[Communication cost ($\alpha=10.0$).]{
            \includegraphics[width=1.0\textwidth]{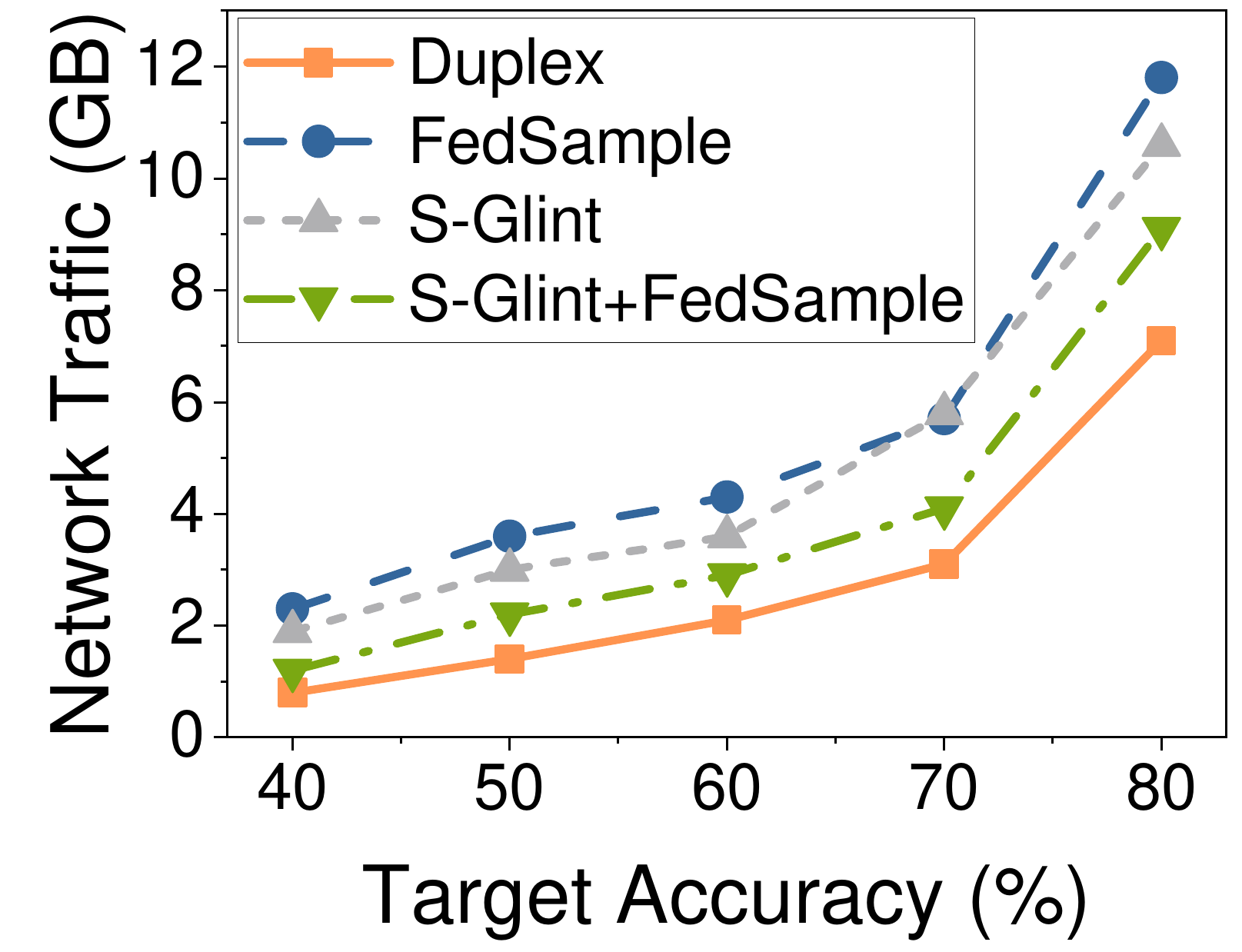}\label{fig:joint-comm}
        }
    \end{minipage}
    \caption{Performance comparison of four methods on Reddit.}\label{fig:joint}
\end{figure}

\subsubsection{Importance of Joint Optimization for Network Topology and Sampling Ratio}\label{subsec:joint_optimize}
Directly combining existing techniques for network topology construction and graph neighbor sampling will lead to significant performance degradation, as the two techniques are interdependent yet not jointly optimized. 
In DFGL, network topology and graph sampling ratio jointly impact the volume of node embeddings exchanged across workers during training, which determine the overall communication cost.
Furthermore, varying network topologies change the interconnection schemes among local subgraphs on workers, leading to distinct optimal graph sampling strategies for each configuration. For instance, a dense topology may amplify the benefits of high sampling ratios by increasing the shared
information but incurs significant communication overhead. Besides, a sparse topology can reduce communication costs but might diminish the benefits of high sampling ratios due to reduced data diversity and quality.

Existing works always assume that sampling strategies
are agnostic to network topology, treating communication as a homogeneous process among workers. Consequently, prior methods optimize \emph{either} network topology \emph{or} graph sampling in isolation, with no framework jointly addressing both at the same time. However, in P2P networks, communication patterns inherently vary with topology, and sampling strategies that ignore this interdependence can result in inefficiencies and degraded training performance. Therefore, jointly optimization of network topology and graph sampling ratios is crucial yet challenge in DFGL. To validate this insight, we conduct experiments to compare the performance of four approaches: 1) S-Glint \cite{liu2022s}, which constructs a sparse network topology without graph neighbor sampling, 2) FedSample \cite{chen2021fedgraph}, which assigns appropriate graph sampling ratios to workers without considering the impact of network topology, 3) S-Glint+FedSample, which is a direct combination of S-Glint and FedSample, optimizing network topology and sampling ratios separately, and 4) \textsc{Duplex}, which coordinates topology construction and graph sampling through a unified decision-making process.

The results presented in Fig. \ref{fig:joint-acc} clearly demonstrate that \textsc{Duplex}, S-Glint and FedSample achieve superior training performance compared to S-Glint+FedSample.
For example, at training round $60$, \textsc{Duplex}, S-Glint and FedSample improve model accuracy by $2.1\%$, $1.8\%$ and $3.6\%$, respectively, over S-Glint+FedSample. Besides, as shown in Fig. \ref{fig:joint-comm}, while S-Glint+FedSample reduces communication overhead in DFGL compared to S-Glint or FedSample, significant optimization space remains. By jointly optimizing network topology and graph sampling ratios, \textsc{Duplex} reduces network traffic by $22.1\%$ compared to S-Glint+FedSample. These results highlight the importance of joint optimization in leveraging the benefits of both network topology construction and graph sampling in DFGL.

\begin{figure}[t]\centering
    \includegraphics[width=0.35\textwidth]{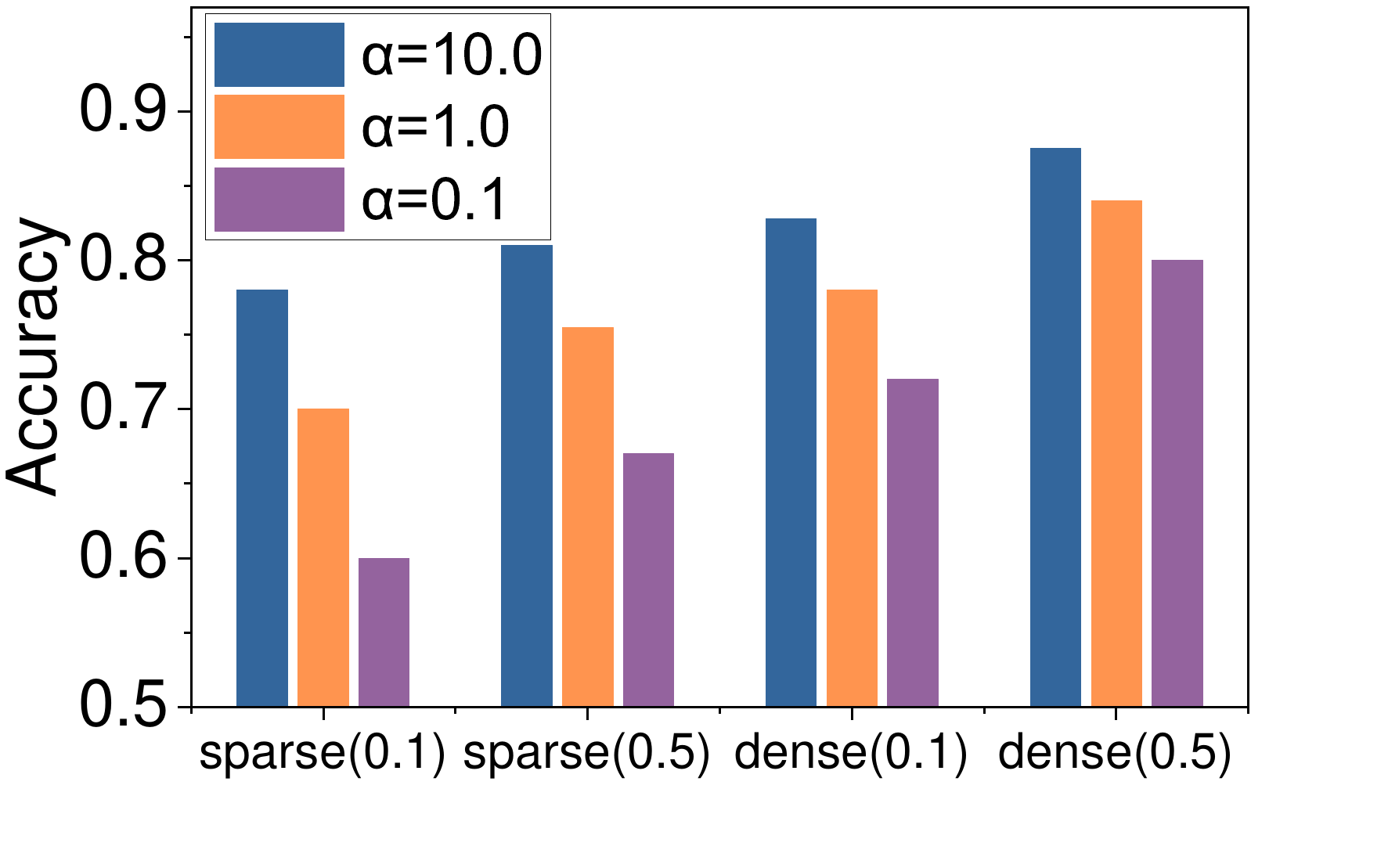}
    \caption{Harm of Non-IID data on model accuracy.}\label{fig:motiv_noniid_acc}
\end{figure}

\subsection{Practical Challenges in \textsc{Duplex}}\label{sec:challenges}
In DFGL, graph data distributed across workers are usually non-IID, which significantly degrades training performance. To illustrate this challenge, we conduct experiments using three non-IID settings, \ie, $\alpha=0.1$ (high non-IIDness), $\alpha=1.0$ (middle non-IIDness) and $\alpha=10.0$ (low non-IIDness). As shown in Fig. \ref{fig:motiv_noniid_acc}, model accuracy declines sharply as non-IIDness increases from low to high. For example, on the sparse topology with a graph sampling ratio of 0.5, model accuracy decreases from 81.2\% to 67.1\%. Furthermore, the extent of accuracy degradation varies with network topology and graph sampling ratios. For example, dense topologies with higher sampling ratios exhibit greater resilience to non-IID data compared to sparse topologies with lower ratios. These findings underscore that network topologies and sampling strategies differentially influence robustness to data heterogeneity.
Therefore, optimizing both components in \textsc{Duplex} is critical to improving training performance on non-IID data.

Besides, due to workers' mobility, communication link instability, and bandwidth competition among multiple applications, the available bandwidth for each worker varies dynamically over time \cite{wang2022accelerating}. Consequently, the fixed
network topology and graph sampling ratio may not be efficient all the time as training progresses under dynamic network conditions. Therefore, it is crucial yet challenging for \textsc{Duplex} to adaptively adjust the configurations $\langle \mathbf{A}, \mathbf{R} \rangle$ in response to time-varying communication speeds of workers.

%% file: content/framework1.tex
This section outlines the design of \textsc{Duplex} and elaborates on how it enhances communication efficiency and training performance in DFGL, especially under dynamic network conditions and non-IID graph data. First, we provide an overview of \textsc{Duplex} in Section~\ref{subsec:overview}. Next, we detail the methods employed for the joint optimization of network topology and graph sampling ratios (Section~\ref{sec:configuration_update}). In addition, we describe the procedures for local GCN training (Section~\ref{subsec:local_GCN_training}) and model aggregation (Section \ref{subsec:model_aggregation}), both of which are executed based on the optimized network topology and sampling ratios.





\subsection{Overview of \textsc{Duplex}}\label{subsec:overview}
We consider a typical DFGL setting comprising a set of workers $\mathcal{M}=\{1,2,...,m \}$, where each worker $i\in\mathcal{M}$ maintains a local subgraph $G_{i}=(V_i, E_i)$ as part of the global graph $G=(V, E)$. 
Besides, a logical control worker (\ie, the coordinator) is responsible for collecting global information about training statuses and network conditions \cite{zhou2021communication, xu2021decentralized}.
Note that the coordinator differs significantly from the PS in FGL because it does not aggregate local models, and hence will not become the communication bottleneck. Since the size of such information (\eg, $100$--$300$ KB \cite{lyu2019optimal}) is
much smaller than that of model parameters, it is reasonable to ignore the associated costs (\eg, network traffic and communication time) for information collection. Furthermore, any worker can act as the coordinator.
The training procedure of \textsc{Duplex} involves multiple rounds, each encapsulating three main steps: {\large\textcircled{\footnotesize 1}} \textbf{Configuration Update.} At the beginning of each round, the coordinator adaptively constructs the network topology and determines the graph sampling ratios according to the information collected from workers (\eg, training statuses and network conditions). 
{\large\textcircled{\footnotesize 2}} \textbf{Local GCN Training.} After receiving the updated configurations of topology and ratios from the coordinator, each worker iteratively updates the local GCN model over its local subgraph for $\tau$ iterations. In each iteration, workers sample a subset of graph nodes according to their respective graph sampling ratios for stochastic gradient
descent, rather than using the full set of graph nodes. {\large\textcircled{\footnotesize 3}} \textbf{Model Aggregation.} Once local GCN training is completed, each worker exchanges its local GCN model parameters with its neighbors in the constructed network topology. Finally, each worker aggregates the received model parameters from its neighbors.

\subsection{Configuration Update}\label{sec:configuration_update}
We first introduce the consensus distance, which is a crucial metric in the coordinator's decision-making process on non-IID graph data (Section~\ref{subsec:consensus_distance}). Then, we present the problem of joint optimization for ne\textbf{T}work t\textbf{O}pology and graph sa\textbf{M}pling r\textbf{A}tio\textbf{S}, termed TOMAS, aiming to minimize communication cost while ensuring satisfactory training performance in dynamic network environments (Section~\ref{subsec:problem_formulation}). Finally, we design an efficient algorithm to solve this problem in Sections~\ref{subsec:DDPG_agent} and \ref{subsec:learning-driven_algorithm}.

\subsubsection{Consensus Distance}\label{subsec:consensus_distance}
Since the graph data on workers is non-IID, local GCN models trained on different workers exhibit significant heterogeneity. To guide the coordinator in jointly optimizing the network topology and graph sampling ratios, we introduce the consensus distance \cite{kong2021consensus, koloskova2019decentralized}, a key metric in \textsc{Duplex} for quantifying the heterogeneity of data distribution through the Euclidean distance among local models.
Specifically, worker $i$'s consensus distance at round $k$ is defined as:
\begin{equation}\label{consensus}
   C_{i}^{(k)}=\|\omega_{i}^{(k)}-\overline{\omega}^{(k)}\|_{2},
\end{equation}
where $\omega_{i}^{(k)}$ represents the local model parameters of worker $i$, and $\overline{\omega}^{(k)}=\frac{1}{m}\sum_{j=1}^{m}\omega_{j}^{(k)}$ denotes the average of all workers' model parameters. The global consensus distance at round $k$ is then:
\begin{equation}\label{consensus2}
   C^{(k)}=\frac{1}{m}\sum\limits_{i=1}^{m}C_{i}^{(k)}.
\end{equation}
Analogous to weight divergence in the PS architectures, consensus distance correlates with data distribution skewness \cite{liao2023adaptive}. Minimizing $C^{(k)}$ can improve training performance (\eg, model accuracy and convergence rate) by reducing discrepancies between local models \cite{hsieh2020non}.

\begin{figure}[t]\centering
    \begin{minipage}[t]{0.492\linewidth}\centering
        \subfigure[Global consensus distance (CD).]{\centering
            \label{fig:consensus_noniid}
            \includegraphics[width=1.0\textwidth]{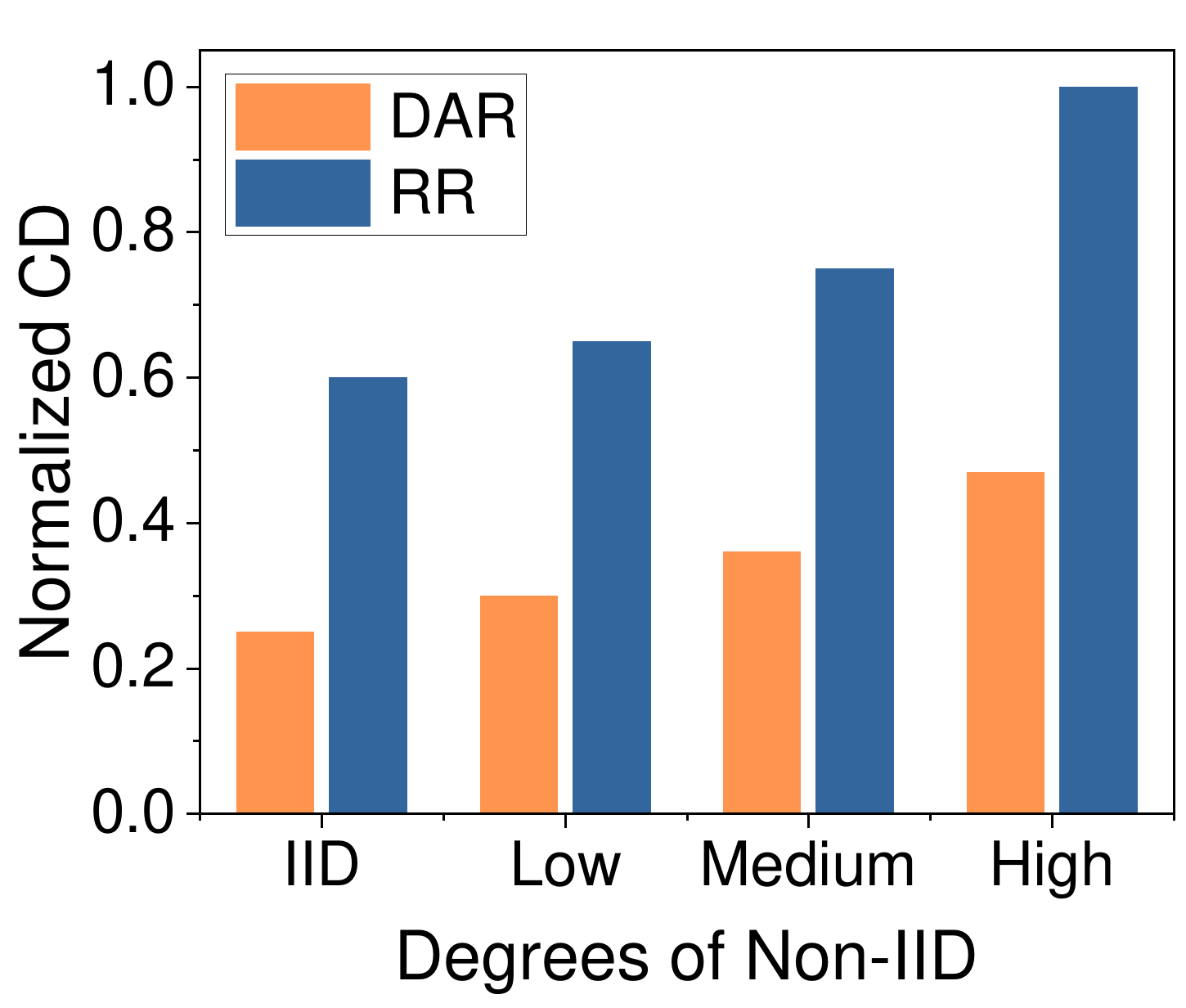}
        }
    \end{minipage}
    \begin{minipage}[t]{0.492\linewidth}\centering
        \subfigure[Model accuracy.]{\centering
            \label{fig:acc_noniid_3}
            \includegraphics[width=1.0\textwidth]{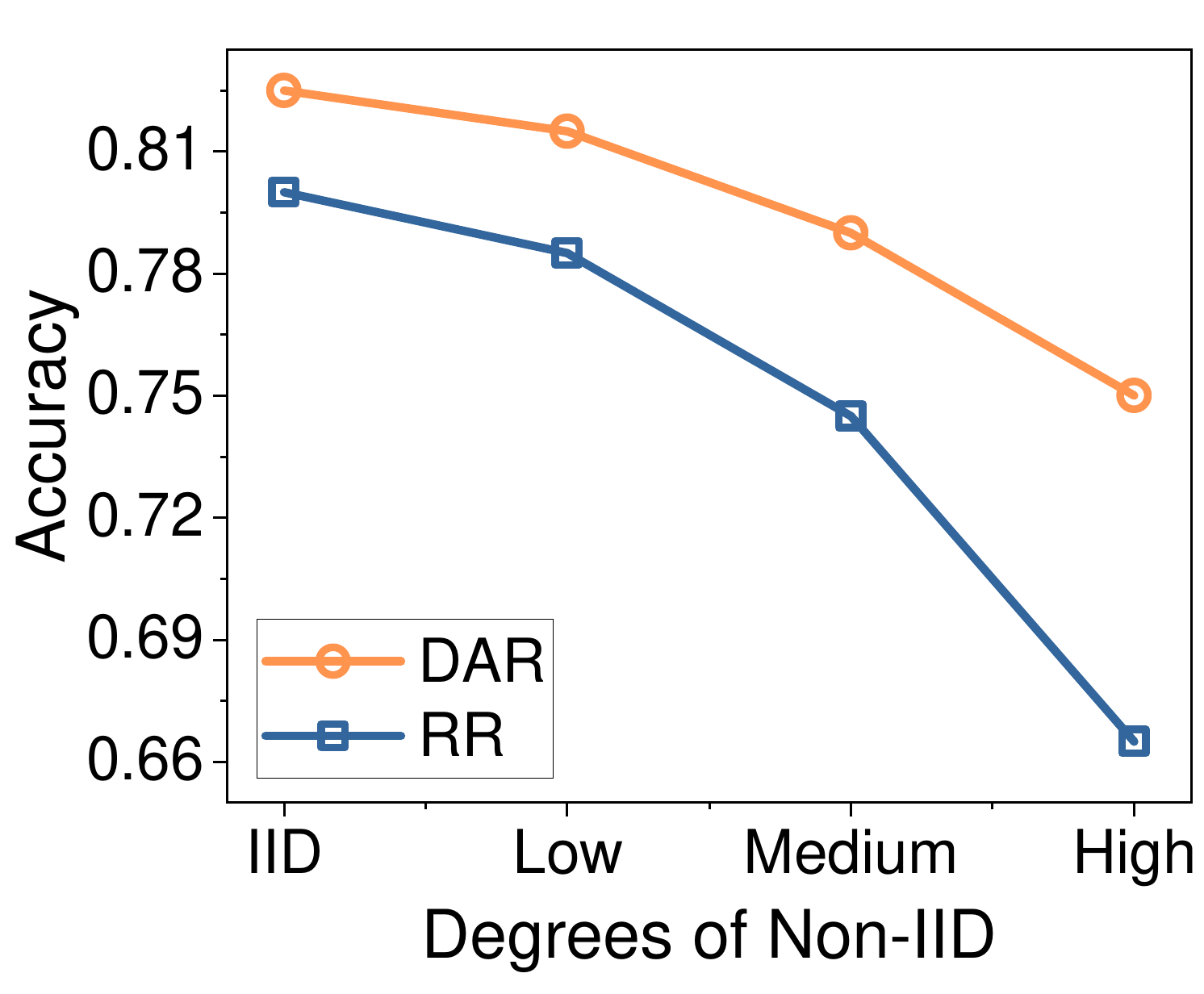}
        }
    \end{minipage}
    \caption{Global consensus distance and model accuracy on the random ring (RR) and the distribution-aware ring (DAR) topology with different non-IID degrees.}\label{}
\end{figure}

In DFGL with non-IID data, workers exchange \emph{graph node embeddings} and \emph{local GCN parameters} with their neighbors to achieve global consensus. Consequently, the global consensus distance becomes a critical metric for capturing the combined impact of network topology and graph sampling ratios. A natural method to minimize this consensus distance is to prioritize exchanges of model parameters and node embeddings between workers with significant pairwise Euclidean distances in their model parameters, thereby improving training performance. To validate this insight, we follow the experimental settings in Section~\ref{subsec:motivations}, training GraphSage under two distinct network topologies, \ie, random ring and distribution-aware ring, respectively. The random ring topology \emph{randomly} arranges workers in a ring structure, with each worker assigned a fixed graph sampling ratio of $0.5$. On the distribution-aware ring topology, workers are greedily connected in each round to the neighbor with the largest pairwise Euclidean distance between their local model parameters. Each worker $i$ is dynamically assigned a sampling ratio of $\frac{0.5 C_{i}^{(k)}}{C^{(k)}}$. We simulate non-IID data  using Dirichlet distribution parameters $\alpha=\{10.0, 1.0, 0.1\}$ to represent low, medium, and high skewness levels.
As shown in Fig. \ref{fig:consensus_noniid}, the distribution-aware ring topology achieves a over $50\%$ lower global consensus distance compared to the random ring topology. Furthermore, Fig. \ref{fig:acc_noniid_3} demonstrates that the distribution-aware approach improves model accuracy by $3.1\%$--$8.5\%$ across various non-IID settings while maintaining robustness under high non-IID conditions ($\alpha=0.1$). These results underscore the necessity of jointly optimizing network topology and sampling ratios guided by consensus distance in DFGL.

\subsubsection{Problem Formulation}\label{subsec:problem_formulation}
The P2P network topology in \textsc{Duplex} at round $k$ is represented by a
symmetric adjacency matrix $\mathbf{A}^{(k)}=[a_{i,j}^{(k)}\in \{0, 1\}]_{m\times m}$, where $a_{i,j}^{(k)}=1$ indicates that worker $i$ and $j$ can communicate with each other at round $k$, and $a_{i,j}^{(k)}=0$ otherwise. The neighbor set of worker $i$ at round $k$ is defined as $\mathcal{N}_{i}^{(k)}=\{j \in \mathcal{M} \mid  a_{i,j}^{(k)}=1\}$. For graph data, let $\mathcal{N}(v)$ denote the one-hop neighbors of node $v \in V_i$ (where $V_i$ is the node set of worker $i$), and let $\mathcal{S}(v) \subseteq \mathcal{N}(v)$ represent the subset of nodes sampled for GCN training. The graph sampling ratio $r_i$ for worker $i$ is defined as the average sampling rate across all nodes in $V_i$:
\begin{equation}\label{sampling ratio}
    r_{i}=\frac{1}{|V_i|}\sum_{v\in V_i}\frac{|\mathcal{S}(v)|}{|\mathcal{N}(v)|}.
\end{equation}

In DFGL, the bandwidth available to each worker is limited and fluctuates dynamically over time due to communication capacity constraints and bandwidth competition among multiple applications. We denote the bandwidth for worker $i$ at round $k$ as $b_{i}^{(k)}=(b_i^{(k,in)},b_i^{(k,out)})$, where $b_i^{(k,in)}$ and $b_i^{(k,out)}$ represent the inbound and outbound bandwidths, respectively. For simplicity, all communication links connected to a worker share the bandwidth equally \cite{wang2022accelerating}. The bandwidth of a link between worker $i$ and worker $j$
is the minimum of the link's inbound and outbound bandwidth:
\begin{equation}\label{bandwidth}
    b_{i,j}^{(k)}=\min \left\{     \frac{b_{i}^{(k,out)}}{|\mathcal{N}_{i}^{(k)}|},\frac{b_{j}^{(k,in)}}{|\mathcal{N}_{j}^{(k)}|} \right\}.
\end{equation}
Due to link asymmetry, note that the bandwidth $b_{i,j}^{(k)}$ and $b_{j,i}^{(k)}$ may be different.

When workers train GCN models synchronously, the global round time, which is the duration between two successive rounds, is given by:
\begin{equation}\label{round_time}
    t^{(k)}=\max\limits_{i\in M}t_i^{(k)},
\end{equation}
where $t_i^{(k)}$ is the round time for worker $i$ at round $k$. This time comprises $t_i^{(k)}=t_i^{(k,cp)}+t_i^{(k,com)}$, where $t_i^{(k,cp)}$ is the local computing time for GCN model updating and $t_i^{(k,com)}$ is the communication time with neighboring workers. As demonstrated in Section \ref{sec:comm_bottleneck}, round time $t_i^{(k)}$ is dominated by communication time $t_i^{(k, com)}$. In particular, $t_i^{(k,com)}$ 
can be calculated as follows:
\begin{equation}\label{round_time}
    t_i^{(k,com)}=\max\limits_{j\in \mathcal{N}_i^{(k)}} \frac{r_i^{(k)}\cdot\mathcal{E}_{i, j}}{b_{i,j}^{(k)}} + \max\limits_{j\in \mathcal{N}_i^{(k)}} \frac{|\omega|}{b_{i,j}^{(k)}},
\end{equation}
where $r_{i}^{(k)}$ is worker $i$'s sampling ratio at round $k$, $\mathcal{E}_{i, j}$ is the total size of node embeddings sent from worker $i$ to worker $j$ (assuming no graph sampling), and $|\omega|$ is the size of GCN model parameters. The communication time $t_i^{(k,com)}$ depends on both the network topology $\mathbf{A}^{(k)}$ (which impacts bandwidth $b_{i, j}^{(k)}$) and the graph sampling ratio $r_{i}^{(k)}$. Therefore, joint optimization of these components is crucial for enhancing communication efficiency. 

Accordingly, we formulate the TOMAS problem as follows:
\centerline{$\mathop{\min}\limits_{\left\{ \left\langle \mathbf{A}^{(k)}, \mathbf{R}^{(k)} \right\rangle \mid k\in\left[1, K\right]\right\}} \sum_{k=1}^{K} t^{(k)}$}
\begin{equation}\label{equ:minimization}
    \st
    \begin{cases}
    C^{(k)} \leq C_{\max}^{(k)}, &\forall k\\
    \frac{1}{m}\sum_{i\in\mathcal{M}} f_{i}(\omega_{i}^{(K)}) \leq \mathcal{F},\\
    a_{i,j}^{(k)}\in \{0,1\}, &\forall i,j\in \mathcal{M},\forall k\\
    0 < r_i^{(k)}\leq 1, &\forall i\in \mathcal{M}, \forall k\\
    \end{cases}
\end{equation}
where $K$ is the total number of training rounds, and $\langle \mathbf{A}^{(k)}, \mathbf{R}^{(k)} \rangle$ denotes the coordinated configuration of the network topology and graph sampling ratios for workers at round $k$. 
The first inequality ensures the global consensus distance $C^{(k)}$ at each round $k$ does not exceed the threshold $C_{\max}^{(k)}$ ($C_{\max}^{(k)}$ is given in Section~\ref{subsec:DDPG_agent}). The second inequality guarantees the convergence of model training, where $\frac{1}{m}\sum_{i\in\mathcal{M}} f_{i}(\omega_{i}^{(K)})$ is the average training loss of all workers at the final round $K$ and $\mathcal{F}$ is a convergence threshold of the loss value. 
The objective is to minimize the total completion time $\sum_{k=1}^{K}t^{(k)}$ of model training.

In fact, the problem formulated in Eq. \eqref{equ:minimization} is hard to solve directly, since it is a nonlinear mixed integer programming problem \cite{papadimitriou1982complexity}. 
Moreover, deriving precise closed-form expressions to describe the influence of network topology and graph sampling ratios on training loss is intractable \cite{chen2021fedgraph}. Hence, instead of designing a heuristic algorithm, we resort to a learning-driven approach to jointly optimize network topology and graph sampling ratios. We believe the Deep Reinforcement Learning (DRL) based method would be effective in solving the TOMAS problem, since DRL can effectively perceive complex relationships between dynamic network states and training performance. However, due to the varied ways of DRL for different problems, it is crucial to adopt a suitable DRL method for the TOMAS problem. By carefully comparing potential DRL methods, we select Deep Deterministic Policy Gradient (DDPG) \cite{lillicrap2015continuous}, as it can efficiently and effectively handle continuous action spaces, which aligns well with the continuity of the graph sampling ratio.
Next, we design a customized DDPG agent tailored to TOMAS’s requirements (Section \ref{subsec:DDPG_agent}) and describe the methodology of the learning-driven algorithm (Section \ref{subsec:learning-driven_algorithm}).

\subsubsection{DDPG Agent Design}\label{subsec:DDPG_agent}
The DDPG agent in \textsc{Duplex} consists of three parts, \ie, state space, action space and reward function.

\noindent\textbf{State space:}
The agent state at round $k$ is defined as $s^{(k)}=(\mathbf{b}^{(k)},\mathbf{T}^{(k)},\mathcal{E}^{(k)},\mathcal{C}^{(k)},\mathbf{F}^{(k)})$. Here, $\mathbf{b}^{(k)}=\{b_{i}^{(k)} \mid \forall i \in \mathcal{M}\}$ and $\mathbf{T}^{(k)}=\{t_{i}^{(k)} \mid \forall i\in \mathcal{M}\}$ represent the available bandwidth and round time for all workers at round $k$, respectively.
The set $\mathcal{E}^{(k)} = \{\mathcal{E}^{(k)}_{i,j} \mid \forall i, j\in \mathcal{M}\}$ denotes the sizes of node embeddings exchanged between any two workers at round $k$ (with graph sampling).
$\mathcal{C}^{(k)}=\{C_{i,j}^{(k)} \mid \forall i,j\in\mathcal{M}\}$ represents the Euclidean distances between all pairs of local model parameters at round $k$, where $C_{i,j}^{(k)}=\|\omega_{i}^{(k)}-\omega_{j}^{(k)}\|_{2}$. Finally, $\mathbf{F}^{(k)}=\{f_{i}(\omega_{i}^{(k)}) \mid \forall i \in \mathcal{M}\}$ 
denotes the local training losses on all workers at round $k$.

\noindent\textbf{Action space:}
The action space at round $k$ can be represented as the coordinated configuration 
$\sigma^{(k)}=\langle \mathbf{A}^{(k)},\mathbf{R}^{(k)} \rangle$, where $\mathbf{A}^{(k)}$ denotes the adjacency matrix of the network topology and $\mathbf{R}^{(k)}=\{r_{i}^{(k)} \mid \forall i\in \mathcal{M}\}$ denotes the set of graph sampling ratios for all workers at round $k$.

\noindent\textbf{Reward function.}
At round $k$, the coordinator computes a reward $u^{(k)}$ after executing the action $\sigma^{(k)}$. The reward should be positively correlated with the system objective \cite{xu2018experience}. That is, with less round time, smaller consensus distance and lower training loss, the reward $u^{(k)}$ gotten would be greater. 
We define the reward function as:
\begin{equation}\label{rewardfunction}
    u^{(k)}=- \chi(\frac{t^{(k)}}{\Bar{t}^{(k-1)}}-1)+
    \varrho(C_{\max}^{(k)}-C^{(k)}) + \varphi^{(\mathcal{F}-\Bar{f}^{(k)})},
\end{equation}
where 
$\chi$, $\varrho$, $\varphi$ are three positive constants. $\Bar{t}^{(k)}$ is the moving average time used to alleviate the impact of data jitter:
\begin{equation}
    \Bar{t}^{(k)}=\Upsilon t^{(k)}+(1-\Upsilon)\Bar{t}^{(k-1)},
\end{equation}
where $\Upsilon\in(0,1)$. $\Bar{f}^{(k)}$ 
denotes the average local training loss of workers at round $k$.
$C_{\max}^{(k)}$ is the exponential moving average of the gradient norm \cite{kong2021consensus}:
\begin{equation}
    C_{\max}^{(k)} = (1-\beta)C_{\max}^{(k-1)} + \frac{\beta}{m}\sum_{i=1}^{m}\|g_{i}^{(k)}\|_{2},
\end{equation}
where $\frac{1}{m}\sum_{i=1}^{m}\|g_{i}^{(k)}\|_{2}$ denotes the average gradient norm of local model updates at round $k$, and $\beta\in[0,1]$.
The first term of the reward function
encourages fast training. 
The longer training time the round $k$ takes, the less reward $u^{(k)}$ will be obtained. 
The second and third terms evaluate the descent of consensus distance and training loss, respectively. The closer to the given threshold of consensus distance $C_{\max}^{(k)}$ and loss value $\mathcal{F}$, the more reward will be received. 
It is worth noting that the actual $\overline{\omega}^{(k)}$ in Eq. \eqref{consensus} is not available in practice. 
Thus, we follow \cite{wang2022accelerating} to estimate the global consensus distance $C^{(k)}$ as follows:
\begin{equation}
\hat{C}^{(k)}=\frac{1}{m^{2}}\sum_{i\in\mathcal{M}}\sum_{j\in\mathcal{M}}(1-a_{i,j}^{(k)})\hat{C}_{i,j}^{(k)},
\end{equation}
where $\hat{C}_{i,j}^{(k)}=\min_{q\in\mathcal{M}\textbackslash\{i,j\}}({C}_{i,q}^{(k)} + {C}_{j,q}^{(k)})$ is the estimated Euclidean distance between worker $i$ and $j$'s local model parameters.

\subsubsection{Learning-Driven Algorithm}\label{subsec:learning-driven_algorithm}
The basic idea of our algorithm is to maintain an actor network $\pi(s|\theta^{\pi})$ and a critic network $Q(s,\sigma|\theta^{Q})$ with parameters $\theta^{\pi}$ and $\theta^{Q}$, respectively. Meanwhile, we maintain two target networks $\pi^{\prime}(s|\theta^{\pi^{\prime}})$ and $Q^{\prime}(s,\sigma|\theta^{Q^{\prime}})$, which have the same structures as $\pi(s|\theta^{\pi})$ and $Q(s,\sigma|\theta^{Q})$.
The actor network and the critic network can be implemented using a Deep Neural Network and a Deep Q-Network \cite{mnih2015human}, respectively. Additionally, the replay buffer $B$ is employed to store historical transitions defined as $(s^{(k)},\sigma^{(k)},u^{(k)},s^{(k+1)})$. 
We update the actor network and the critic network using a mini-bath of transitions sampled from the replay buffer. 
Given a state $s^{(k)}$, the actor network generates an action:
\begin{equation}\label{action}
    \sigma^{(k)}=\pi(s^{(k)}|\theta^{\pi}).
\end{equation}
Then, the critic network would return a $Q$ value $Q(s^{(k)},\sigma^{(k)}|\theta^{Q})$ when provided with a state $s^{(k)}$ and an action $\sigma^{(k)}$ as input.
In order to update the critic network, we should also calculate the target $Q$ value:
\begin{equation}\label{targetQ}
    y^{(k)}=u^{(k)}+\gamma Q^{\prime}(s^{(k+1)},\pi^{\prime}(s^{(k+1)}|\theta^{\pi^{\prime}})|\theta^{Q^{\prime}}),
\end{equation}
where $\gamma$ is the discount factor for the future rewards. Then, the Temporal-Difference error (TD-error) is obtained:
\begin{equation}\label{TDerror}
    \delta^{(k)}=y^{(k)}-Q(s^{(k)},\sigma^{(k)}|\theta^{Q}).
\end{equation}
The critic network can be updated using the TD-error through gradient descent.
Besides, to update the actor network, we should obtain the policy gradient as:
\begin{equation}\label{policygradient}
    \nabla_{\theta^{\pi}}Q(s^{(k)},\sigma^{(k)})=\\
    \nabla_{\sigma}Q(s,\sigma|\theta^{Q})|_{u^{(k)}}^{s^{(k)}}\cdot\nabla_{\theta^{\pi}}\pi(s|\theta^{\pi})|_{s^{(k)}}.
\end{equation}
Accordingly, the actor network can be updated using the policy gradient with gradient ascent.

\begin{algorithm}[t]
    \SetAlgoNoLine
    \SetAlgoNoEnd
    \caption{Control algorithm on coordinator} \label{alg:control}
        Initialize actor network $\pi$, critic network $Q$, target networks $\pi^{\prime}$ and $Q^{\prime}$ and replay buffer $B$\; \label{initialize}
        Receive initial state $s^{(1)}$ from workers\; \label{receive_s}
        \For{$k=1$ to $K$}{
            Obtain action $\sigma^{(k)}$ by Eq. \eqref{action}\; \label{obtain_action}
            Send action $\sigma^{(k)}$ to workers\; \label{send_action}
            Receive $s^{(k+1)}$ from workers\; \label{receive_state}
            Obtain reward $u^{(k)}$ by Eq. \eqref{rewardfunction}\; \label{obtain_reward}
            Put transition $(s^{(k)}, \sigma^{(k)}, u^{(k)}, s^{(k+1)})$ into $B$\;\label{begin_transition}
            \For{$t=1$ to $N$}{
                Sample a mini-batch of transitions from $B$\;
                Compute target $Q$ value by Eq. \eqref{targetQ}\; \label{line:targetQ}
                Compute TD-error by Eq. \eqref{TDerror}\; \label{compute_TD_error}
                Compute policy gradient by Eq. \eqref{policygradient}\; \label{compute_policy}
                Accumulate weight-change for actor network and critic network by Eq.\eqref{updateactornetwork}\; \label{accumulate_weight}
            }
            Update the actor network and the critic network\; \label{update_the_network}
            Update the target networks by Eqs. \eqref{updatetargetactornetwork}\; \label{update_target_networks}
        }
\end{algorithm}
In each round, the coordinator samples transitions $(s^{(k)},\sigma^{(k)},u^{(k)},s^{(k+1)})$ from the buffer to update the actor network and the critic network. The accumulated weight change for the actor network and the critic network is:
\begin{equation}\label{updateactornetwork}
    \begin{cases}
    \Delta_{\theta^{\pi}}:=\Delta_{\theta^{\pi}}+\frac{1}{|B|}\cdot\nabla_{\theta^{\pi}}Q(s^{(k)},\sigma^{(k)}), \\
    \Delta_{\theta^{Q}}:=\Delta_{\theta^{Q}}+\frac{1}{|B|}\cdot\delta^{(k)}\cdot\nabla_{\theta^{Q}}Q(s^{(k)},\sigma^{(k)}),
    \end{cases}
\end{equation}
where $\nabla_{\theta^{\pi}}Q(s^{(k)},\sigma^{(k)})$ is policy gradient defined in Eq. \eqref{policygradient} and $\delta^{(k)}$ is TD-error defined in Eq. \eqref{TDerror}. 

The adaptive control algorithm in \textsc{Duplex} is presented in Algorithm. \ref{alg:control}. At the beginning of the algorithm, the coordinator first initializes the parameters of the actor network and the critic network. To apply an off-policy training method, the target networks $Q^{\prime}(s,\sigma|\theta^{Q^{\prime}})$, $\pi^{\prime}(s|\theta^{\pi^{\prime}})$ and a replay buffer are employed to improve the training speed. 
Subsequently, the coordinator awaits the reception of the initial state from workers (Line \ref{receive_s}). At each training round $k$, the coordinator firstly obtains the action (\ie, sampling ratios and network topology) $\sigma^{(k)}$ by Eq. \eqref{action} (Line \ref{obtain_action}) and sends the action to workers (Line \ref{send_action}). Based on the received action, each worker trains the local GCN model (Section \ref{subsec:local_GCN_training}) and aggregates models from neighbors (Section \ref{subsec:model_aggregation}). Next, a new sate $s^{(k+1)}$ is sent from workers to the coordinator (Line \ref{receive_state}), and the reward associated with the action is obtained (Line \ref{obtain_reward}). Following this, the coordinator samples the transition data from the replay buffer to train the actor and critic networks (Lines \ref{begin_transition}-\ref{update_the_network}). 
Specifically, the target $Q$ value and TD-error are computed by Eqs. \eqref{targetQ} and \eqref{TDerror}, respectively (Lines \ref{line:targetQ}-\ref{compute_TD_error}).
Then, the policy gradient is calculated by the chain rule (Line \ref{compute_policy}).
Concurrently, the weight changes are accumulated for updating the actor network and the critic network (Lines \ref{accumulate_weight}-\ref{update_the_network}). Last, the target actor network and critic network are updated as follows (Line \ref{update_target_networks}):
\begin{equation}\label{updatetargetactornetwork}
    \begin{cases}
    \theta^{\pi^{\prime}}=\xi\theta^{\pi}+(1-\xi)\theta^{\pi^{\prime}},\\
    \theta^{Q^{\prime}}=\xi\theta^{Q}+(1-\xi)\theta^{Q^{\prime}},
    \end{cases}
\end{equation}
where $\xi\in(0,1]$ is the update coefficient of target networks.

\begin{figure*}[t]\centering
    \begin{minipage}[t]{0.242\linewidth}\centering
        \subfigure[Graph sampling ($r_{i}=0.25$).]{\centering
            \label{fig:sampling_0.25}
            \includegraphics[width=0.91\textwidth]{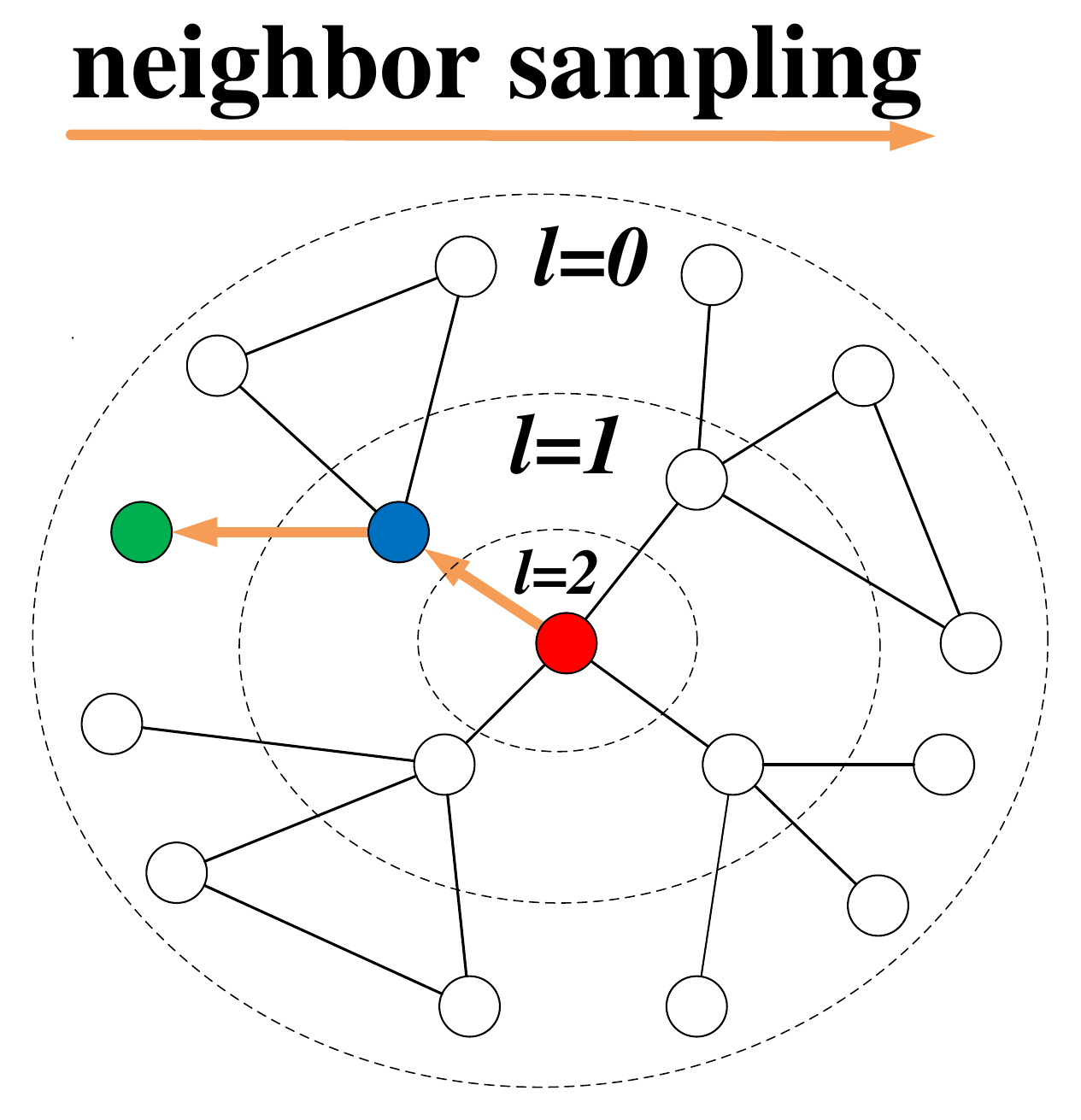}
        }
    \end{minipage}
    \begin{minipage}[t]{0.242\linewidth}\centering
        \subfigure[Graph sampling ($r_{i}=1.0$).]{\centering
            \label{fig:sampling_1.0}
            \includegraphics[width=0.91\textwidth]{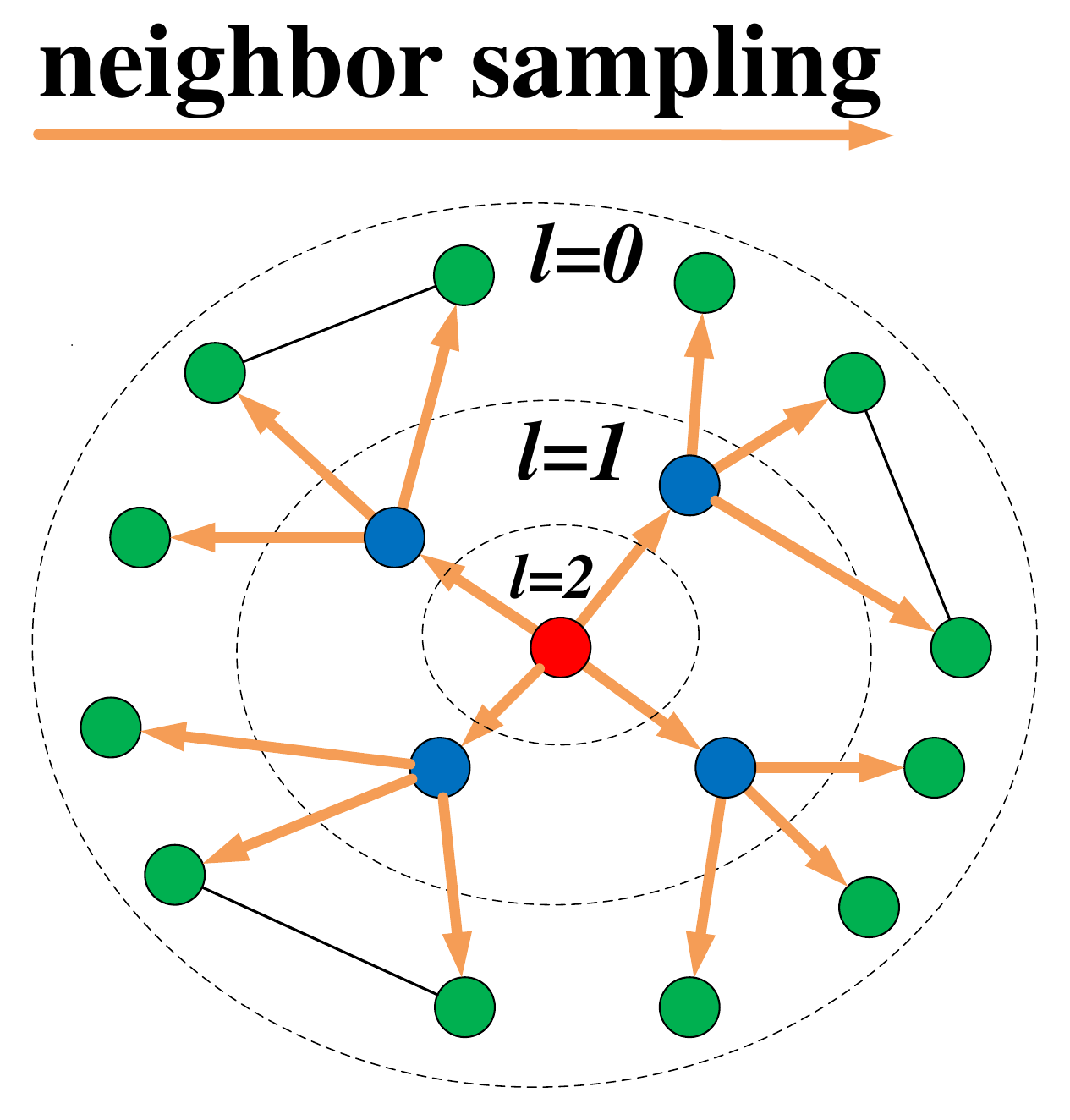}
        }
    \end{minipage}
    \begin{minipage}[t]{0.242\linewidth}\centering
        \subfigure[Graph sampling ($r_{i}=0.5$).]{\centering
            \label{fig:sampling_0.5}
            \includegraphics[width=0.91\textwidth]{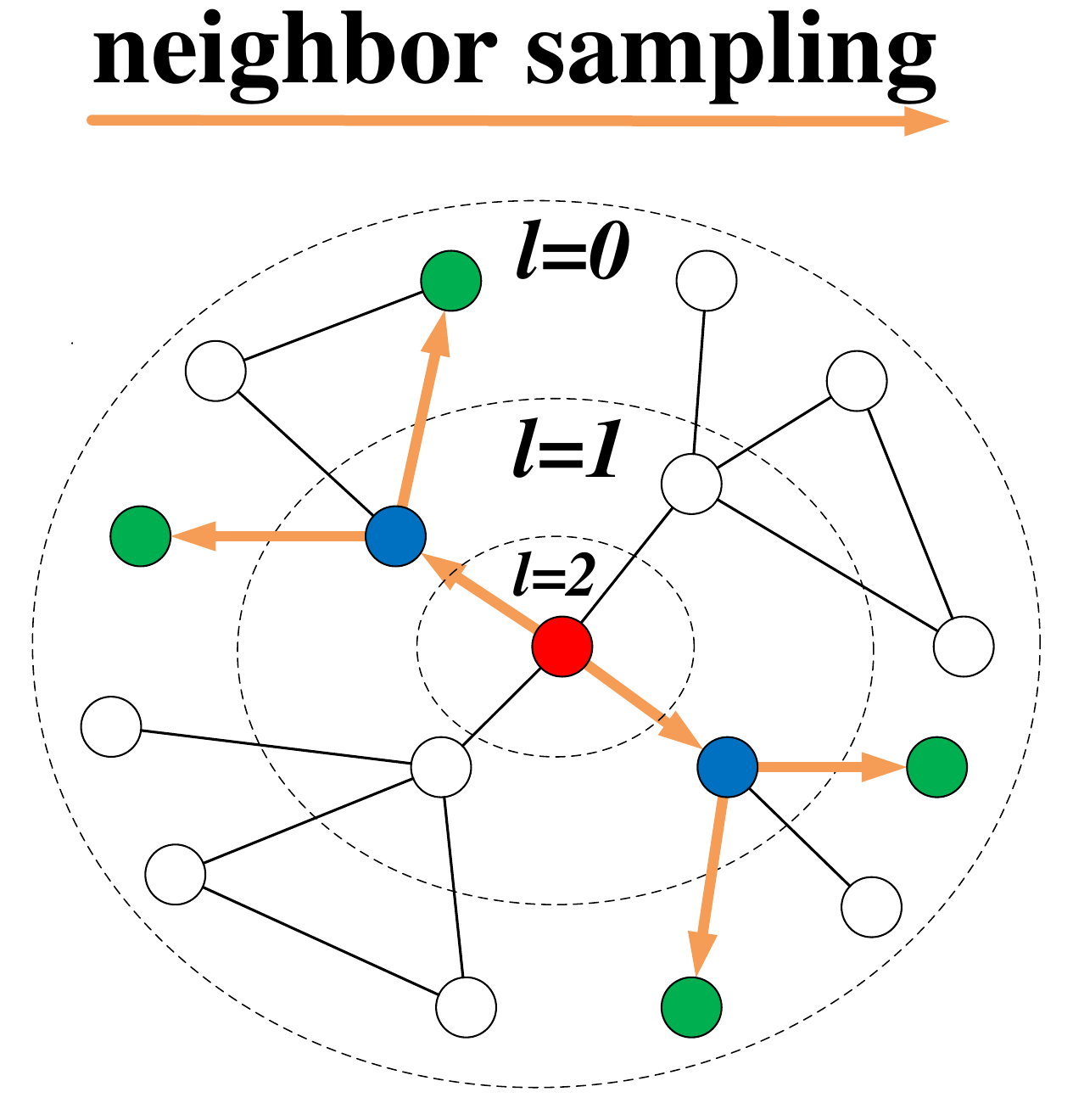}
        }
    \end{minipage}
    \begin{minipage}[t]{0.2425\linewidth}\centering
        \subfigure[Graph aggregation ($r_{i}=0.5$).]{\centering
            \label{fig:aggregate_0.5}
            \includegraphics[width=0.91\textwidth]{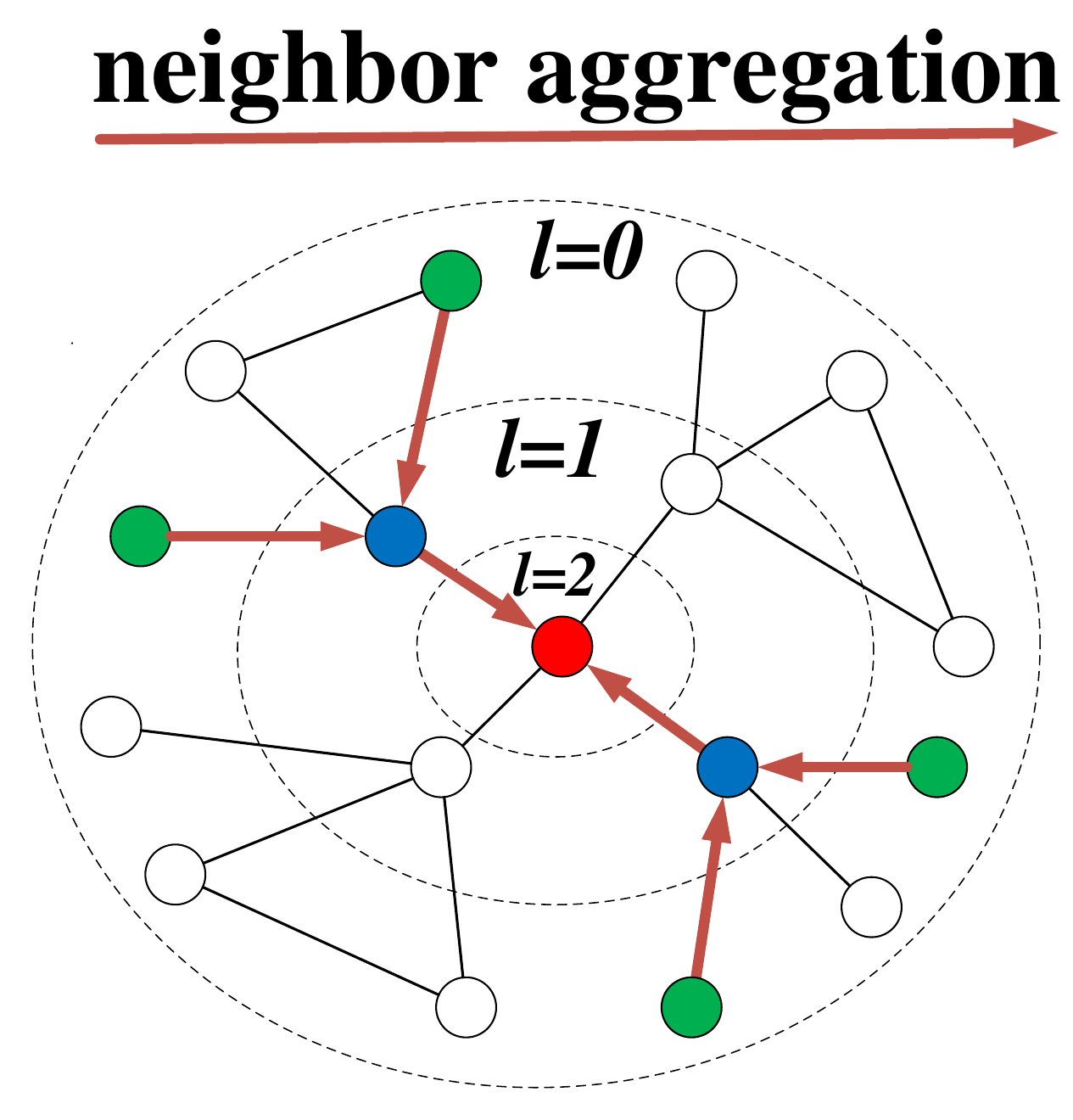}
        }
    \end{minipage}
    \caption{Illustration of graph sampling and aggregation of a 2-layer GCN on a single worker. The green and blue nodes are in the 0-th layer and 1-th layer, respectively, which are sampled by the red node. 
    }\label{fig:sampling_agg}
\end{figure*}
\begin{figure}[t]\centering
    \includegraphics[width=0.405\textwidth]{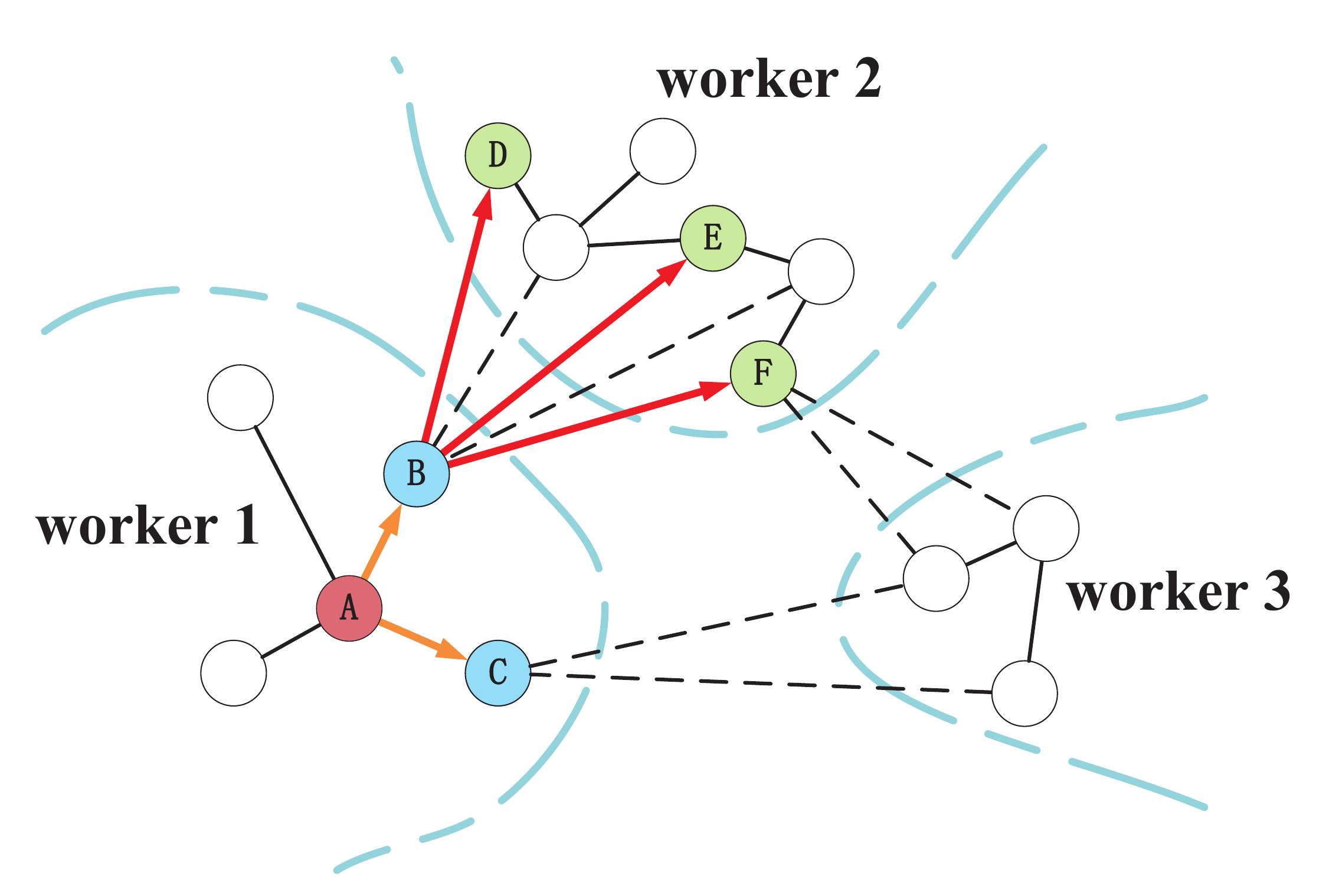}
    \caption{Illustration of graph sampling in \textsc{Duplex} with sampling ratio $r_{1}=0.5$. The black dotted lines indicate external edges among different subgraphs, while the orange and red lines represent internal and external graph neighbor sampling, respectively.}\label{fig:crossWorker}
\end{figure}
\begin{algorithm}[t]
    \caption{Procedure at worker $i$} \label{alg:worker}
        \SetAlgoNoEnd
        \SetAlgoNoLine
        \For{$k=1$ to $K$}{
            Receive $\mathcal{N}_{i}^{(k)}$, $r_{i}^{(k)}$ from the coordinator\;
            Perform local training by \FuncSty{LocalTraining()}\;\label{line:localTrain}
            Send (receive) models to (from) neighbors $\mathcal{N}_{i}^{(k)}$\; \label{line:model_exchange}
            Aggregate models and obtain $\omega_{i}^{(k+1)}$ by Eq. \eqref{modelagg}\; \label{line:aggModel}
            Compute consensus distance $C_{i,j}^{(k)},\forall j\in \mathcal{N}_{i}^{(k)}$\; \label{line:communicate}
            Send consensus distance and loss to the coordinator\;
        }
        \BlankLine
        \SetKwFunction{FTrain}{LocalTraining}
        \SetKwProg{Fn}{Function}{:}{}
        \Fn{\FTrain{}}{ \label{line:localTrainStart}
            \For{each training iteration}{
                Randomly sample a mini-batch of nodes $\mathcal{B}_{i}\subset V_{i}$\; \label{line:sample_mini_batch}
                 $\{V_{i}^{l}\}$,$\{\mathcal{S}^{l}(v)\}$=\FuncSty{GraphSampling($\mathcal{B}_{i}$, $r_{i}^{(k)}$)}\;\label{line:sampling}
                \For{$l=1$ to $L$}{ \label{line:aggStart}
                    \For{each node $v\in V_{i}^{l}$}{
                        $\mathscr{E}_v^{l} = \text{AGG}(\{h_u^{l-1}\mid u\in\mathcal{S}^{l}(v)\})$\;
                        $h_v^{l}=\mathbf{U}^{l}(h_v^{l-1}\Vert \mathscr{E}_v^{l})$\;
                        
                    }
                } \label{line:aggEnd}
                Compute training loss $\mathcal{L}_{\mathcal{B}_i}$ by Eqs. \eqref{eq:predict} and \eqref{loss_of_GCN}\; \label{line:loss_gcn}
                Compute model gradient and update GCN model parameters by gradient descent\;\label{line:localTrainEnd}
            }
        }
        \BlankLine
        \SetKwFunction{FSam}{GraphSampling}
        \SetKwProg{Fn}{Function}{:}{}
        \Fn{\FSam{$\mathcal{B}_{i}$, $r_{i}$}}{ \label{line:splStart} \label{line:graph_sampling_begin}
            Initialize node set $V_{i}^{L}=\mathcal{B}_{i}$\; \label{line:node_L}
            \For{$l=L$ to $1$}{ \label{line:sample_1}
                Initialize node set $V_i^{l-1}=V_i^{l}$\;
                \For{each node $v\in V_{i}^{l}$}{
                    Randomly sample a subset of nodes $\mathcal{S}^{l}(v)$ from $v$'s neighbors $\mathcal{N}(v)$ based on $r_{i}^{(k)}$\;
                    Update node set $V_{i}^{l-1}=V_{i}^{l-1}\cup\mathcal{S}^{l}(v)$\;\label{line:sample_L}
                }
            }
            \textbf{return} $\{V_{i}^{l} \mid l\in[1, L]\}$, $\{\mathcal{S}^{l}(v) \mid l\in[1,L]\}$\;
            \label{line:splEnd}
        }
\end{algorithm}
\subsection{Local GCN Training}\label{subsec:local_GCN_training}
After receiving updated configurations of network topology and graph sampling ratios from the coordinator, each worker $i\in \mathcal{M}$ iteratively updates its local GCN model on its local subgraph via stochastic gradient descent (SGD) for $\tau$ iterations. The local GCN training procedure on each worker $i$ is detailed in Algorithm \ref{alg:worker}.

Within each iteration, worker $i$ first randomly samples a mini-batch of graph nodes $\mathcal{B}_i$ from its local subgraph $G_i$ (Line \ref{line:sample_mini_batch}). It then performs graph sampling to randomly select a subset of $L$-hop neighbors (for an $L$-layer GCN) for each node in $\mathcal{B}_i$, guided by the sampling ratio $r_{i}^{(k)}$ (Lines \ref{line:sampling}, \ref{line:graph_sampling_begin}-\ref{line:splEnd}). These sampled nodes are used for training the local GCN model. Figs. \ref{fig:sampling_0.25}-\ref{fig:sampling_0.5}
illustrate this graph sampling process for various sampling ratios under isolated conditions (\eg, no inter-worker communication). Specifically, given an $L$-layer GCN, all nodes in $\mathcal{B}_i$ constitute the node set at layer $L$ (Line \ref{line:node_L}). From layer $L$ to layer $1$, a subset of neighbors for each node in the current layer is randomly sampled to populate the node set for the preceding layer (Lines \ref{line:sample_1}-\ref{line:sample_L}). However,
due to the existence of external graph edges connecting nodes across workers (\eg, black dotted lines in Fig. \ref{fig:crossWorker}), the sampled graph nodes by worker $i$ may reside on other workers. As shown in Fig. \ref{fig:crossWorker}, considering a system with three workers, Worker $1$ and Worker $2$ are neighbors in the network topology while Worker 3 is isolated. Worker 1 initially samples nodes B and C (connected to node A) during graph sampling. Subsequently, the $2$-hop neighbors of node A (\ie, nodes D, E and F on Worker 2) are further sampled. Since Worker 3 is non-adjacent to Worker 1, it contributes no nodes to this sampling process. 

Upon completing the graph sampling, each worker iteratively transforms low-level node embeddings into higher-level ones from layer $1$ to layer $L$ using GC operations defined in Eq.~\eqref{eq:gc} (Lines \ref{line:aggStart}-\ref{line:aggEnd}). Since the sampled node set $\mathcal{S}^{l}(v)$ may include nodes from worker $i$'s neighboring workers, worker $i$ should request corresponding node embeddings from those workers.
After performing the GC operation on the top layer $L$, the training loss $\mathcal{B}_i$ for the mini-batch $\mathcal{B}_{i}$ is computed using Eqs. \eqref{eq:predict} and \eqref{loss_of_GCN} (Line \ref{line:loss_gcn}). Finally, the local GCN model parameters $\omega_{i}^{(k)}$ are updated through gradient descent (Line \ref{line:localTrainEnd}):
\begin{equation}
    \omega_{i}^{(k)} = \omega_{i}^{(k)} - \eta\cdot \nabla_{\omega_{i}^{(k)}} \mathcal{L}_{\mathcal{B}_i},
\end{equation}
where $\eta$ is the learning rate, and $\nabla_{\omega_{i}^{(k)}} \mathcal{L}_{\mathcal{B}_i}$ denotes the gradient of the loss with respect to the model parameters.

\subsection{Model Aggregation}\label{subsec:model_aggregation}

Upon performing local GCN training, workers send (receive) their local GCN parameters to (from) neighbors in the network topology (Line \ref{line:model_exchange}). Each worker then aggregates the received parameters to update its local model (Line \ref{line:aggModel}). Let $\mathbf{P}^{(k)}=[\mathbf{P}_{i, j}^{(k)}]_{m\times m}$ denote the mixing weight matrix for model aggregation at round $k$. The update rule for worker $i$ is as follows: 
\begin{equation}\label{modelagg}
    \omega_{i}^{(k+1)} = \omega_{i}^{(k)} + \sum_{j\in\mathcal{N}_{i}^{(k)}} \mathbf{P}_{i,j}^{(k)}(\omega_{j}^{(k)}-\omega_{i}^{(k)}).
\end{equation}
By carefully setting $\mathbf{P}_{i, j}^{(k)}$ according to the network topology $\mathbf{A}^{(k)}=[a_{i, j}^{(k)}]_{m\times m}$, the local models on workers will converge to a stationary point \cite{xu2021decentralized}. Specifically, the degree matrix $\mathbf{D}^{(k)}=[\mathbf{D}^{(k)}_{i, j}]_{m\times m}$ is a diagonal matrix, with $\mathbf{D}_{i, i}^{(k)}$ equal to the row sum of $\mathbf{A}^{(k)}$. The Laplacian matrix is defined as $\mathbf{L}^{(k)}=\mathbf{A}^{(k)}-\mathbf{D}^{(k)}$. Boyd \etal \cite{xiao2004fast} proved that the optimal convergence time for a given network topology can be derived by setting $\mathbf{P}_{i, j}^{(k)}$ as follows:
\begin{equation}
    \mathbf{P}_{i, j}^{(k)}=
    \begin{cases}
        \frac{2}{\lambda_{2}(\mathbf{L}^{(k)})+\lambda_{m}(\mathbf{L}^{(k)})}, & a_{i, j}^{(k)} =1\\
        0, & \text{otherwise}
    \end{cases}
\end{equation}
where $\lambda_{m}(\mathbf{L}^{(k)})$ denotes the $m$-th smallest eigenvalue of matrix $\mathbf{L}^{(k)}$.

Additionally, based on the received model parameters from neighbors, each worker $i$ computes the consensus distance $C_{i,j}^{(k)}$ between its model and those of its neighbors:
\begin{equation}
   C_{i,j}^{(k)}=\|\omega_{i}^{(k)}-\omega_{j}^{(k)}\|_{2}, \forall j\in \mathcal{N}_{i}^{(k)}.
\end{equation}
Each worker sends these consensus distances along with the training loss to the coordinator. These training statutes help the coordinator to adjust the network topology and graph sampling ratios for the next round.

\subsection{Privacy Protection}
Due to external graph edges connecting nodes across workers, sampled graph nodes may require cross-worker node embedding exchanges during GC operations, raising privacy concerns for raw node features. To prevent the leakage of local raw node features (\ie, $h_v^{(0)}$), \textsc{Duplex} restricts cross-worker GC operations to layers 2 through $L$ and enforces intra-worker GC operations on layer 1. Specifically, on the first layer, each worker $i$ aggregates only local raw node features within its subgraph $G_i$. For node $v$ in $G_i$, the first-layer embedding $h_v^{1}$ is computed as:
\begin{equation}
    \begin{gathered}
        \mathscr{E}_v^{1} = \text{AGG}(\{h_u^{0}\mid u\in\mathcal{S}^{0}(v) \cap G_i\}),\\
        h_v^{1}=\mathbf{U}^{1}(h_v^{0}\Vert \mathscr{E}_v^{0}),
    \end{gathered}
\end{equation}
where $\mathcal{S}^{0}(v)$ denotes the sampled neighbors of node $v$. No cross-worker communication occurs here. Starting from layer 2, workers exchange embeddings $h_u^l$ ($l\geq 1$) with neighbors. For exampple, if worker $i$ requires $h_u^l$ (from worker $j$) to compute $h_v^{l+1}$, worker $j$ sends $h_u^l$ to worker~$i$. This method can protect raw feature data of node $u$. Specifically, it is infeasible for worker $i$ to further infer the local raw node feature $h_u^{0}$ from the embedding $h_u^{l}$ ($l > 0$).
This is because the aggregation function $\text{AGG}(\cdot)$ (\eg, element-wise mean or sum of multiple vectors) in each layer, which is an irreversible process, destroys information about individual $h_u^{0}$. Besides, each layer applies nonlinear transformations (\eg, ReLU) in the GC operation, further obfuscating $h_u^{0}$. This analysis is consistent with privacy arguments in previous federated graph learning literature \cite{chen2021fedgraph}.
Therefore, \textsc{Duplex} ensures privacy-preserving collaboration by sharing only post-aggregation embeddings while maintaining model efficacy.

%% file: content/evaluation.tex

\subsection{Experimental Setup}
\textbf{Experimental Platform.} 
Our experiments are conducted on a physical platform including 50 NVIDIA Jetson devices \cite{jetson}, which comprise 20 Jetson TX2s, 25 Jetson NXs, and 5 Jetson AGXs. The detailed technical specifications of these devices are listed in Table \ref{table:hardware}.
Specifically, the TX2 and NX devices can work in one of four computation modes while the AGX devices have eight modes. Devices (\ie, workers) working in different modes exhibit diverse computing capabilities. On this platform, we build the software infrastructure based on Docker Swarm \cite{merkel2014docker} and the PyTorch deep learning library \cite{paszke2019pytorch}.
Docker Swarm effectively builds a decentralized system and helps detect the status of each device. Besides, communication between devices is established using the socket library \cite{socket}, which enables efficient data transfer through a set of sending and receiving functions. To accurately capture the heterogeneous bandwidths among workers, we follow prior works \cite{jiang2024semi, liu2023yoga} by connecting all devices via wireless links and arranging them at different locations within a building of 400$m^{2}$. Due to different distances between the devices and routers as well as the random channel noise, the measured bandwidth of the devices fluctuates between 5 to 20 Mbps.

\begin{table}[t]
    \centering
    \caption{Development boards used in experiments.} \label{table:hardware}
    \vspace{-2mm}
    \resizebox{85mm}{!}{
        \centering
        \begin{tabular}{c|c|c|c}
            \hline
            Device & GPU & CPU & Performance\\
            \hline
            TX2 & 256-core Pascal (8GB) & 2-core Denver 2 (64bit) & 2 TFLOPS\\
            NX & 384-core Volta (16GB) & 6-core Carmel (64bit) & 21 TOPS\\
            AGX & 512-core Volta (32GB) & 8-core Carmel (64bit) & 32 TOPS\\
            \hline
        \end{tabular}
    }
\end{table}

\begin{table}[t]\label{table:dataset}
    \centering
    \caption{Summary of the graph data statistics.}
    \resizebox{90mm}{!}{
        \centering
        \label{table:dataset}
        \begin{tabular}{ccccc}
            \toprule
            Datasets&\#Nodes&\#Edges&Features&\#Classes\\
            \midrule
            ogbn-arxiv&169,343&1,166,243&128&40\\
            ogbn-products&2,449,029&61,859,140&100&47\\
            Reddit&232,965&114,615,892&602&41\\
            \bottomrule
        \end{tabular} 
    }
\end{table}
\begin{table*}[t]
    \centering
    \caption{Final test accuracy (\%) of \textsc{Duplex} and the baselines on the three datasets.} \label{table:over_all_test_accuracy}
    \resizebox{178mm}{!}{
        \centering
        \begin{tabular}{c|c|ccc|ccc|cccc|cccc}
            \toprule
             \multirow{2}*{\small Datasets} & \multirow{2}*{\small \textsc{Duplex}} & \multicolumn{3}{c|}{\small D-FedGraph} & \multicolumn{3}{c|}{\small D-FedPNS} & \multicolumn{4}{c|}{ \small Glint} & \multicolumn{4}{c}{\small TDGE} \\ [0.5ex]
             &  & \cellcolor{lightgray}ring & \cellcolor{lightgray}sparse & \cellcolor{lightgray}dense & \cellcolor{lightgray}ring &  \cellcolor{lightgray}sparse & \cellcolor{lightgray}dense & \cellcolor{lightgray}0.1 & \cellcolor{lightgray}0.3 & \cellcolor{lightgray}0.5 & \cellcolor{lightgray}0.7 &
             \cellcolor{lightgray}0.1 & \cellcolor{lightgray}0.3 & \cellcolor{lightgray}0.5 & \cellcolor{lightgray}0.7\\ [0.1ex] \\ [-2.5mm]
            \midrule \\ [-3.5mm]

            
            \multicolumn{1}{c|}{ogbn-arxiv} & \textbf{53.21} & \small 42.26 &\small 49.78 & \textbf{52.97} &\small 43.52 &\small 50.29 & \textbf{53.47} &\small 45.33 &\small 50.98 &\small 52.91 & \textbf{53.59} & \small 43.73 & \small 50.21 & \textbf{53.11} & \small 53.74\\ [0.5ex]

            \midrule \\ [-3.5mm]

            
            \multicolumn{1}{c|}{Reddit} & \textbf{88.23} &\small 79.14 &\small 83.51 & \textbf{88.19} &\small 80.85 &\small 83.86 & \textbf{88.36} &\small 83.31 &\small 86.65 & \textbf{88.32} &\small 89.27 & \small 82.65 & \small 84.94 & \small 87.82 & \textbf{88.33}\\ [0.5ex]

            \midrule \\ [-3.5mm]
            
            
            \multicolumn{1}{c|}{ogbn-products} & \textbf{65.57} &\small 57.53 &\small 62.19 &\textbf{65.78} &\small 58.33 &\small 62.74 & \textbf{65.42} &\small 58.37 &\small 62.55 &\small 64.23 & \textbf{65.59} & \small 57.42 & \small 62.13 & \small 63.87 & \textbf{65.35}\\ [0.5ex]
            
            
            \bottomrule
        \end{tabular}
    }
\end{table*}

\noindent\textbf{Datasets and Models.}
We evaluate the performance of \textsc{Duplex} and baselines on three widely-used datasets: ogbn-arxiv, ogbn-products \cite{hu2020open} and Reddit \cite{hamilton2017inductive}, which are summarized in Table \ref{table:dataset}.
\begin{itemize}
    \item The ogbn-arxiv dataset captures the citation network between Computer Science papers on arXiv. Our aim is to categorize each paper node's subject area of its corresponding arXiv paper. 
    \item The ogbn-products dataset, which aims to predict the category of a product, represents a network of products sold on Amazon, with nodes representing individual products and edges indicating frequent co-purchases. 
    \item The Reddit dataset, which aims to predict the community to which a post belongs, consists of a large collection of user-generated content from the social website.
\end{itemize}
\begin{figure*}[t]\centering
    \begin{minipage}[t]{0.325\linewidth}\centering
        \subfigure[ogbn-arxiv]{\centering
                \label{fig:arxiv_accuracy_time}
                \includegraphics[width=1.0\linewidth]
                {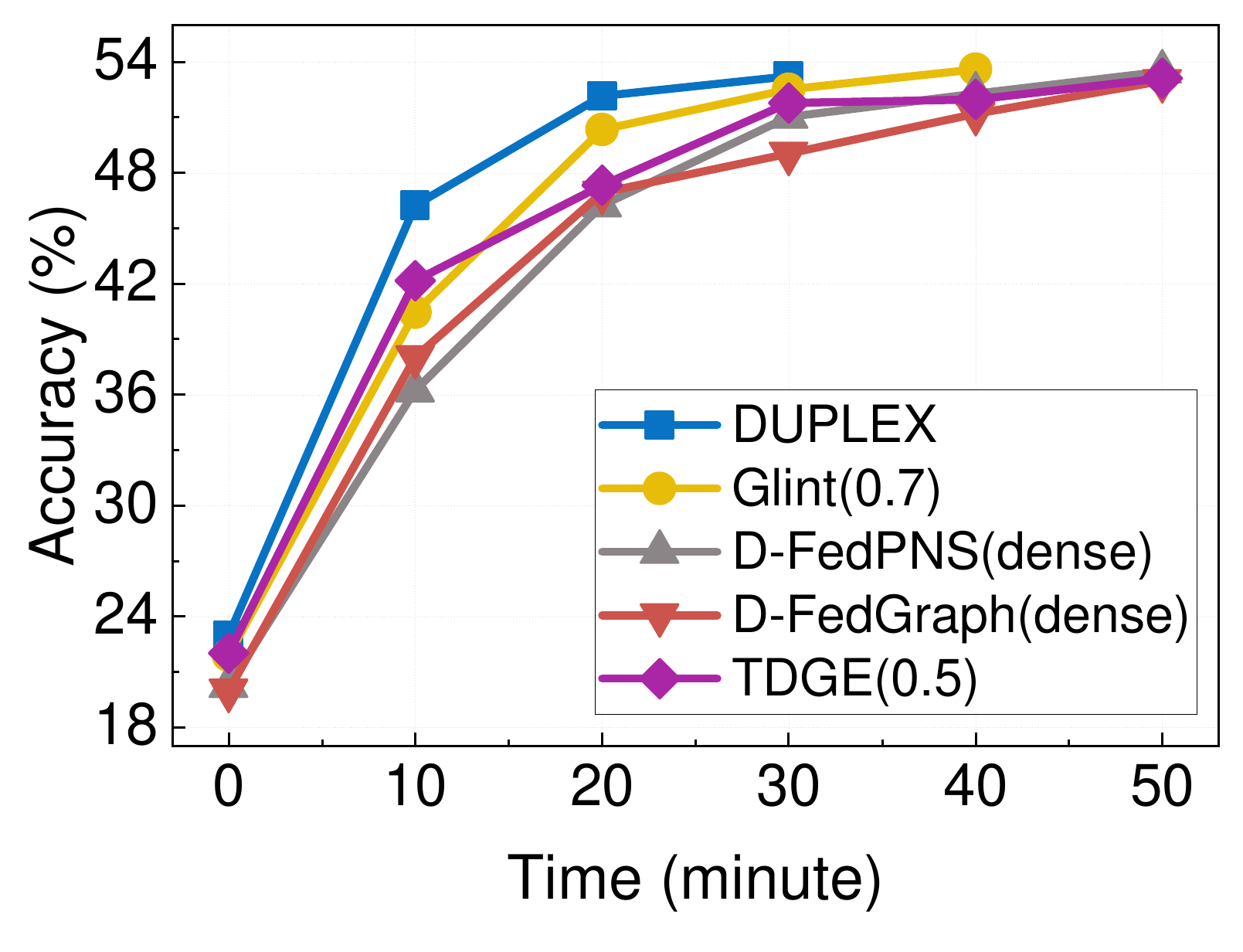}
            }
    \end{minipage}
    \begin{minipage}[t]{0.325\linewidth}\centering
        \subfigure[Reddit]{\centering
                \label{fig:reddit_accuracy_time}
                \includegraphics[width=1.0\linewidth]
                {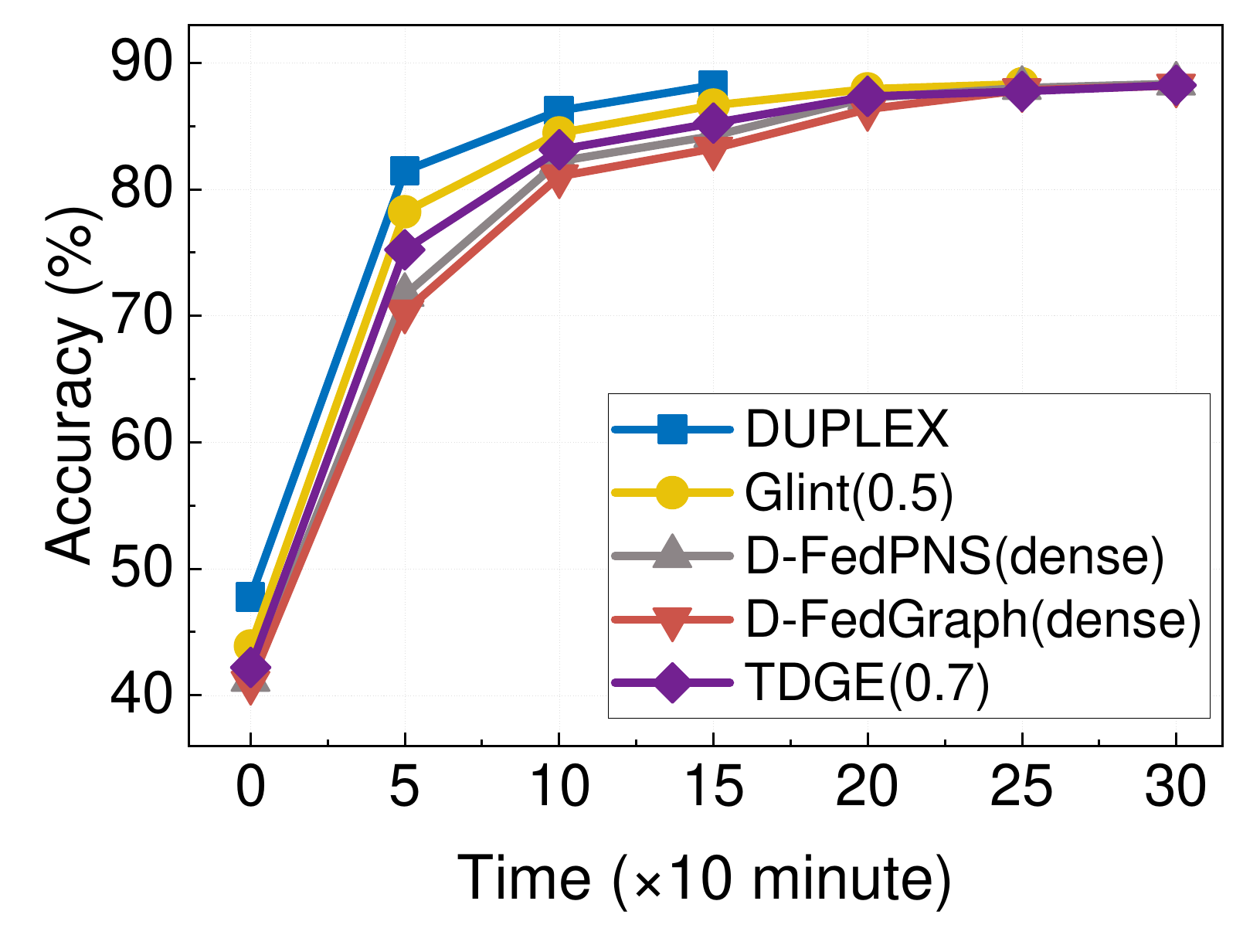}
            }
    \end{minipage}
    \begin{minipage}[t]{0.325\linewidth}\centering
        \subfigure[ogbn-products]{\centering
            \includegraphics[width=1.0\textwidth]
            {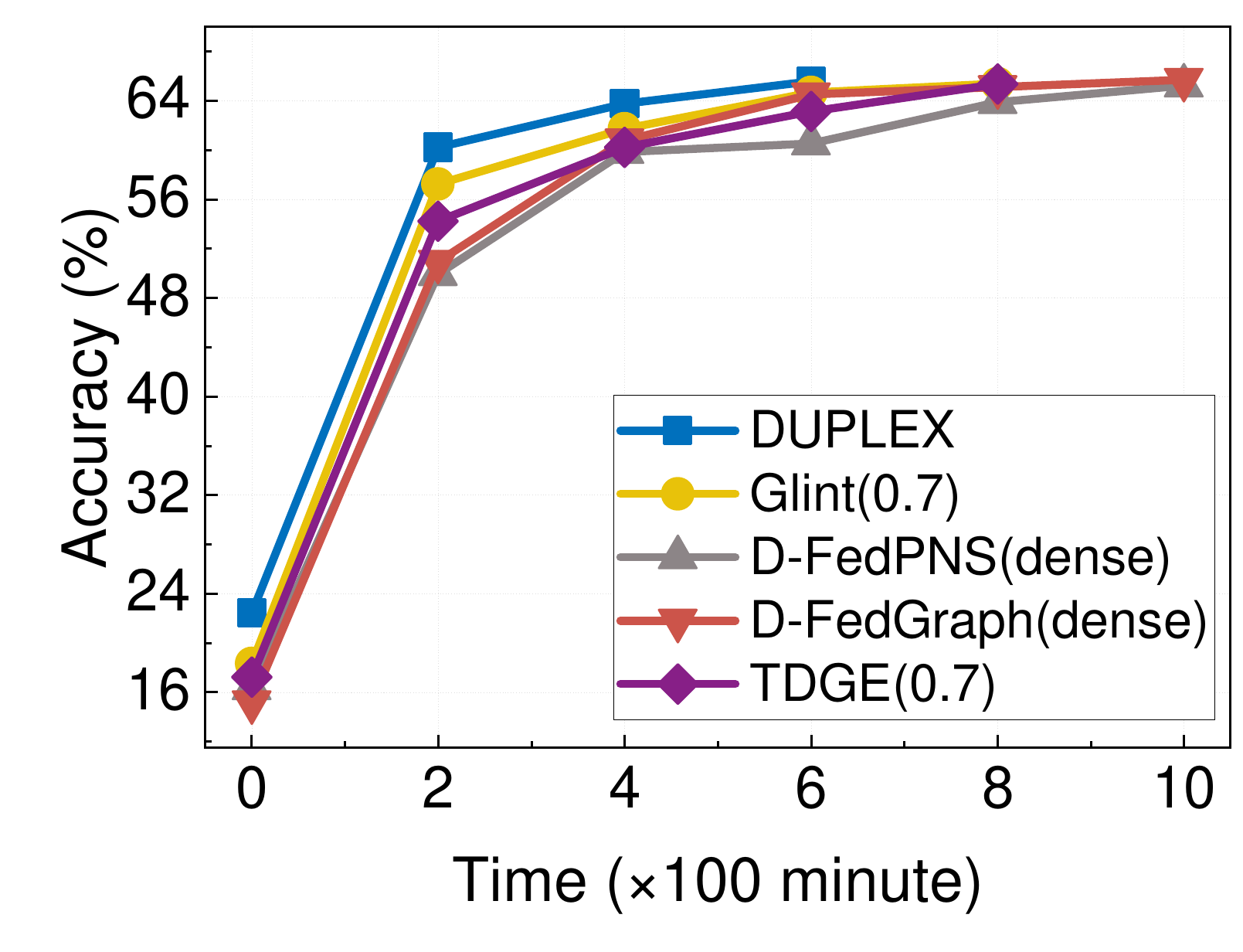}
            \label{fig:ogbn_products_accuracy_time}
        }
    \end{minipage}
    \caption{Time-to-Accuracy of \textsc{Duplex} and the baselines on the three datasets. For a fair comparison, the baselines are selected based on the results in Table \ref{table:over_all_test_accuracy}, where their final test accuracy (bold) is close to that of \textsc{Duplex}.}\label{fig:accuray_time}
\end{figure*}
We perform node classification tasks using two well-known deep graph learning models: a two-layer GCN model \cite{kipf2016semi} on ogbn-arxiv and Reddit, and a two-layer GraphSAGE model \cite{hamilton2017inductive} on ogbn-products. 
Specifically, GCN is a scalable model for learning on graph data.
Besides, GraphSAGE is a general inductive model that builds on GCN to generate node embeddings efficiently \cite{hamilton2017inductive}.\\
\textbf{Non-IID Data Partitioning.} To simulate the non-IID graph data in the real world, we follow the prior work \cite{he2021fedgraphnn} to partition the original training set into 50 parts (\ie, local training sets), each of which is assigned to a specific worker. The graph edges (including internal and external edges) are maintained according to the global graph in the original training set. Specifically, we assume that on every worker, the classes of graph nodes in the local training set follow the Dirichlet distribution. Accordingly, we sample the local training set $\mathcal{D}_{i}\sim Dir_{i}(\alpha)$ from the original training set and allocate $\mathcal{D}_{i}$ to each worker $i$. $\alpha$ determines the degree of non-IID, and a lower value of $\alpha$ generates a high node class distribution shift. 
Unless otherwise specified, we maintain a default setting of $\alpha=10.0$ throughout our experiments.\\
\textbf{Baselines.}
We present a comparative evaluation of \textsc{Duplex} against the following methods: 
\begin{itemize}
    \item \textbf{S-Glint} \cite{liu2022s} is a DFGL framework that effectively constructs a fixed and sparse network topology by evaluating the convergence contributions of workers.
    \item \textbf{TDGE} \cite{duan2024topology} presents a hypercube graph construction method to reduce data heterogeneity by carefully selecting neighbors of each device. In addition, TDGE explores the communication patterns in hypercube topology and proposes a sequential synchronization scheme to reduce communication cost.
    \item \textbf{D-FedPNS} is an alternative of FedPNS \cite{du2022federated}, which is a FGL framework conducting the periodic graph neighbor node sampling and embedding synchronization of cross-worker neighbors. We extend FedPNS to D-FedPNS by forcing P2P communication between workers without changing the periodic sampling strategy in FedPNS.
    \item \textbf{D-FedGraph} is an alternative of FedGraph \cite{chen2021fedgraph}, which is a FGL framework intelligently sampling graph nodes for training by automatically adjusting sampling policies based on deep reinforcement learning. We extend FedGraph to D-FedGraph by forcing P2P communication between workers without changing the sampling policies.
\end{itemize}

In order to provide a comprehensive comparison between the above methods and \textsc{Duplex}, we train D-FedGraph and D-FedPNS on three network topologies: ring topology, sparse topology (each worker has 10 neighbors), and dense topology (each worker has 25 neighbors). 
We formulate various baselines by training D-FedGraph and D-FedPNS on different network topologies, such as D-FedGraph(dense) and D-FedPNS(sparse), representing D-FedGraph is trained based on the dense and D-FedPNS is based on the sparse network topology, respectively. For S-Glint and TDGE, we assign different graph sampling ratios for workers to represent several baselines. For example, Glint(0.1) and TDGE(0.1) represent that each worker in S-Glint and TDEG has a sampling ratio of 0.1.\\
\textbf{Performance Metrics.}
We adopt three metrics to evaluate the performance of different algorithms in our experiments. 
\textcircled{\footnotesize{1}} \emph{Test accuracy}: During each round, the coordinator averages the test accuracy of all local models to compute the overall test accuracy. 
\textcircled{\footnotesize{2}} \emph{Training time}: For fairness, we compare the end-to-end training time of different algorithms to reach the same target test accuracy.
\textcircled{\footnotesize{3}} \emph{Communication cost}: We compare the total communication cost for exchanging graph node embeddings and local models of all workers while achieving the same target test accuracy.
\begin{figure*}[t]\centering
    \begin{minipage}[t]{0.325\linewidth}\centering
        \subfigure[ogbn-arxiv]{\centering
            \label{fig:arxiv_time_accuracy}
            \includegraphics[width=1.0\textwidth]{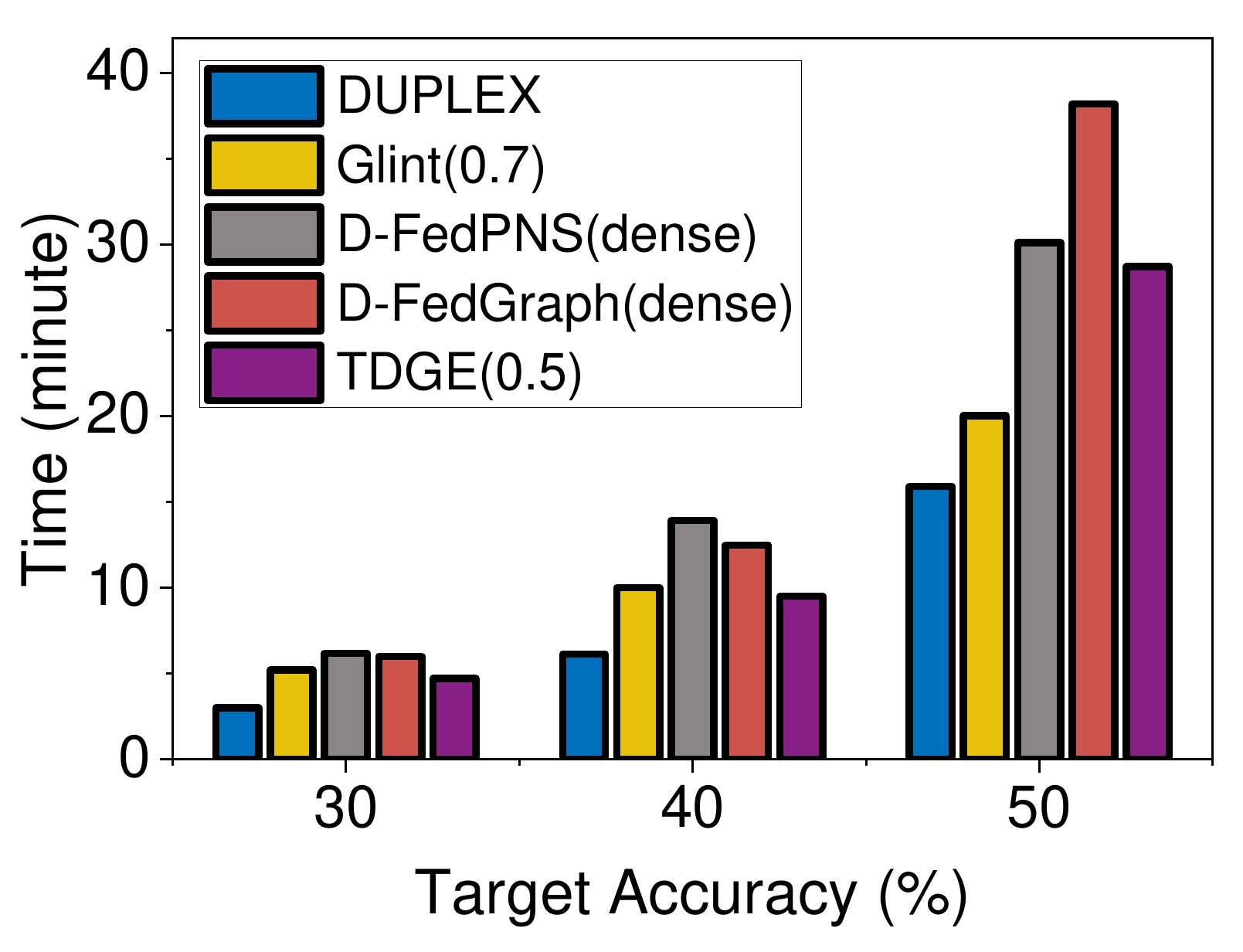}
        }
    \end{minipage}
    \begin{minipage}[t]{0.325\linewidth}\centering
        \subfigure[Reddit]{\centering
            \label{fig:reddit_time_accuracy}
            \includegraphics[width=1.0\textwidth]{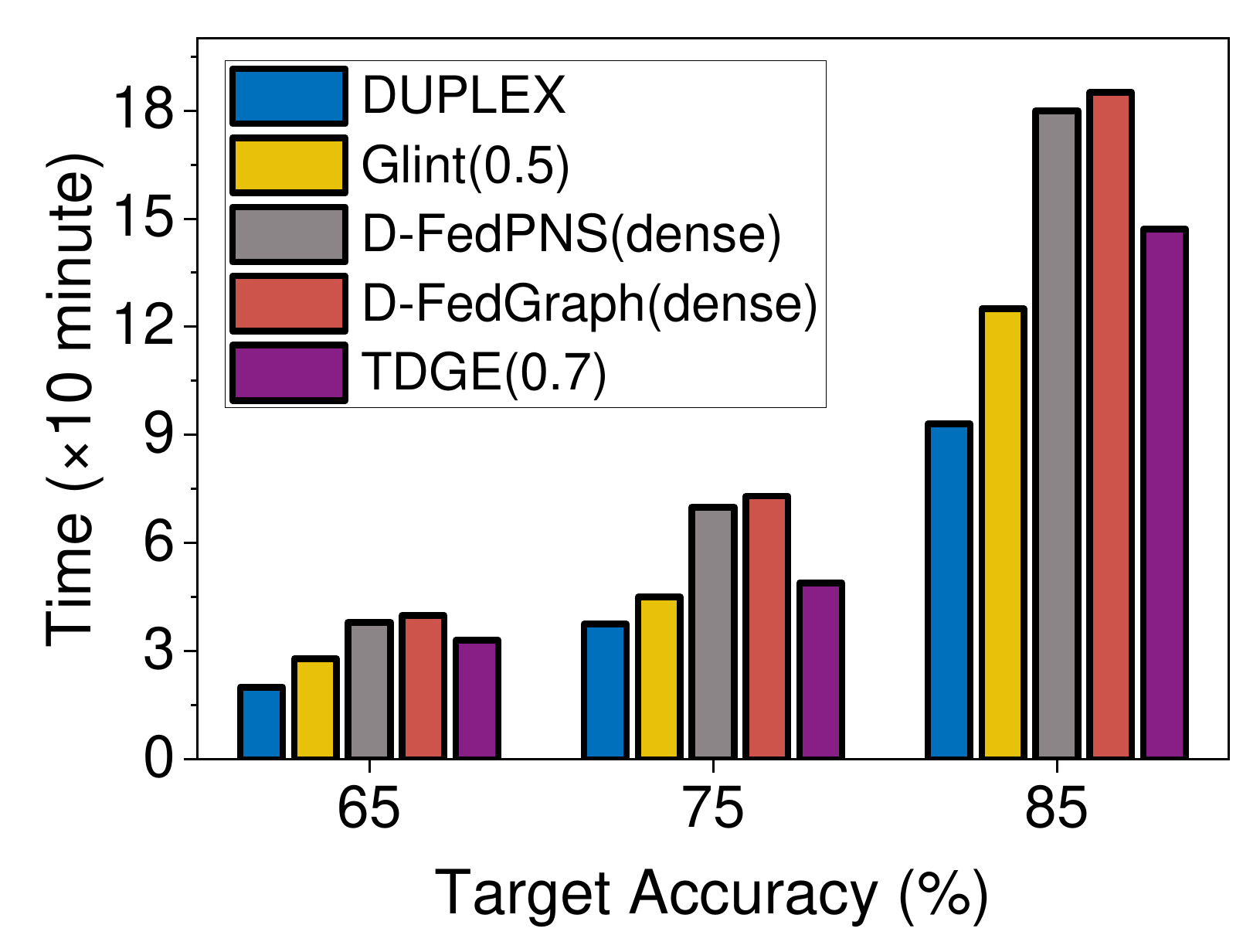}
        }
    \end{minipage}
    \begin{minipage}[t]{0.325\linewidth}\centering
    \subfigure[ogbn-products]{\centering
        \label{fig:ogbn-products_time_accuracy}
        \includegraphics[width=1.0\textwidth]{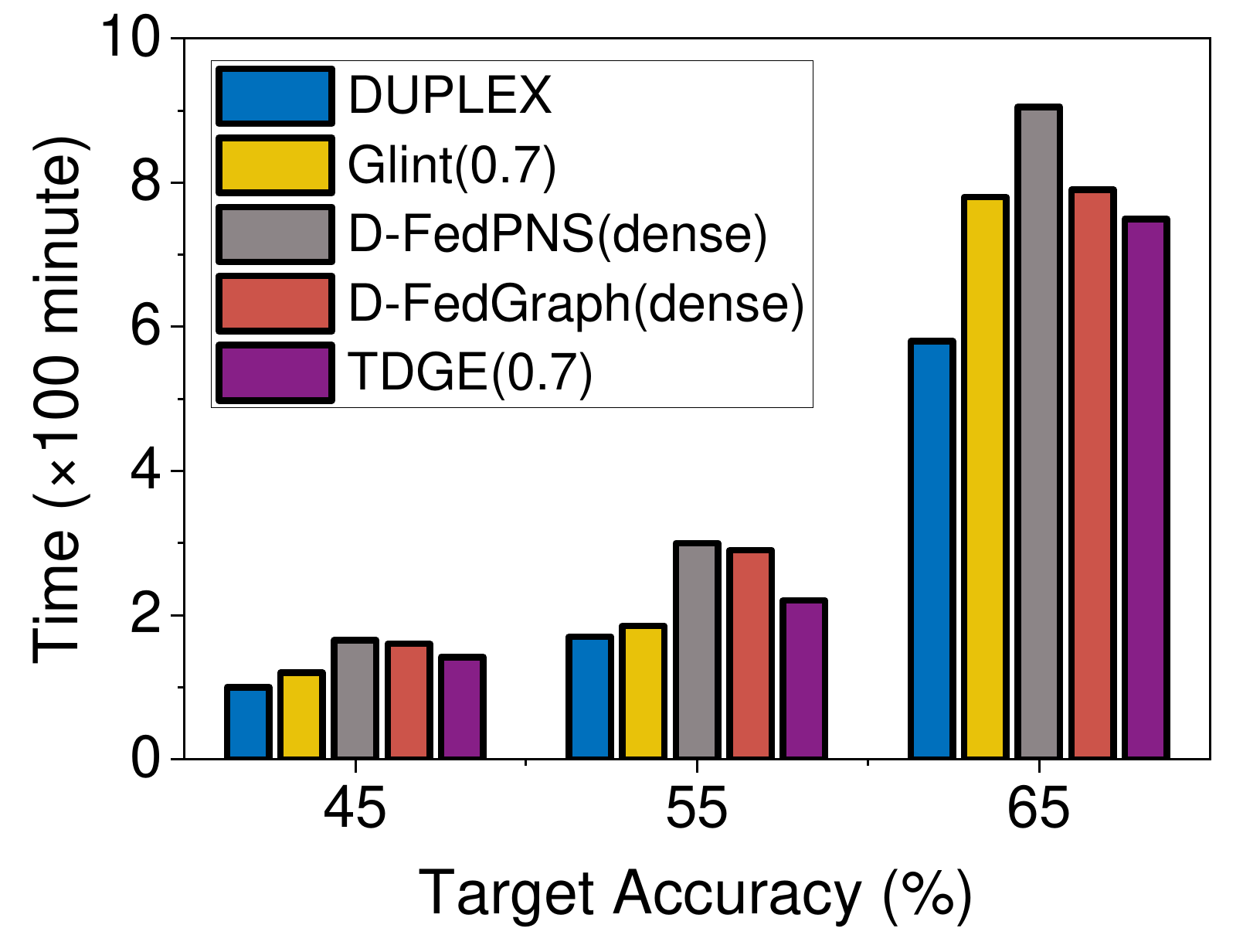}
    }
    \end{minipage}
    \caption{Training time of \textsc{Duplex} and the selected baselines to reach the target accuracy on the three datasets.}\label{fig:time_accuracy}
\end{figure*}
\begin{figure*}[t]\centering
    \begin{minipage}[t]{0.325\linewidth}\centering
        \subfigure[ogbn-arxiv]{\centering
            \label{fig:arxiv_traffic_accuracy}
            \includegraphics[width=1.0\textwidth]{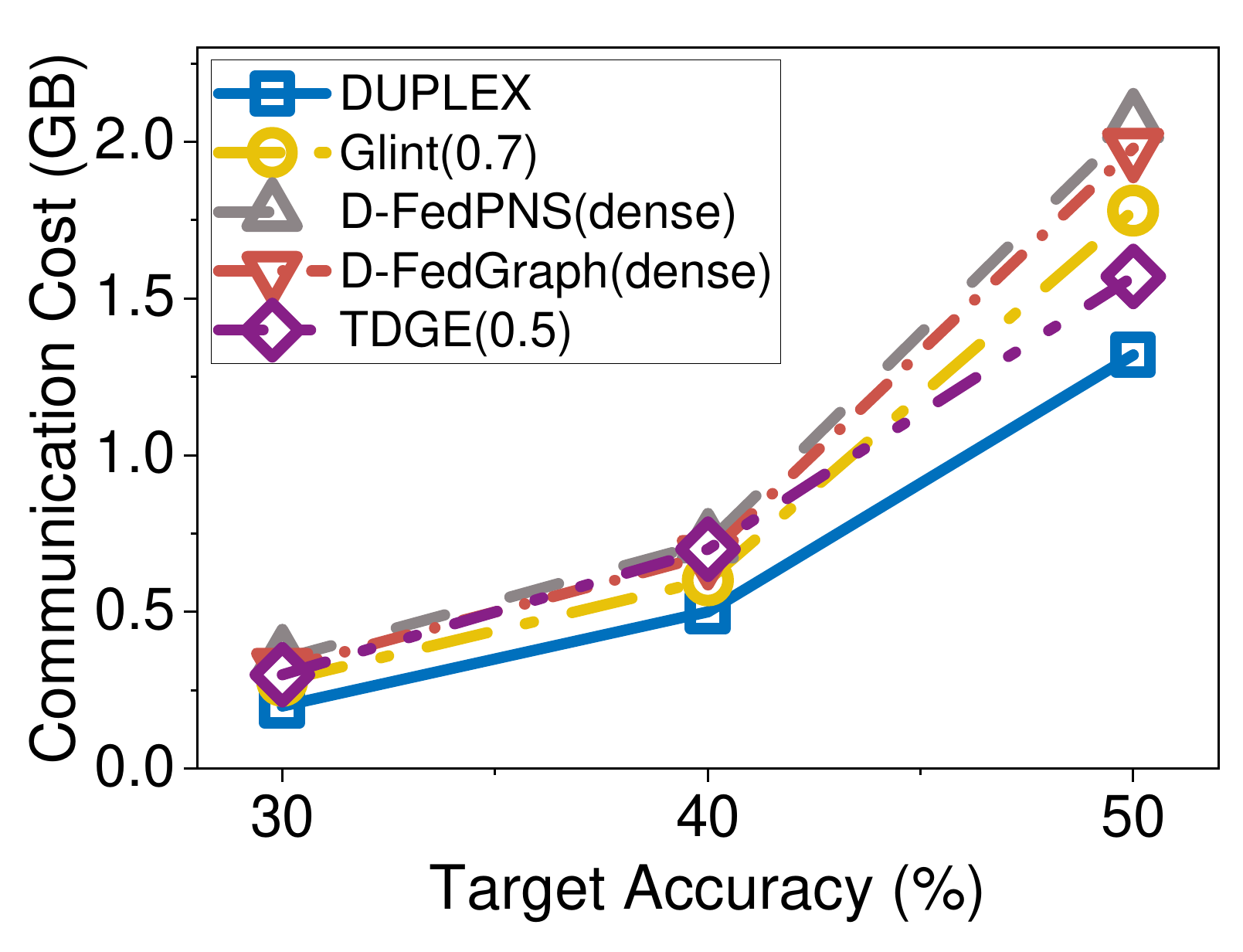}
        }
    \end{minipage}
    \begin{minipage}[t]{0.325\linewidth}\centering
        \subfigure[Reddit]{\centering
            \label{fig:reddit_traffic_accuracy}
            \includegraphics[width=1.0\textwidth]{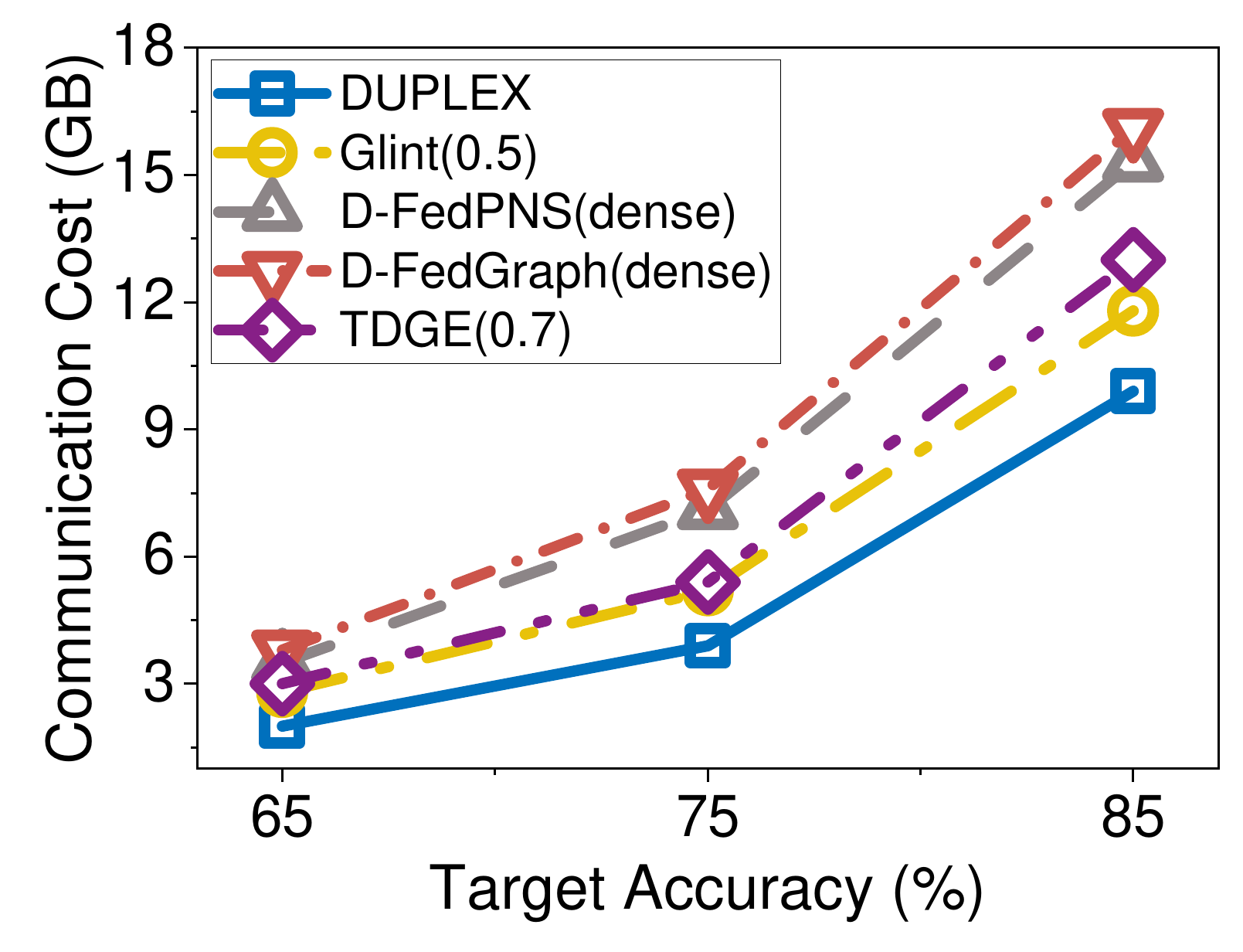}
        }
    \end{minipage}
    \begin{minipage}[t]{0.325\linewidth}\centering
    \subfigure[ogbn-products]{\centering
        \label{fig:ogbn-products_traffic_accuracy}
        \includegraphics[width=1.0\textwidth]{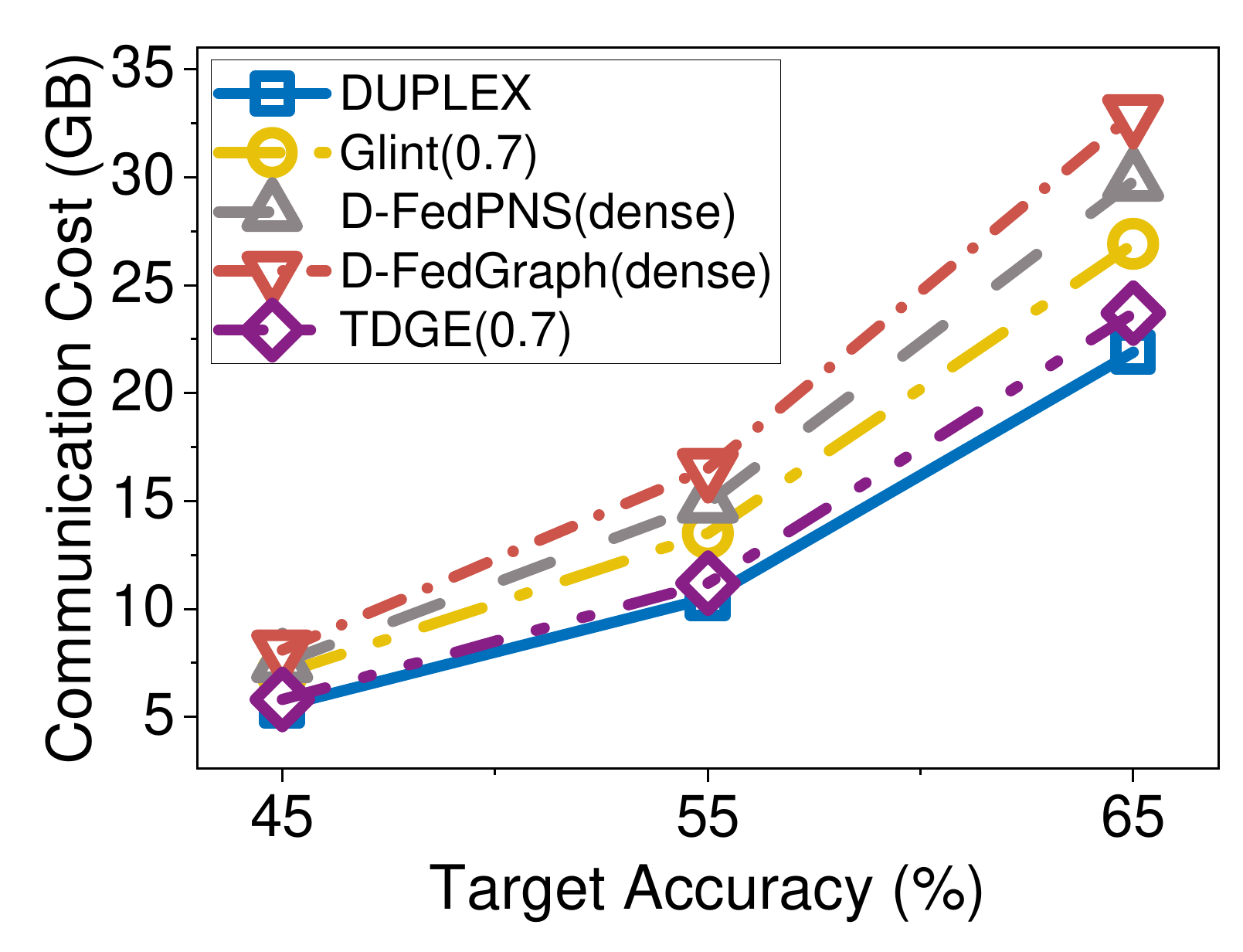}
    }
    \end{minipage}
    \caption{Communication cost of \textsc{Duplex} and the baselines to reach the target accuracy on the three datasets.}\label{fig:traffic_accuracy}
\end{figure*}
\textbf{Parameter Settings.}
We use a hidden feature size of 128 for GCN and 256 for GraphSage. Besides, we adopt the Adam optimizer \cite{kingma2014adam} to optimize the models, with an initial learning rate of 0.01 and the weight decay of $3e^{-4}$. The number of local updates and the training batch size are fixed to 5 and 64 for Reddit, and 10 and 128 for ogbn-arxiv and ogbn-products, respectively. We train the models for a specific number of rounds to ensure convergence: 200 rounds for GCN on ogbn-arxiv, 100 rounds for GCN on Reddit, and 150 rounds for GraphSage on ogbn-products. For the reward function of Eq. \eqref{rewardfunction}, we set $\chi$, $\varrho$ and $\varphi$ as 2, 1 and 10, respectively.

\subsection{Overall Performance}\label{subsec_simulation_results}
\textbf{Test Accuracy.} The test accuracy of \textsc{Duplex} and the baselines on the three datasets are listed in Table \ref{table:over_all_test_accuracy}. The baselines are labeled with different graph neighbor sampling ratios and network topologies in the shadow rows. On ogbn-arxiv, since D-FedGraph(dense), D-FedPNS(dense), Glint(0.7) and TDGE(0.5) achieve comparable test accuracy to \textsc{Duplex}, they are selected for subsequent experiments to compare training time and communication cost with \textsc{Duplex}. For Reddit, we consider D-FedGraph(sparse), D-FedPNS(dense), Glint(0.5) and TDGE(0.7) for further comparison with \textsc{Duplex}. Lastly, D-FedGraph(dense), D-FedPNS(dense), Glint(0.7) and TDGE(0.7) are chosen on ogbn-products for the same reason.
These selections are made based on the similar test accuracy achieved by the baselines compared to \textsc{Duplex}, ensuring a fair evaluation of our proposed algorithm on resource consumption, such as training time and communication cost, to reach the target test accuracy. 

The time-to-accuracy performance of \textsc{Duplex} and the selected baselines on the three datasets are presented in Fig. \ref{fig:accuray_time}.
We observe that the test accuracy of \textsc{Duplex} first steadily increases and then stabilizes on the three datasets. It indicates that the model training of \textsc{Duplex} will converge to a stationary point. 
Impressively, \textsc{Duplex} always achieves the highest accuracy compared to these baselines within the same training time.
For instance, in Fig. \ref{fig:ogbn_products_accuracy_time}, \textsc{Duplex} outperforms Glint(0.7), D-FedGraph(dense), D-FedPNS(dense) and TDGE(0.7) in accuracy by 3.1\%, 10.2\%, 9.7\% and 6.3\%, respectively, after 200 minutes of training on ogbn-products.
Moreover, \textsc{Duplex} always converges faster than the baselines to achieve the same target accuracy. This is because the suboptimal network topology or graph sampling strategy of the baselines cannot fully handle the heterogeneous data distributions in DFGL. 
In contrast, \textsc{Duplex} achieves a faster convergence rate and higher test accuracy with adaptive topology construction and proper sampling ratio assignment.
\begin{table}[t]
    \centering
    \caption{Test accuracy (\%) of \textsc{Duplex} and the baselines on ogbn-arxiv, Reddit and ogbn-products under the communication resource constraint of 1.5GB, 10GB and 20GB, respectively.} \label{table:final_acc}
    \resizebox{80mm}{!}{
        \centering
        \begin{tabular}{cccc}
            \toprule
            Datasets & ogbn-arxiv & Reddit & ogbn-products \\ [0.3ex]
            
            \midrule & \\ [-2.8mm]

            D-FedGraph & \small 49.32 & \small 83.98 & \small 61.37 \\ [0.4ex]

            D-FedPNS &  \small 49.89 & \small 84.32 & \small 62.19 \\ [0.4ex]

            Glint & \small 51.53 & \small 85.72 & \small 62.91 \\ [0.4ex]

            TDGE & \small 50.36 & \small 85.52 & \small 61.73 \\ [0.4ex]

            \textsc{Duplex} & \textbf{53.23} & \textbf{88.25} & \textbf{65.58} \\ [0.3ex]
            
            
            \bottomrule
        \end{tabular}
    }
\end{table}

To further evaluate the model performance of \textsc{Duplex} and selected baselines, we train models under the communication resource constraint of 1.5GB, 10GB and 20GB on ogbn-arxiv, Reddit and ogbn-products, respectively. As shown in Table \ref{table:final_acc}, \textsc{Duplex} can achieve the highest accuracy. 
For example, \textsc{Duplex} achieves the test accuracy improvement by about 4.2\%, 3.6\%, 2.7\% and 3.9\% on ogbn-products, compared with D-FedGraph, D-FedPNS, Glint and TDGE, respectively.\\
\textbf{Time Cost.} In order to gain a deeper understanding of resource consumption associated with the baselines and \textsc{Duplex}, we conduct a set of experiments to record the completion time when they attain the same target accuracy on the three datasets. The simulation results are depicted in Fig. \ref{fig:time_accuracy}, revealing that \textsc{Duplex} outperforms the baselines in terms of reducing the completion time required to attain the target accuracy. For instance, as illustrated in Fig. \ref{fig:reddit_time_accuracy}, 
\textsc{Duplex} takes 93 minutes to achieve the target accuracy of 85\% for GCN on Reddit, while D-FedGraph(dense), D-FedPNS(dense), Glint(0.5) and TDGE(0.7) take 185 minutes, 180 minutes, 121 minutes and 147 minutes, respectively. 
Besides, for GraphSage on ogbn-products as shown in
Fig. \ref{fig:ogbn-products_time_accuracy}, \textsc{Duplex} speeds up training by about 1.4$\times$, 1.3$\times$, 1.6$\times$ and 1.3$\times$, compared with Glint, D-FedGraph(dense), D-FedPNS(dense) and TDGE(0.7), respectively. These improvements highlight the efficiency gains enabled by \textsc{Duplex} in terms of accelerating the training process of DFGL by optimizing the network topology and the
graph sampling ratio.\\
\textbf{Communication Cost.} We measure the required communication cost of \textsc{Duplex} and the baselines to attain the same target accuracy on the three datasets. The results in Fig. \ref{fig:traffic_accuracy} show that \textsc{Duplex} outperforms the baselines in terms of communication cost. Specifically, by Fig. \ref{fig:arxiv_traffic_accuracy}, \textsc{Duplex} reduces the communication cost by 26\%, 33\%, 36\% and 16\%, respectively, compared with Glint(0.7), D-FedGraph(dense), D-FedPNS(dense) and TDGE(0.5) on ogbn-arxiv to reach the accuracy of 50\%. Besides, by Fig. \ref{fig:reddit_traffic_accuracy}, \textsc{Duplex} reduces the communication cost on Reddit by 16\%, 39\%, 35\% and 24\%, respectively, compared with Glint, D-FedGraph(dense), D-FedPNS(dense) and TDGE(0.7) to reach the accuracy of 85\%. 
These results indicate that \textsc{Duplex} is highly scalable to effectively reduce communication cost across a variety of different datasets.
\begin{figure}[t]\centering
    \begin{minipage}[t]{0.49\linewidth}\centering
        \subfigure[Reddit.]{\centering
            \label{fig:noniid-acc-reddit}
            \includegraphics[width=1.0\textwidth]{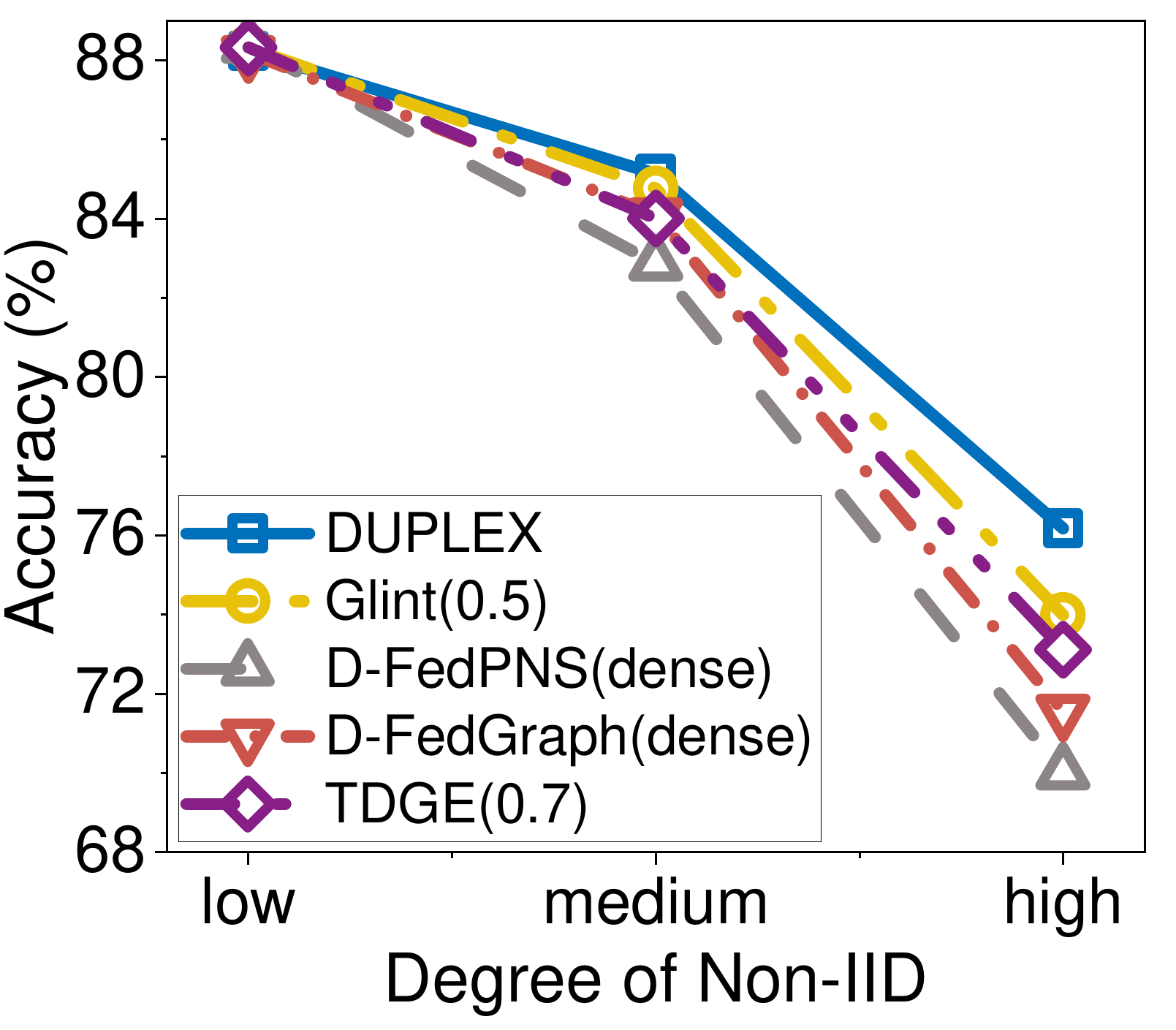}
        }
    \end{minipage}
    \begin{minipage}[t]{0.49\linewidth}\centering
        \subfigure[ogbn-products.]{\centering
            \label{fig:noniid-acc-products}
            \includegraphics[width=1.0\textwidth]{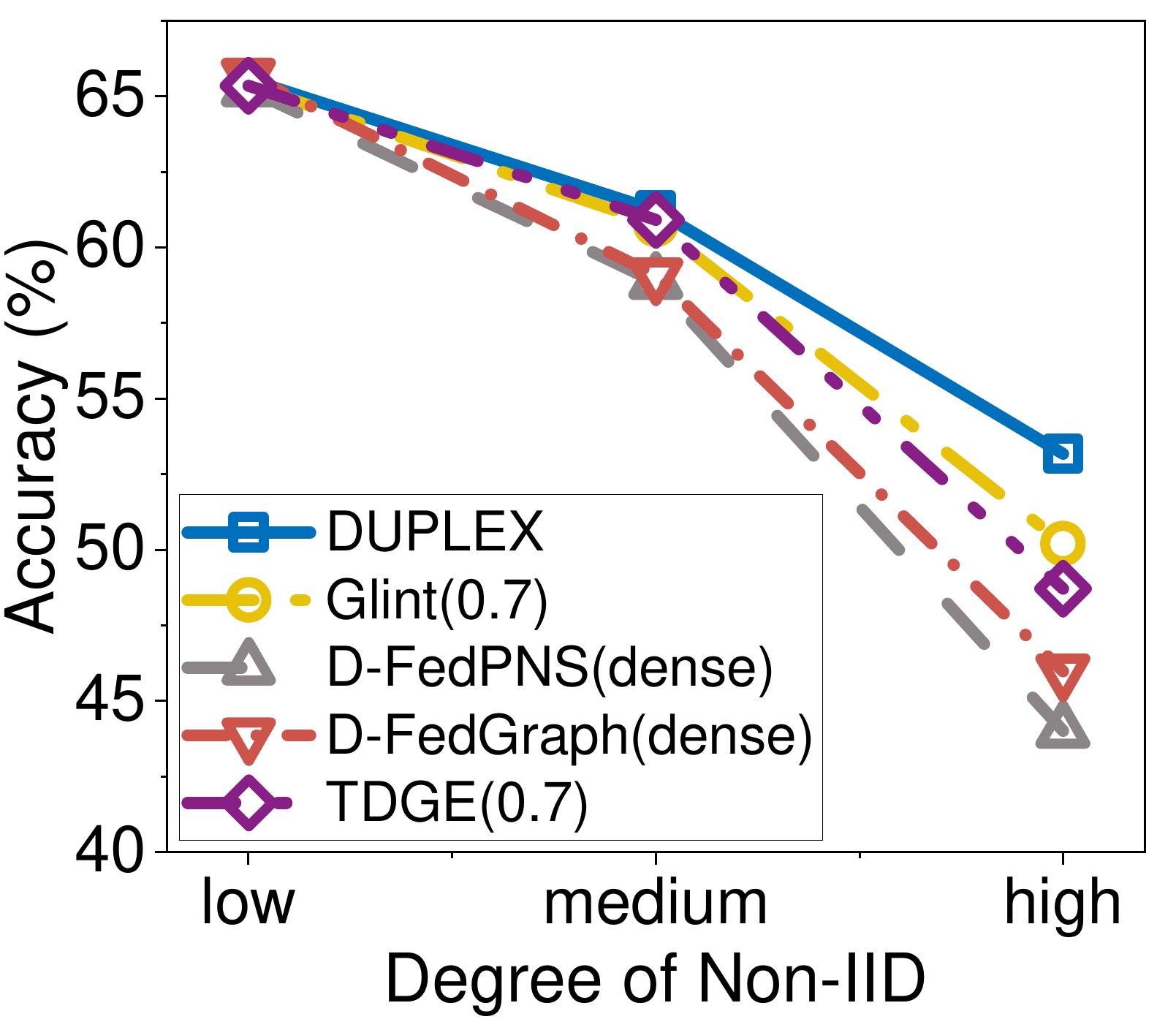}
        }
    \end{minipage}
    \caption{Impact of non-IID data on model performance.}\label{fig:noniid-acc}
\end{figure}
\begin{figure}[t]\centering
    \begin{minipage}[t]{0.49\linewidth}\centering
        \subfigure[Reddit.]{\centering
            \label{fig:noniid-comm-reddit}
            \includegraphics[width=1.0\textwidth]{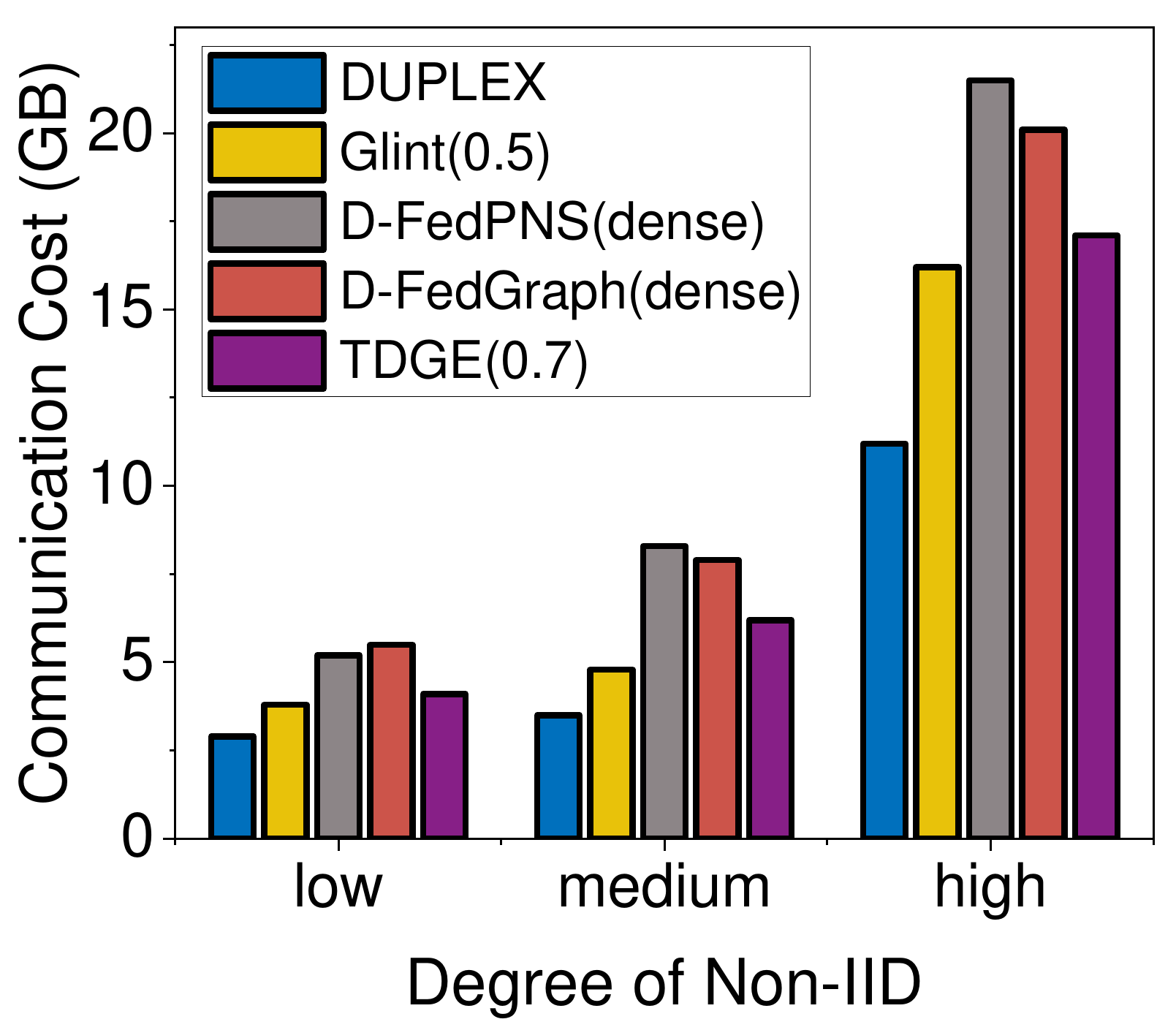}
        }
    \end{minipage}
    \begin{minipage}[t]{0.49\linewidth}\centering
        \subfigure[ogbn-products.]{\centering
            \label{fig:noniid-comm-products}
            \includegraphics[width=1.0\textwidth]{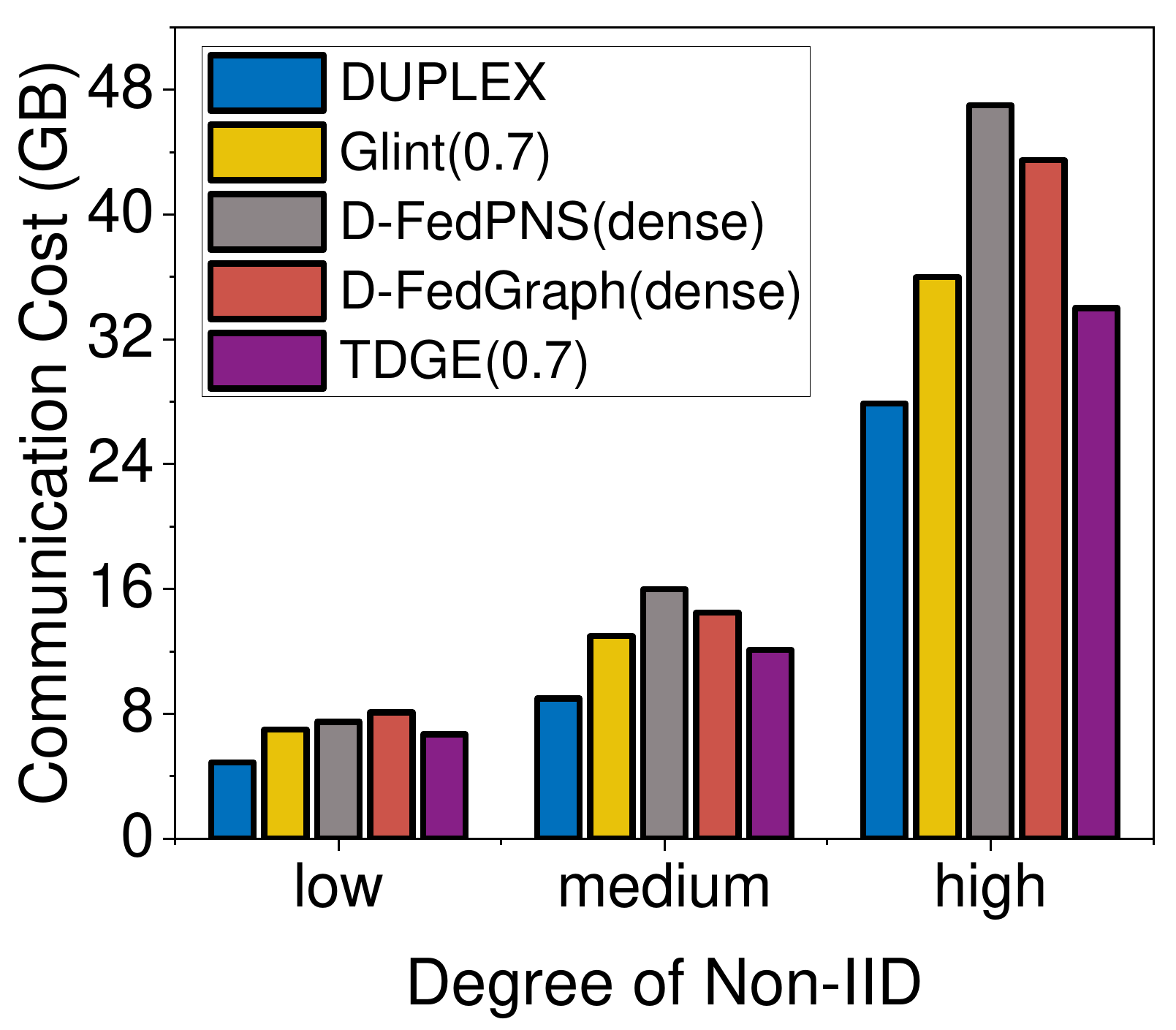}
        }
    \end{minipage}
    \caption{Impact of non-IID data on communication cost.}\label{fig:noniid-comm}
\end{figure}
\begin{table}[t]
    \centering
    \caption{Test accuracy (\%) of native \textsc{Duplex} and the breakdown versions on the three datasets.} \label{table:abla_test_accuracy}
    \resizebox{89mm}{!}{
        \centering
        \begin{tabular}{c|c|cccccc}
            \toprule
             \multirow{2}*{Datasets} & \multirow{2}*{\textsc{Duplex}} & \multicolumn{6}{c}{breakdown versions of \textsc{Duplex}} \\ [0.5ex]
             &  & \cellcolor{lightgray}ring & \cellcolor{lightgray}sparse & \cellcolor{lightgray}dense & \cellcolor{lightgray}0.3 & \cellcolor{lightgray}0.5 & \cellcolor{lightgray}0.7 \\ [0.1ex] \\ [-2.5mm]
            \midrule \\ [-3.5mm]

            
            \multicolumn{1}{c|}{ogbn-arxiv} & \textbf{53.21} & 43.45 & 48.97 & \textbf{52.81} & 48.23 & 51.87 & \textbf{53.25} \\ [0.5ex]

            \midrule \\ [-3.5mm]

            
            \multicolumn{1}{c|}{Reddit} & \textbf{88.23} & 77.84 & 82.19 & \textbf{87.96} & 84.63 & 87.07 & \textbf{88.86} \\ [0.5ex]

            \midrule \\ [-3.5mm]
            
            
            \multicolumn{1}{c|}{ogbn-products} & \textbf{65.57} & 55.26 & 61.96 & \textbf{65.49} & 63.23 & \textbf{65.23} & 66.23 \\ [0.5ex]
            
            
            \bottomrule
        \end{tabular}
    }
\end{table}
\begin{table}[t]
    \centering
    \caption{Test accuracy (\%) of different versions of \textsc{Duplex} on ogbn-arxiv and Reddit under the communication resource constraint of 1.5GB and 10GB, respectively.} \label{table:abla_final_acc}
    \resizebox{83mm}{!}{
        \centering
        \begin{tabular}{c|c|cc}
            \toprule
            Datasets & \textsc{Duplex} & \textsc{Duplex}(dense) & \textsc{Duplex}(0.7) \\ [0.3ex]
            
            \midrule & \\ [-2.8mm]

            ogbn-arxiv & \textbf{53.25} & 48.01 & 49.78 \\ [0.3ex]

            Reddit & \textbf{88.25} & 83.58 & 86.12 \\ [0.4ex]
            
            
            \bottomrule
        \end{tabular}
    }
\end{table}
\subsection{Impact of non-IID data}
In this section, we evaluate the performance of \textsc{Duplex} and the baselines under different degrees of data heterogeneity. Specifically, we train models on three Dirichlet distributions with $\alpha=\{10.0, 1.0, 0.1\}$, where a smaller $\alpha$ denotes a higher non-IID degree. As shown in Fig. \ref{fig:noniid-acc}, \textsc{Duplex} achieves more robust performance on non-IID data compared to the baselines. Although all algorithms suffer from model accuracy degradation with a higher degree of data heterogeneity, \textsc{Duplex} performs significantly more stably than the baselines. For example, as the non-IID degree increases (with $\alpha$ decreasing from $10.0$ to $0.1$) on ogbn-products, the model accuracy of Glint(0.7), D-FedPNS(dense), D-FedGraph(dense) and TDGE(0.7) decreases dramatically by 15.2\%, 21.2\%, 19.7\% and 16.6\%, respectively, while that of \textsc{Duplex} only degrades by 11.4\%. 

Besides, we observe that \textsc{Duplex} also improves resource efficiency on non-IID data. Fig. \ref{fig:noniid-comm} shows the communication costs for model training to reach the target accuracy of 70\% and 45\% on Reddit and ogbn-products, respectively. With an increase of non-IID degrees, all algorithms incur higher communication costs. However, the cost of \textsc{Duplex} grows much more slowly than that of the baselines. Specifically, under different non-IID degrees, \textsc{Duplex} saves about 22.5\%-40.6\% in communication costs compared to the baselines. These results suggest that the joint optimization of network topology and graph sampling, guided by consensus distance and training loss, enables \textsc{Duplex} to effectively alleviate the negative impact of non-IID data.
\begin{figure}[t]\centering
    \begin{minipage}[t]{0.49\linewidth}\centering
        \subfigure[ogbn-arxiv.]{\centering
            \label{fig:abla_arxiv_time_accuracy}
            \includegraphics[width=0.9998\textwidth]{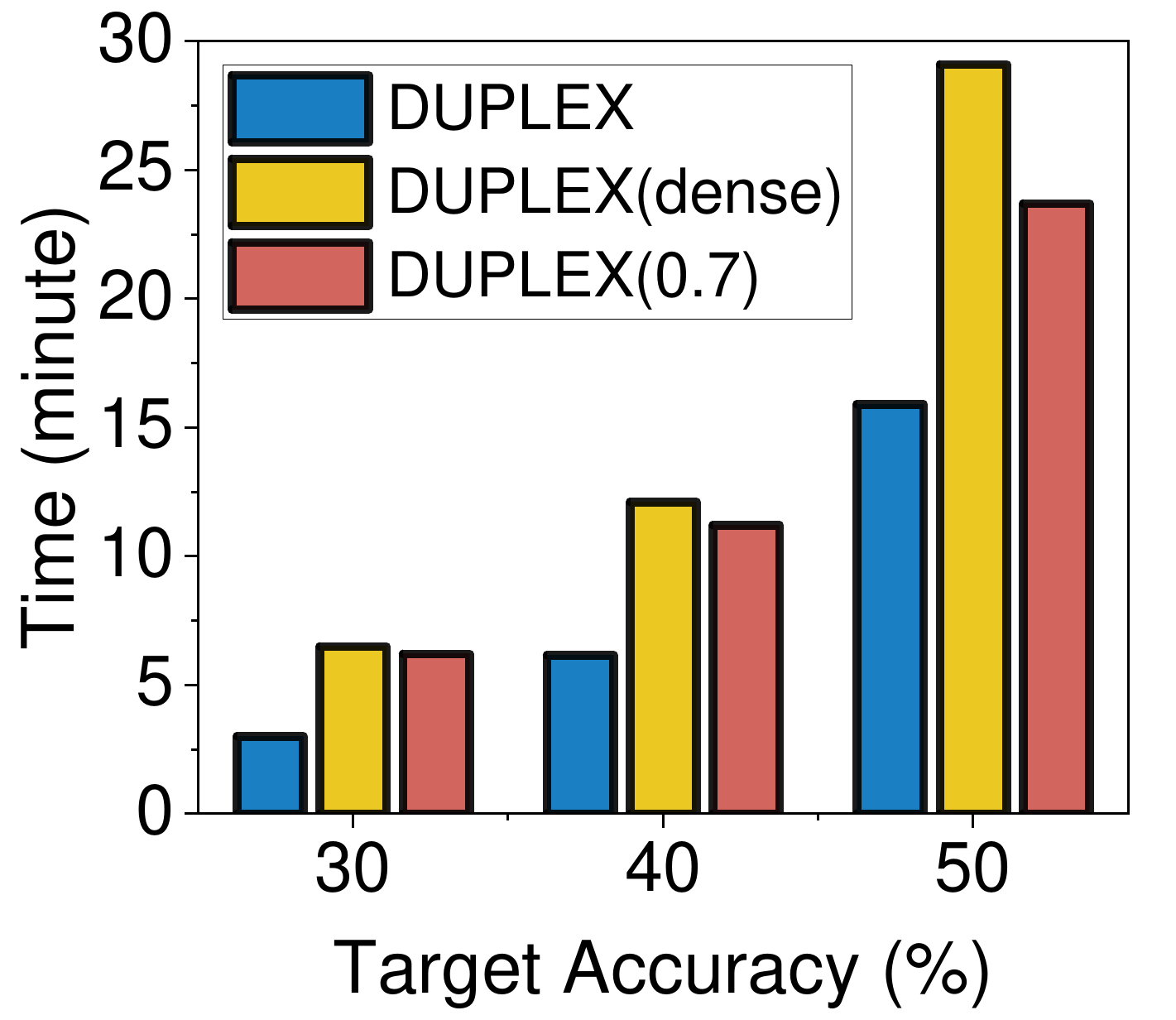}
        }
    \end{minipage}
    \begin{minipage}[t]{0.49\linewidth}\centering
        \subfigure[Reddit.]{\centering
            \label{fig:abla_reddit_time_accuracy}
            \includegraphics[width=0.9998\textwidth]{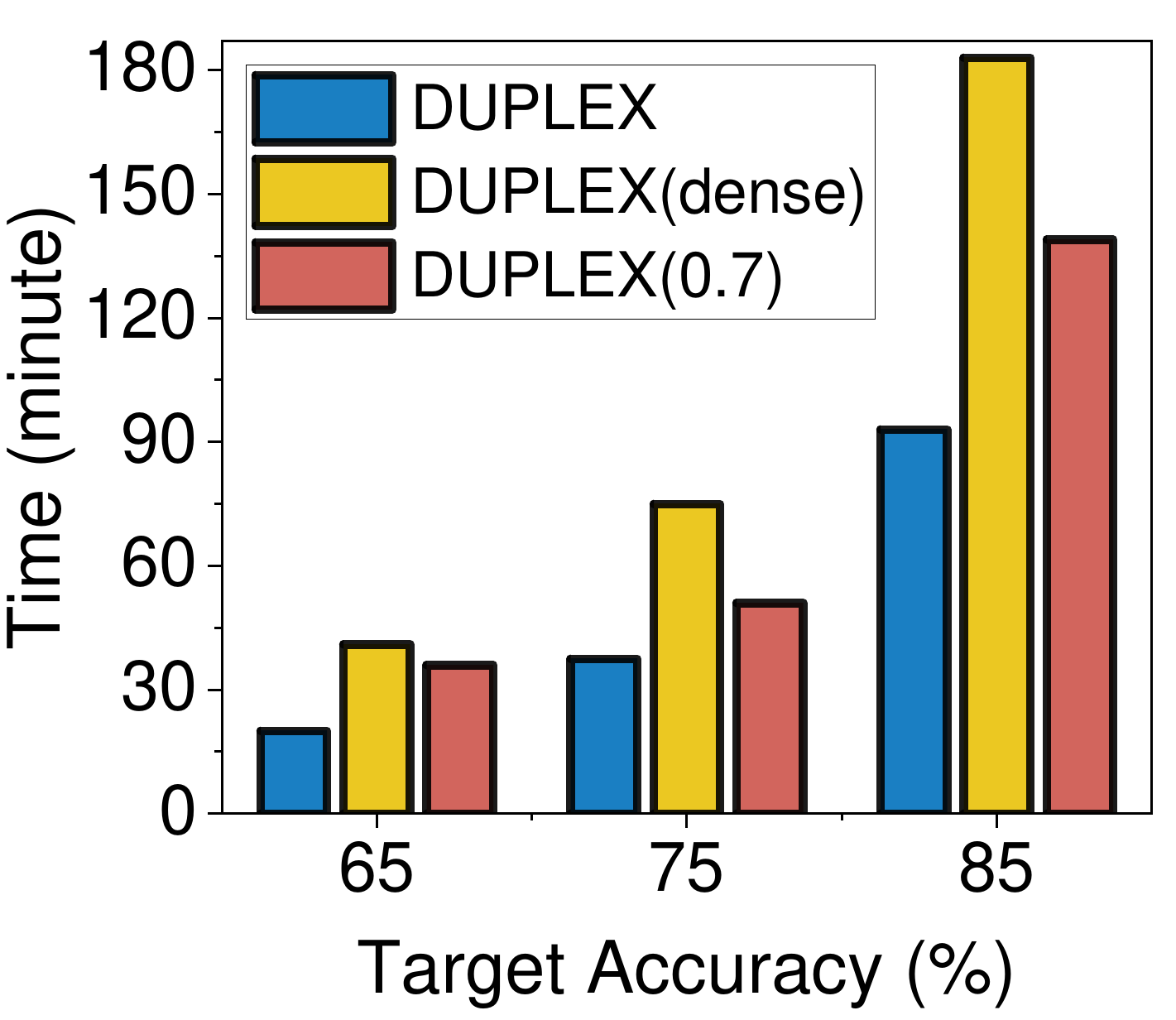}
        }
    \end{minipage}
    \caption{Training time of different versions of \textsc{Duplex} to reach the target accuracy. The breakdown versions of \textsc{Duplex} are selected from Table \ref{table:abla_test_accuracy}, where their final test accuracy (bold) is close to that of native \textsc{Duplex}.}\label{fig:abla_time_accuracy}
\end{figure}
\begin{figure}[t]\centering
    \begin{minipage}[t]{0.49\linewidth}\centering
        \subfigure[ogbn-arxiv.]{\centering
            \label{fig:abla_arxiv_commun_accuracy}
            \includegraphics[width=0.9998\textwidth]{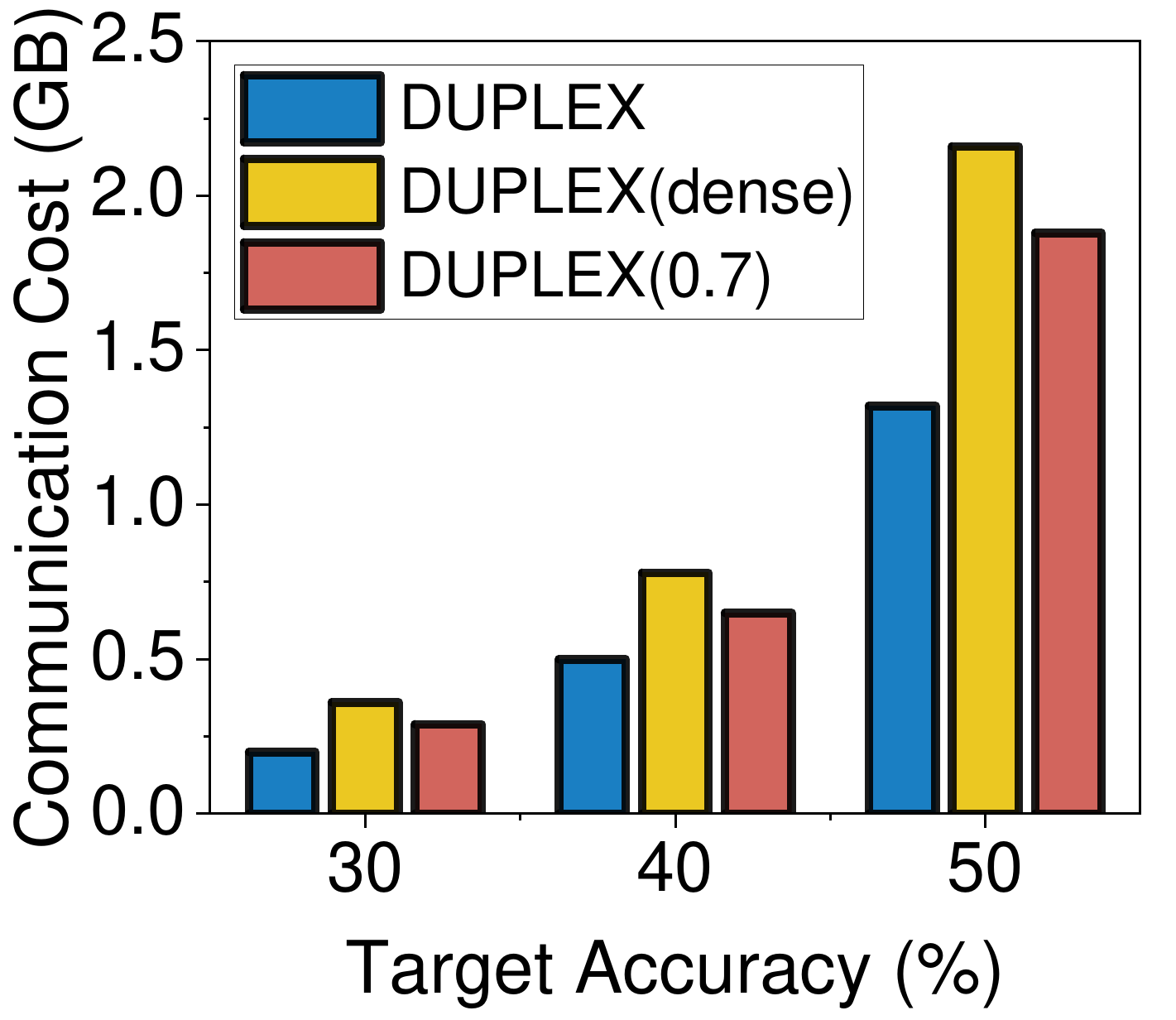}
        }
    \end{minipage}
    \begin{minipage}[t]{0.49\linewidth}\centering
        \subfigure[Reddit.]{\centering
            \label{fig:abla_reddit_commun_accuracy}
            \includegraphics[width=0.9998\textwidth]{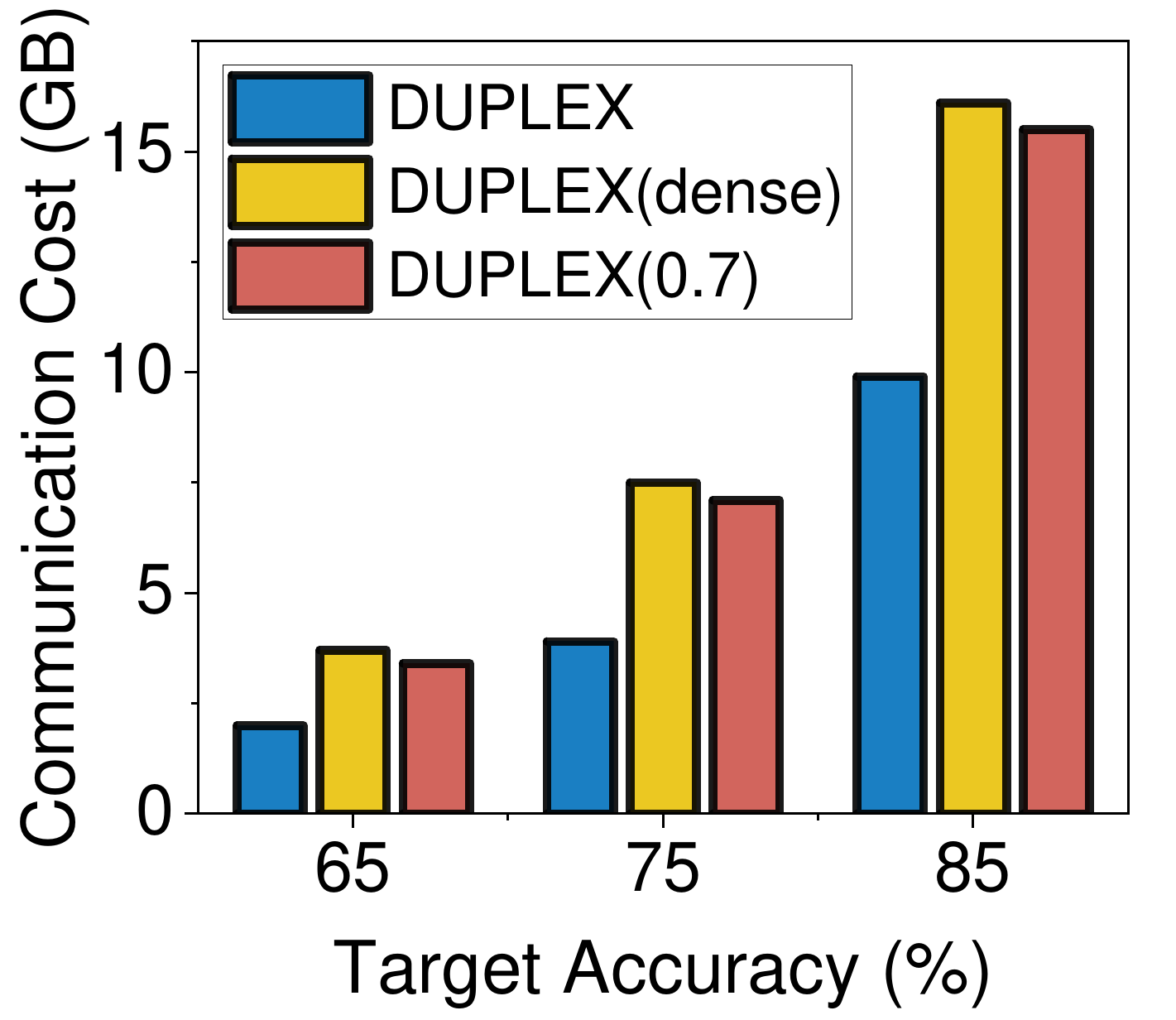}
        }
    \end{minipage}
    \caption{Communication cost of different versions of \textsc{Duplex} to reach the target accuracy.}\label{fig:abla_commun_accuracy}
\end{figure}

\subsection{Ablation Study}
In this section, we implement several breakdown versions of \textsc{Duplex} to evaluate the effectiveness of topology construction and graph neighbor sampling in \textsc{Duplex}, respectively.
\begin{itemize}
    \item \textbf{\textsc{Duplex} with fixed network topology.} We disable the adaptive network topology construction in \textsc{Duplex}, in which workers communicate with others based on the fixed network topology without changing the graph sampling ratios in native \textsc{Duplex}. We formulate various
    breakdown versions of \textsc{Duplex}, \ie, \textsc{Duplex}(ring), \textsc{Duplex}(sparse) and \textsc{Duplex}(dense), by training \textsc{Duplex} on different network topologies.

    \item \textbf{\textsc{Duplex} with fixed graph sampling ratio.} We disable the adaptive graph neighbor sampling in \textsc{Duplex}, in which workers conduct graph neighbor sampling according to the fixed sampling ratio and communicate with others based on the network topology output by deep reinforcement learning. We formulate various breakdown versions of \textsc{Duplex}, \ie, \textsc{Duplex}(0.3), \textsc{Duplex}(0.5) and \textsc{Duplex}(0.7), by training \textsc{Duplex} according to different sampling ratios.
\end{itemize}

\noindent\textbf{Test Accuracy.}
Table \ref{table:abla_test_accuracy} shows the test accuracy of native \textsc{Duplex} (\ie, \textsc{Duplex} with optimized network topology construction and graph neighbor sampling) and the breakdown versions on the three datasets. Without adaptive network topology construction or graph sampling, the test accuracy decreases significantly on all datasets. For example, on Reddit, \textsc{Duplex} with the sparse topology and \textsc{Duplex} with fixed graph sampling ratio of 0.3 achieve 6.04\% and 3.60\% lower final accuracy than that of native \textsc{Duplex}. Besides, as shown in Table \ref{table:abla_final_acc}, under the same communication cost constraint (1.5GB on ogbn-arxiv and 10GB on ogbn-products), native \textsc{Duplex} is able to visibly improve test accuracy by 2.13\%-5.24\% on ogbn-arxiv and Reddit, compared with the breakdown versions of \textsc{Duplex}, \ie, \textsc{Duplex}(dense) and \textsc{Duplex}(0.7). These results indicate that the joint optimization of network topology and graph sampling strategy contributes to the accuracy improvements of \textsc{Duplex}.

\begin{figure*}[t]\centering
    \includegraphics[width=1.0\textwidth]{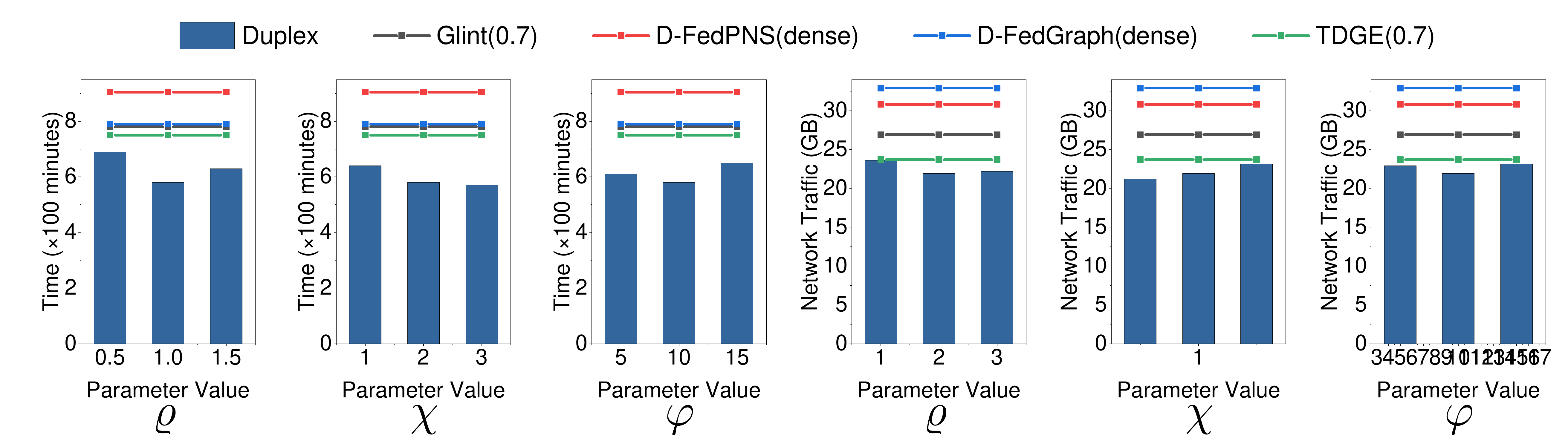}
    \caption{\textsc{Duplex} outperforms baselines with different parameters.}\label{fig:sensitivity}
\end{figure*}

\noindent\textbf{Training Time and Communication Cost.}
As illustrated in Fig. \ref{fig:abla_time_accuracy}, the training time for breakdown versions of \textsc{Duplex} to reach the target accuracy on ogbn-arxiv (50\%) and Reddit (85\%) increases 49.1\%-96.8\%, compared to native \textsc{Duplex}. Besides, Fig. \ref{fig:abla_commun_accuracy} evaluates the effect of the network topology construction and graph sampling on the communication cost. For instance, compared to native \textsc{Duplex}, \textsc{Duplex}(dense) and \textsc{Duplex}(0.7) needs 63.6\% and 42.4\% more communication cost to reach the target accuracy of 50\% on ogbn-arxiv. By jointly constructing the dynamic network topology and assigning proper graph sampling ratios to workers, \textsc{Duplex} not only effectively accelerates training but also reduces communication costs.

\noindent\textbf{Analysis.} The performance degradation caused by fixed network topologies or sampling ratios stems from three key limitations. First, fixed topologies or sampling ratios cannot adapt to fluctuating network bandwidth across workers. For example, slow or congested workers become bottlenecks, delaying synchronization and model training. Besides, dynamic topologies and sampling ratios enable broader exchanges of model parameters and node embeddings by periodically reconnecting workers, fostering model diversity and avoiding local optima. Fixed topologies and sampling ratios restrict information flow, leading to suboptimal convergence and model accuracy on non-IID data. In addition, fixed ratios and topologies under-share data points in early training phases and over-share instances in later phases.

\subsection{Sensitivity Analysis}
To evaluate the impact of the parameters $\chi$, $\varrho$ and $\varphi$ in the reward function, we vary each parameter independently while fixing the others to default values. $\chi$ penalizes the reward as the round time becomes longer. A larger $\chi$ provides a more severe penalize on increase of round time.
$\varrho$ prioritizes the descent of consensus distance. A larger $\varrho$ promotes reduction of discrepancies between workers' local models, improving the training performance on non-IID data. 
$\varphi$ prioritizes the minimizing of the training loss. A larger $\varphi$ promotes the convergence of local models. Fig. \ref{fig:sensitivity} shows that \textsc{Duplex} outperforms baselines consistently in terms of completion time and network traffic to reach the target accuracy across different $\chi$ (\eg, 1, 2, 3), $\varrho$ (\eg, 0.5, 1, 1.5), and $\varphi$ (\eg, 5, 10, 15). However, some extreme values of these parameters, \eg, 10 for $\chi$ and $\varrho$, $50$ for $\varphi$, lead to significant performance degradation. Therefore, we recommend a
conservative parameter setting (\eg, $\chi=2$, $\varrho=1$, $\varphi=10$) for most scenarios.

\subsection{Scalability Evaluation}
To evaluate the scalability of \textsc{Duplex} in complex graph processing, we train the GraphSAGE model on a complex graph dataset, \ie, ogbn-mag\footnote{Ogbn-mag is \emph{a heterogeneous academic network} comprising four types of entities, \ie, papers (736,389 nodes), authors (1.13 million nodes), institutions (8,740 nodes), and fields of study (59,965 nodes), interconnected by four types of directed relationships, \ie, authors affiliating with institutions, authors writing papers, papers citing other papers, and papers belonging to research fields. The primary task is 349-class classification, where the goal is to predict the venue of each paper based on its content, citations, authors, and institutional affiliations. With over 21 million edges and 1.93 million nodes, ogbn-mag poses challenges in handling heterogeneous information fusion, scalability, and long-range dependencies in academic networks. 
} \cite{hu2020open}, which is a large-scale and heterogeneous academic graph derived from a subset of the Microsoft Academic Graph (MAG). We test the completion time and network traffic required for different methods to reach the same target accuracy (\ie, 42\%) at different worker scales (\eg, 50, 100, 200, 300, 400, 500).

\begin{figure}[t]\centering
    \begin{minipage}[t]{0.485\linewidth}\centering
        \subfigure[Completion time vs. Number of workers.]{
            \includegraphics[width=1.0\textwidth]{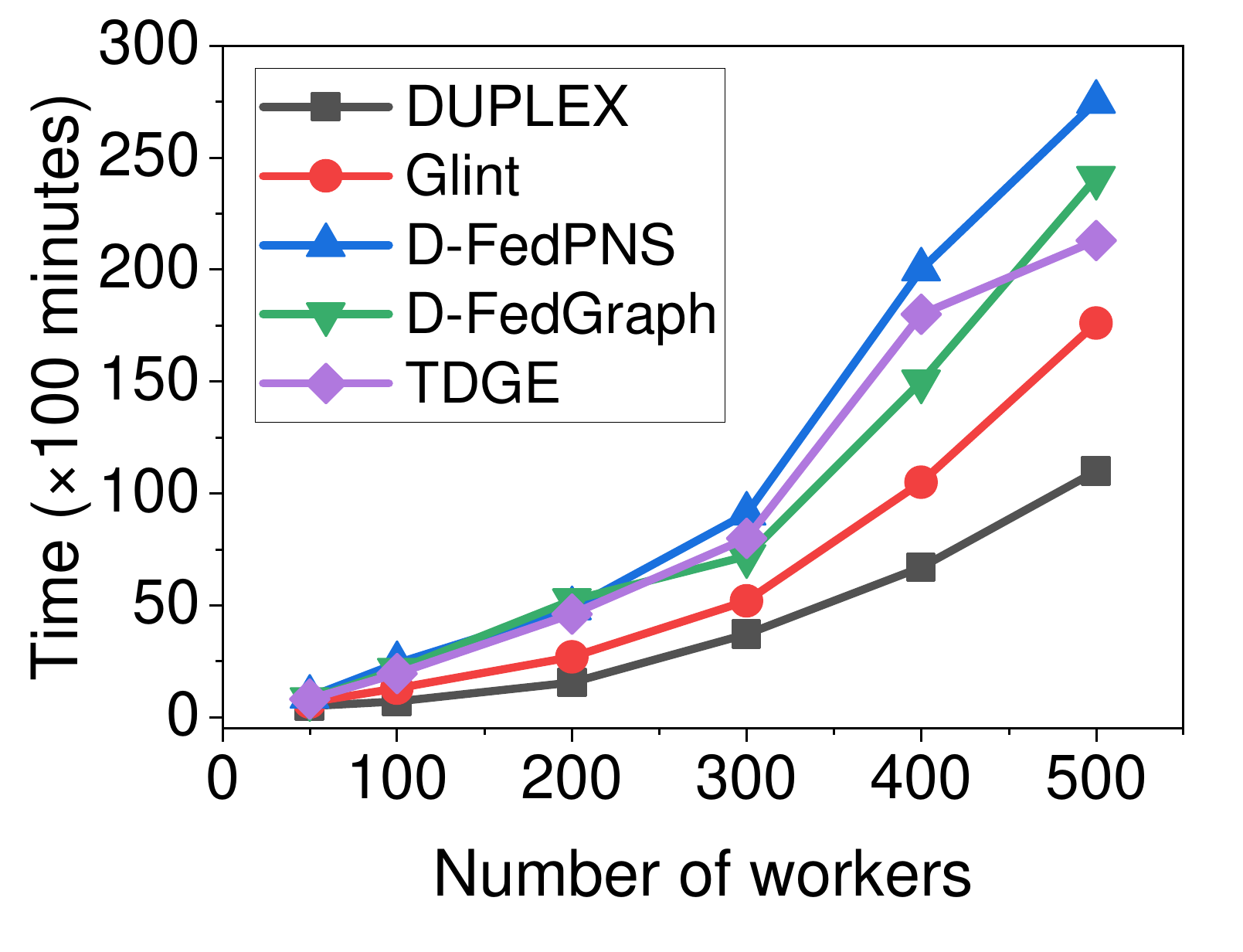}\label{fig:scalability_time}
        }
    \end{minipage}
    \begin{minipage}[t]{0.485\linewidth}\centering
        \subfigure[Network traffic vs. Number of workers.]{
            \includegraphics[width=1.0\textwidth]{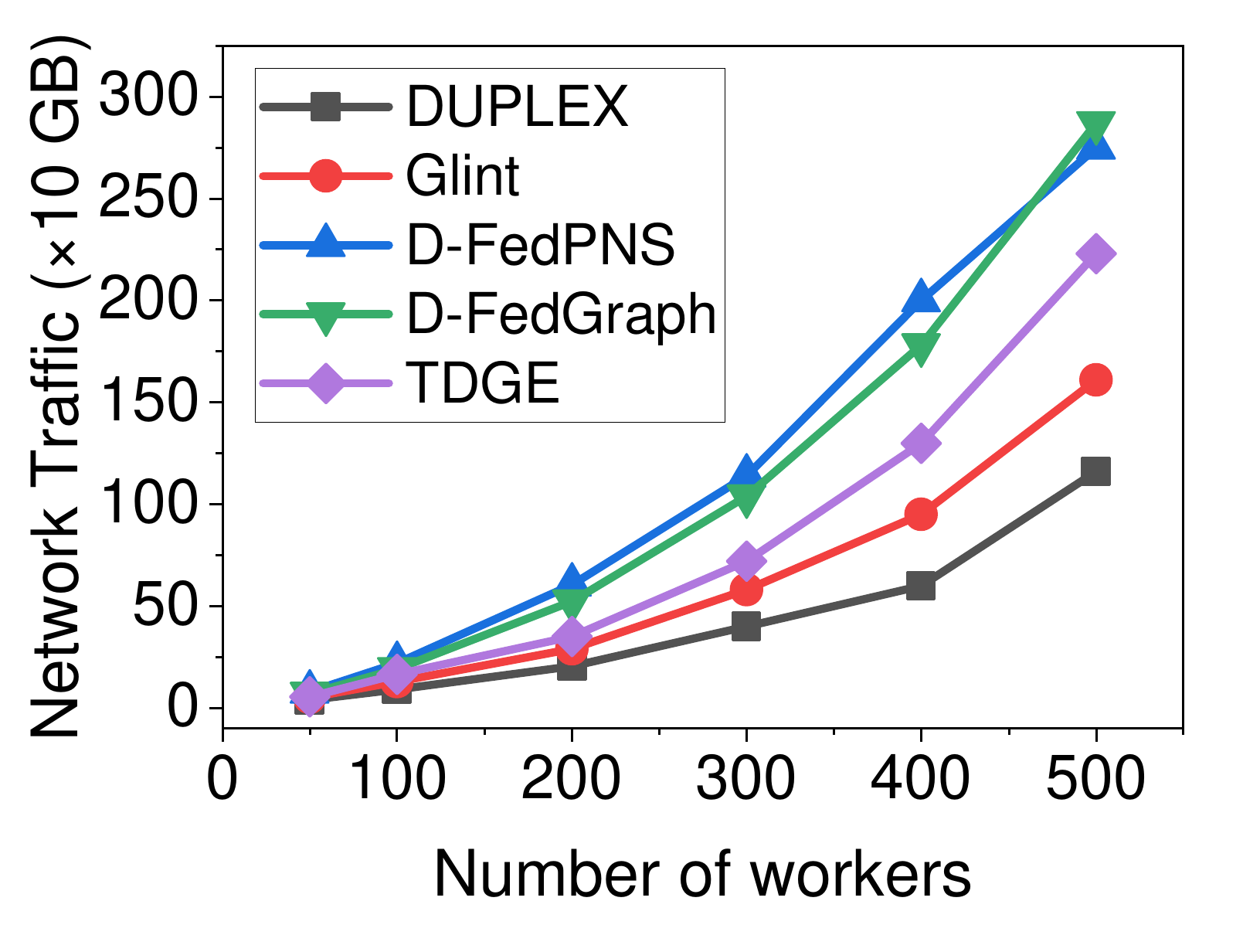}\label{fig:scalability_comm}
        }
    \end{minipage}
    \caption{Completion time and network traffic on different worker scales.}\label{fig:scalability}
\end{figure}

As shown in Fig. \ref{fig:scalability}, the results show that the completion time and network traffic required for \textsc{Duplex} increase much slowly with the increasing scale of workers, compared to baselines. For instance, when the worker scale increases from 50 to 500, the completion time/network traffic growth rate of \textsc{Duplex} is 22.1$\times$/28.3$\times$, while the growth rates of Glint, D-FedGraph, TDGE, and D-FedPNS are 25.9$\times$/31.1$\times$, 29.8$\times$/40.2$\times$, 26.1$\times$/40.5$\times$, and 30.2$\times$/34.4$\times$, respectively. These results highlight the scalability of \textsc{Duplex} in complex graph processing with large-scale workers.

%% file: content/related.tex
Graph Neural Networks (GNNs) \cite{corso2024graph, sharma2024survey} have revolutionized the processing of graph-structured data, where data samples are interconnected through graph topologies. 
Unlike traditional neural networks such as Convolutional Neural Networks \cite{zhao2024review}, which are tailored for grid-like data structures, GNNs excel at extracting neighborhood information from the underlying graph topology. 
This capability enables them to efficiently capture and learn the intricate, structured knowledge inherent in graph data. 
However, training high-quality GNNs necessitates access to vast amounts of graph data, which in real-world scenarios is often distributed across numerous edge devices \cite{liu2024federated}. 
Privacy concerns pose significant barriers to aggregating this data centrally, as data holders are reluctant to share their private information.\\
\textbf{Federated Graph Learning.} FGL \cite{liu2024federated, huang2024federated}
has emerged as a promising solution to this challenge through a distributed and collaborative training paradigm where each data holder (\ie, the worker) participates in the training process using their local graph data. 
This approach not only preserves data privacy but also enhances security by sharing only model parameters rather than raw data. 
Current research in FGL can be broadly classified into three scenarios: 1) each data holder owns only a single node within the global graph \cite{he2023hierarchical}; 
2) each data holder has a subgraph of a global graph \cite{chen2024fedgl, wan2024federated, lei2023federated};
and 3) each data holder possesses multiple independent graphs \cite{feng2024federated}. 
The second scenario is the most prevalent in real-world applications and has become the mainstream focus in FGL \cite{liu2024federated}. 
The optimization efforts in this article are also based on this scenario. 
In this context, while each data holder manages their own subgraph, edges exist that connect different subgraphs, representing structural relationships between data holders. 
The primary objective is for each data holder to leverage their subgraph data to collaboratively train the global GNN model, enhancing overall performance through shared learning while maintaining data privacy.\\
\textbf{Decentralized Federated Graph Learning.} 
One of the primary concerns in traditional worker-server-based FGL is the single point of failure and limited scalability \cite{beltran2023decentralized}. To address this, Pei \etal \cite{pei2021decentralized} proposed a new decentralized federated graph learning framework (D-FedGNN), which allows multiple data holders to train GNN models in a peer-to-peer network without a centralized server. Another method, SpreadGNN \cite{he2022spreadgnn}, extends FGL to realistic serverless and multi-task settings, and utilizes a novel optimization algorithm to ensure the convergence of GNN models. 
Moreover, DCI-PFGL \cite{xie2023dci} introduces a cross-institutional framework for IoT service recommendation, employing anonymized graph feature embeddings and smart contracts on a blockchain to maintain privacy while addressing data heterogeneity. 
Efficient communication is crucial for scaling decentralized FGL, with recent studies emphasizing communication cost reduction.
For example, Glint \cite{liu2021glint}  and S-Glint \cite{liu2022s} are novel DFGL frameworks that leverage traffic throttling and flow scheduling to alleviate communication bottlenecks in DFGL tasks, leading to enhanced service quality for graph learning.\\
\textbf{Graph Neighbor Sampling.} The recent advancements in graph neighbor sampling within FGL address several pressing challenges, including computational efficiency, communication overhead, and representation accuracy.
For example, FedGraph \cite{chen2021fedgraph} uses deep reinforcement learning to dynamically assign a distinct neighbor sampling ratio for each worker in large-scale federated graphs, addressing issues such as privacy leakage and training overhead. 
This method dynamically balances training accuracy and speed, making it viable for privacy-sensitive and resource-intensive applications.
Du \etal \cite{du2022federated} optimized neighbor sampling intervals to balance convergence accuracy with runtime efficiency. 
By periodically sampling neighbors, this algorithm mitigates resource constraints without compromising model accuracy.
FedAAS \cite{li2024historical}, developed by Li \etal, utilizes historical embeddings to enhance sampling efficiency by focusing on nodes with significant impact, reducing communication costs while maintaining model accuracy in FGL. 
This method addresses scalability concerns inherent in large graph networks.
Additionally, FedAIS \cite{li2024federated} focuses sampling efforts on high-importance nodes in the graph to adapt to structural heterogeneity. 
This importance-based sampling method improves both the efficiency and scalability of FGL by concentrating on influential nodes.
However, these works mainly focus on PS-based FGL, which suffers from the single point of failure problem as the PS may be down.\\
\textbf{Topology Construction.}
A key objective in decentralized federated learning (DFL) is achieving efficient communication and high model performance without a parameter server, where the topology of the peer-to-peer network plays a pivotal role. 
For example, cluster topologies are advantageous in leveraging data distribution similarities, which help reduce communication costs while maintaining performance comparable to centralized models.
Wireless ring networks present another approach, using carefully designed consensus coefficients and network coding to balance learning efficacy with communication latency. 
DFGL, a special variant of DFL, shares the same goals as DFL but faces unique challenges. Specifically, DFGL requires extensive communication of graph node embeddings among workers during training, leading to high communication overhead.
To address this, Liu \etal \cite{liu2022s} proposed constructing a fixed network topology for worker communication, selectively transmitting only critical information and thus significantly reducing network traffic. 
Krasanakis \etal \cite{krasanakis2022p2pgnn} also advanced DFGL by developing a peer-to-peer GNN model for decentralized node classification. 
Their method integrate asynchronous diffusion and graph-based communication mechanisms, reducing dependency on centralized data exchanges while sustaining model accuracy.


\noindent\textbf{Non-IID Handling Strategies.} The non-IID problem in FGL manifests in two principal dimensions, \ie, data distribution heterogeneity and graph topology heterogeneity. The former encompasses heterogeneity in node/edge attributes (\eg, feature skews in social network graphs) and labels distributions (\eg, class imbalance in molecular property prediction tasks). The later arises from structural divergences, including variations in node degrees, edge weights, or global connectivity patterns. Most FGL research focuses on data distribution heterogeneity \cite{tan2022towards, liu2024federated}, providing some personalized FL-inspired solutions such as regularized local loss, knowledge distillation,  clustering, and personalized aggregation. For instance, FedAlign \cite{lin2020improving} introduces an optimal transport distance-based regularization term between local and global models to minimize model divergence.  InfoFedSage \cite{guo2023information} employs a PS-side generative module coupled with worker-side information bottleneck regularization to alleviate the non-IID problem. GCFL \cite{xie2021federated} partitions workers into clusters using Graph Isomorphism Network (GIN) \cite{xu2018powerful}, enabling cluster-specific model aggregation.
However, existing works' solutions primarily adopt the PS architecture and fail to address the high communication overhead in DFGL. Our framework bridges these gaps through consensus distance-guided joint optimization of network topology and graph sampling, improving both training performance and communication efficiency in non-IID DFGL environments.

%% file: content/limitations.tex
\textsc{Duplex} relies on a logical coordinator to collect global states and compute configurations. While the coordinator is fully different from the PS that needs to aggregate local model parameters and update the global model, its transient failure could temporarily disrupt adaptation of network topology and graph sampling. To eliminate single-point dependencies, we are integrating a leader election protocol (\eg, Raft consensus) among workers to dynamically re-elect a coordinator during failures.
Further, we are exploring decentralized consensus algorithms (\eg, blockchain-inspired Proof-of-Stake) to distribute coordinator responsibilities across workers. This aligns with our goal of minimizing centralization while retaining the efficiency benefits of coordinator-guided topology and sampling optimization.
In addition, the synchronous training paradigm in \textsc{Duplex} simplifies deployment but may delay aggregation in heterogeneous edge environments due to stragglers. To address this, we plan to extend \textsc{Duplex} with asynchronous aggregation protocols and staleness-aware model exchange strategies to accommodate heterogeneous worker capabilities. Specifically, stragglers with stale model updates (beyond a staleness threshold) will asynchronously trigger local aggregation with neighbors, decoupling them from the global synchronization barrier. Besides, the coordinator will prioritize workers with lower staleness for configuration updates, ensuring timely adaptation of network topology and sampling ratios while accommodating slower workers.

%% file: content/conclusion.tex
In this work, we propose an efficient and effective DFGL framework, named \textsc{Duplex}, which integrates network topology construction and graph neighbor sampling to address the challenge of non-IID graph data and reduce communication costs. 
We also propose an efficient learning-driven algorithm to adaptively determine the optimal topology and sampling ratios for workers, considering resource budgets and data distributions, simultaneously.
We build a simulated DFGL environment to evaluate the performance of \textsc{Duplex} via extensive experiments. The results significantly demonstrate the effectiveness and efficiency of \textsc{Duplex}.